\documentclass[fleqn,usenatbib]{mnras}

\usepackage{newtxtext,newtxmath}
\usepackage[T1]{fontenc}
\usepackage{ae,aecompl}
\usepackage{graphicx}
\usepackage{changepage}
\usepackage{adjustbox}
\usepackage{rotating}

\usepackage{graphicx}	% Including figure files
\usepackage{lscape}
\usepackage{longtable}
\usepackage{mathtools}

\title[The non-linear IRRC of low-z galaxies]{The non-linear infrared-radio correlation of low-z galaxies:  implications for redshift evolution, a new radio SFR recipe, and how to minimize selection bias}

\author[D. Cs. Moln\'{a}r et al.]{
D\'{a}niel Cs. Moln\'{a}r,$^{1,2,3}$\thanks{E-mail: daniel.molnar@inaf.it}
Mark T. Sargent,$^{1}$
Sarah Leslie,$^{4,3}$
Benjamin Magnelli,$^{5}$
\newauthor
Eva Schinnerer,$^{3}$
Giovanni Zamorani,$^{6}$
Jacinta Delhaize,$^{7}$
Vernesa Smol\v{c}i\'{c},$^{8}$
\newauthor
Kre\v{s}imir Tisani\'{c},$^{8}$
Eleni Vardoulaki$^{5,9,10}$
\\
$^{1}$Astronomy Centre, Department of Physics \& Astronomy, University of Sussex, Brighton, BN1 9QH, England\\
$^{2}$INAF - Osservatorio Astronomico di Cagliari, Via della Scienza 5, I-09047 Selargius (CA), Italy\\
$^{3}$MPI for Astronomy, K\"{o}nigstuhl 17, D-69117 Heidelberg, Germany\\
$^{4}$Leiden Observatory, Leiden University, PO Box 9513, NL-2300 RA Leiden, the Netherlands\\
$^{5}$Argelander Institut f\"ur Astronomie, Universit\"at Bonn, Auf dem H\"ugel 71, Bonn, D-53121, Germany\\
$^{6}$INAF--Osservatorio Astronomico di Bologna, via P. Gobetti 93/3, 40129 Bologna, Italy\\
$^{7}$Department of Astronomy, University of Cape Town, Private Bag X3, Rondebosch 7701, South Africa\\
$^{8}$Department of Physics, Faculty of Science, University of Zagreb,  Bijeni\v{c}ka cesta 32, 10000  Zagreb, Croatia\\
$^{9}$Max-Planck-Institut f\"{u}r Radioastronomie, Auf dem H\"{u}gel 69, 53121 Bonn, Germany\\
$^{10}$Th\"{u}ringer Landessternwarte, Sternwarte 5, 07778 Tautenburg, Germany
}

\date{Accepted 2021 March 05. Received 2021 March 05; in original form 2019 December 07}

\pubyear{2021}

\begin{document}
\label{firstpage}
\pagerange{\pageref{firstpage}--\pageref{lastpage}}
\maketitle

% Abstract of the paper
\begin{abstract}
The infrared-radio correlation (IRRC) underpins many commonly used radio luminosity--star formation rate (SFR) calibrations. In preparation for the new generation of radio surveys we revisit the IRRC of low-$z$ galaxies by (a) drawing on the best currently available IR and 1.4 GHz radio photometry, plus ancillary data over the widest possible area, and (b) carefully assessing potential systematics. We compile a catalogue of $\sim$9,500 z\,$<$\,0.2 galaxies and derive their 1.4\,GHz radio ($L_{\mathrm{1.4}}$), total IR, and monochromatic IR luminosities in up to seven bands, allowing us to parameterize the wavelength-dependence of monochromatic IRRCs from 22--500\,$\mu$m. For the first time for low-$z$ samples, we quantify how poorly matched IR and radio survey depths bias measured median IR/radio ratios, $\overline{q}_{\mathrm{TIR}}$, and discuss the level of biasing expected for low-z IRRC studies in ASKAP/MeerKAT fields. For our subset of $\sim$2,000 high-confidence star-forming galaxies we find a median $\overline{q}_{\mathrm{TIR}}$ of 2.54 (scatter: 0.17\,dex). We show that $\overline{q}_{\mathrm{TIR}}$ correlates with $L_{\mathrm{1.4}}$, implying a non-linear IRRC with slope 1.11$\pm$0.01. Our new $L_{\mathrm{1.4}}$--SFR calibration, which incorporates this non-linearity, reproduces SFRs from panchromatic SED fits substantially better than previous IRRC-based recipes. Finally, we match the evolutionary slope of recently measured $\overline{q}_{\mathrm{TIR}}$--redshift trends without having to invoke redshift evolution of the IRRC. In this framework, the redshift evolution of $\overline{q}_{\mathrm{TIR}}$ reported at GHz frequencies in the literature is the consequence of a partial, redshift-dependent sampling of a non-linear IRRC obeyed by low-$z$ {\it and} distant galaxies.
\end{abstract}
% Select between one and six entries from the list of approved keywords.
% Don't make up new ones.

%%%%%%%%%%%%%%%%%%%%%%%%%%%%%%%%%%%%%%%%%%%%%%%%%%

%%%%%%%%%%%%%%%%% BODY OF PAPER %%%%%%%%%%%%%%%%%%

\section{Introduction}

The infrared and radio synchrotron continuum luminosities are observed to be closely related in star-forming galaxies  \citep{kruit71, kruit73, Jong85, Helou85, condon92, yun01}. Since the far-infrared (FIR; 25 -- 1000 $\mu$m) emission is predominantly generated by star formation (SF) activity \citep{kennicutt98, charlot2000}, this so-called infrared-radio correlation (IRRC) implies that radio power in most galaxies is also related to SF. The IRRC has been used to establish a radio-based star formation rate (SFR) calibration (e.g. \citealt{condon92,murphy11}). The main advantages of the radio synchrotron continuum over other SF tracers are (i) the fact that it is unattenuated by interstellar dust, and hence does not require appropriate corrections, (ii) the high angular resolution that is achievable in interferometric observations with radio telescope arrays, and (iii) especially with next generation telescopes, superb sensitivity and survey speed. However, despite abundant literature on the topic \citep[e.g.][]{voelk89,helou93,bell03,lacki10a,schleicher2013}, the detailed physics shaping the IRRC remain poorly understood from the theoretical perspective.

In order to better leverage the aforementioned strengths of radio continuum emission as an SF tracer, numerous studies in the past decade have sought to improve its accuracy by calibrating it against other, theoretically better-established SF tracers \citep[e.g.][]{hodge08,brown17,davies17,gurkan18,read18,duncan20}, or examined the variation of the IRRC with other galaxy properties, such as stellar mass \citep[e.g.][]{magnelli15,delvecchio21} or galaxy type \citep{moric10,roychowdhury12,nyland17}. An especially frequently debated aspect of the IRRC is its (non-)evolution with redshift \citep[e.g.][]{garrett02,appleton04,garn09,jarvis10,sargent10a,sargent10b,mao11,smith14,magnelli15,calistro-rivera17,delhaize17,molnar18,delvecchio21}. A majority of these studies compare their results to the classical works of \citet{yun01} and \citet{bell03}, since these are considered to be the main reference points for the low-z IRRC. The overall IRRC properties, such as the slope and dispersion of the relation, proved to be broadly consistent between these two studies.  With evidence for non-linearity at low IR luminosities supported by the findings of \citet{yun01}, \cite{bell03} provided a refined luminosity-dependent radio -- SFR calibration.
However, both of these cornerstone papers use the 60 and 100\,$\mu$m photometry from the Infrared Astronomical Satellite \citep[IRAS;][]{neugebauer84} to estimate IR luminosities, and thus lack the now standard spectral energy distribution (SED) fitting approach, and had to rely on the shallow but wide radio coverage of the NRAO VLA Sky Survey \citep[NVSS;][]{condon98}.

 \begin{figure}
 \includegraphics[width=0.45\textwidth]{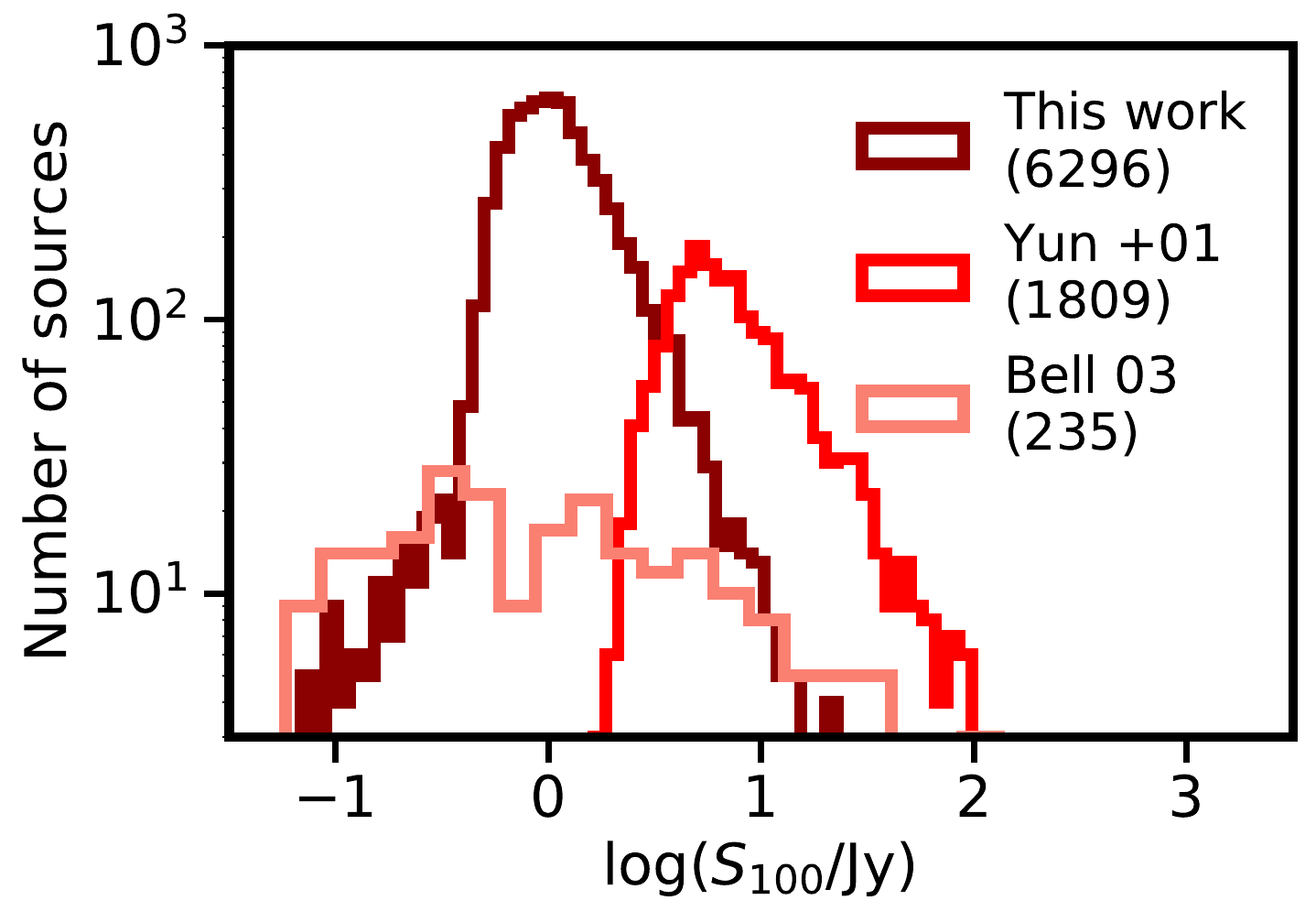}
 \includegraphics[width=0.45\textwidth]{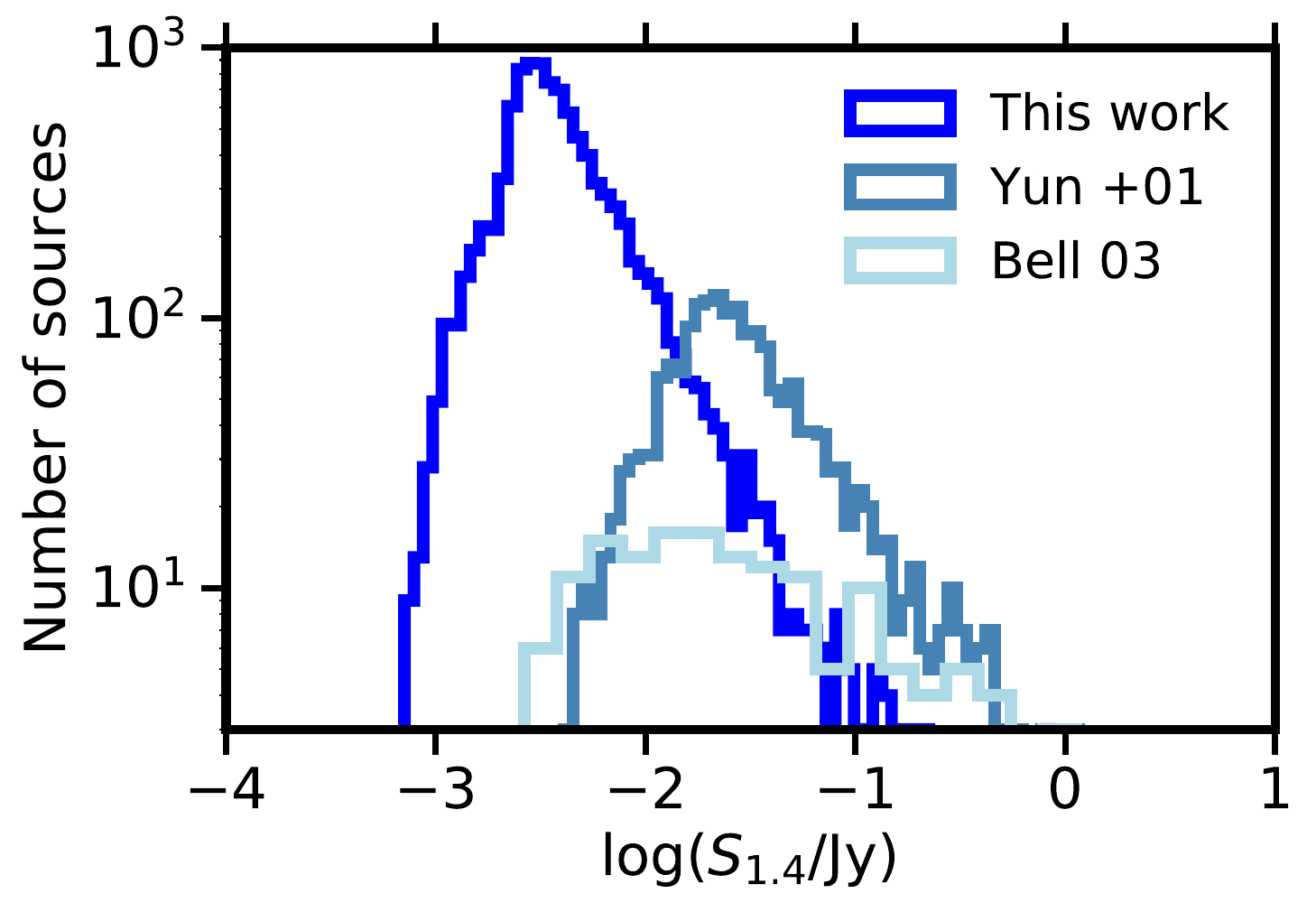}
 \caption{Comparison of 100\,$\mu$m and 1.4\,GHz radio flux densities in our catalogue to the flux distributions of the \protect\cite{yun01} and \protect\cite{bell03} samples. Number counts are on a logarithmic scale to ease comparison between the samples.}
 \label{fig::yun_bell_us_flux}
 \end{figure}

Since the publication of the seminal \cite{yun01} and \cite{bell03} works, deeper radio and IR measurements and better overall IR photometric coverage have become available, mainly thanks to the Faint Images of the Radio Sky at Twenty centimetres survey \citep[FIRST;][]{becker95,helfand15}, the Herschel Space Observatory (\citealt{pilbratt10} hereafter, Herschel) and the Wide-field Infrared Survey Explorer \citep[WISE;][]{wright10}. Furthermore, as discussed in \cite{sargent10a}, IR- and radio-selection effects can bias median IR-radio ratio measurements. Avoiding such biases requires a careful approach to sample selection, and this has so far almost exclusively been discussed in the context of redshift evolution, but much less when it comes to calibrating radio-based SFR measurements on low-redshift samples. Meanwhile, a new generation of deeper and wider surveys on modern radio telescopes -- e.g., SKA pathfinders, the Karl G. Jansky Very Large Array \citep[JVLA;][]{perley11} and the Low Frequency Array \citep[LOFAR;][]{harleem13}, and SKA precursors, the Australian Square Kilometre Array Pathfinder \citep[ASKAP;][]{johnston07,deboer09} and the Meer Karoo Array Telescope \citep[MeerKAT;][]{booth09} -- is providing a more complete census of radio emission from star-forming galaxies both in the local and distant Universe. In preparation for this next generation of studies, it is thus timely to revisit the low-$z$ IRRC. To this end we use the aforementioned FIRST, Herschel and WISE observations and other ancillary data. We also define a highly pure star-forming galaxy (SFG) sample through careful separation of SFGs and active galactic nuclei (AGN), and we perform SED fits that exploit IR photometry covering a broader wavelength range, to assemble a large z $<$ 0.2 sample with an eye to quantifying potential systematics due to methodology and/or selection effects.

\begin{figure}
\includegraphics[width=0.45\textwidth]{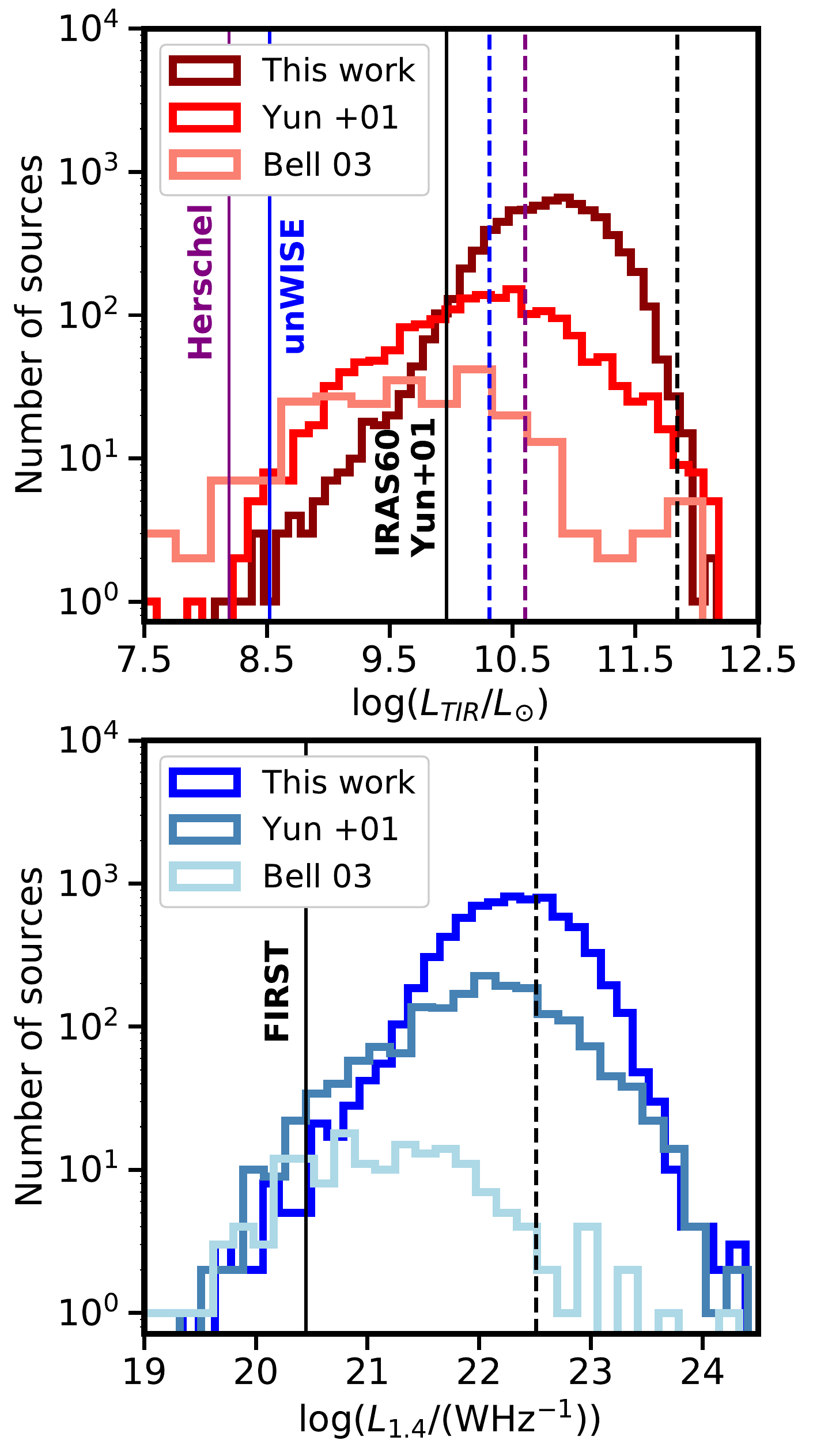}
\caption{Total infrared and 1.4 GHz radio luminosity distributions of the \protect\cite{yun01} and \protect\cite{bell03} samples in comparison to our catalogue (for details on the calculations of radio and IR luminosities, see Sect. \ref{sect::sed_fits}). Solid (dashed) lines show the various luminosity limits at z $=$ 0.01 (z $=$ 0.1) derived from Fig. \ref{fig::tir_lim}. Number counts are on a logarithmic scale to ease comparison between the samples.}
\label{fig::yun_bell_us}
\end{figure}

This work presents an initially $\sim$5 times larger and $\sim$4 times deeper data set than the one used in \cite{yun01}. Fig. \ref{fig::yun_bell_us_flux} shows a comparison of IR and radio flux densities in the catalogues of \cite{bell03}\footnote{The 100\,$\mu$m flux densities from \protect\cite{bell03} were derived using the published 60\,$\mu$m IR luminosities and IRAS 60 and 100\,$\mu$m flux density ratios.}, \cite{yun01} and our work. We probe fainter sources than \citet{yun01} both in the radio and IR, while we have a similar coverage to \citet{bell03} at 100 $\mu$m. However, comparisons of both 1.4 GHz radio continuum ($L_{\rm 1.4GHz}$) and total IR ($L_{\rm TIR}$; 8 -- 1000 $\mu$m) luminosities\footnote{We converted the publicly available 1.4\,GHz luminosities of \protect\cite{yun01} and \protect\cite{bell03} to flux densities assuming a radio spectral slope of $-0.7$. Total IR luminosities for the sample in \citet{yun01} were calculated by first using the published 60 and 100\,$\mu$m flux densities and Eq. (2) and (3) in \citet{yun01} to obtain FIR luminosities, and then multiplying them by $\sim$ 2, the average offset between FIR and TIR luminosities \citep[see e.g.][]{bell03}.}, seen in Fig. \ref{fig::yun_bell_us}, reveal that we only substantially increase the number of high-luminosity objects at $\log(L_{\rm 1.4GHz} / \mathrm{W Hz^{-1}}) \geqslant 21$ and $\log(L_{\rm TIR} / L_{\odot})\geqslant 10$, respectively. The primary reason for this is the $\sim$2.6 times larger area covered by \cite{yun01}. Our catalogue with IR and radio luminosity measurements for 9,645 galaxies is publicly available (for details see Appendix \ref{app::pr_descr}) to support follow-up studies investigating the low-z IRRC's dependence on various galaxy parameters and thus gaining insights into the physics regulating the correlation.

In Sect. \ref{sect::dat} we describe the archival data products used, and the catalogue construction process. Sect. \ref{sect::methods} details the calculation of IR and radio luminosities, the identification of AGN and SFG sources, and gives a brief summary of all the data products used in our analysis. In Sect. \ref{sect::res} we characterise the properties of both the monochromatic and bolometric IRRCs of low-$z$ galaxies, and we demonstrate and quantify sensitivity related selection effects. Based on this, in Sect. \ref{sect::disc}, we discuss the implications of our findings for the radio--SFR calibration, and for interpreting the observed redshift evolution of the IRRC.

Throughout this paper, we use a flat $\Lambda$CDM cosmology with $\Omega_M = 0.3$ and $\mathrm{H_0 = 70 \, km\, s^{-1}\, Mpc^{-1}}$. Star formation rates reported assume a \citet{chabrier03} initial mass function\footnote{For conversion factors between our assumption and other widely-used IMF models see e.g. \citet{madau14}} (IMF).

\section{Data}
\label{sect::dat}

The starting point for our sample construction is the Sloan Digital Sky Survey (SDSS) Data Release 12 \citep{alam15}, which provides positions and redshifts of nearly 470 million unique optical sources over roughly 1/3 of the Celestial sphere. For our analysis, we selected SDSS galaxies with spectroscopic (58\%) or, where unavailable, photometric (42\%) redshifts\footnote{We restrict ourselves to high-quality photometric redshifts by imposing the criteria photoErrorClass\,=\,1, nnCount\,$>$\,95, and 0\,$<\,\mathrm{zErr}\,<$\,0.03. The SDSS spectroscopic and photometric catalogues were combined using the fluxID, which -- for each spectroscopic source -- identifies the corresponding SDSS photometric object (objID) that contributes most to the spectrum.} below $z$\,=\,0.2, corresponding to a look-back time of $\lesssim$2.4\,Gyr. This selection results in an optical parent sample of 3,001,410 galaxies for which we identify infrared and radio counterparts as described in Sects. \ref{sect::ir_data} and \ref{sect::rad_data}, respectively. Table \ref{tab::sensitivity} gives a summary of the sensitivity and area covered by these archival data. We note that limiting the sample to galaxies at $z$\,=\,0.1 to further minimise the impact of evolutionary trends within this redshift range leaves the measured IRRC properties unchanged within 1 $\sigma$.

\begin{table}
\centering
\caption{Sensitivities and sky coverage of surveys used to construct our catalogue. The 5\,$\sigma$ depths quoted for the PACS and SPIRE Point Source Catalogs (PPSC and SPSC, respectively) represent their median tabulated 5\,$\sigma$ flux uncertainties. PPSC 100\,$\mu$m depth, marked by ($^*$) in the table, is most likely underestimated, due to relatively low number of detections permitting a statistically less robust noise simulation. We increased the nominal 1 mJy FIRST detection limit by 22\% to reflect the scaling we applied to FIRST fluxes in our sample to compensate missing large scale flux in our low-z sample (see Sect. \ref{sect::rad_data}). All other values are taken from the corresponding data release papers cited in the text.}
\label{tab::sensitivity}
\begin{tabular}{ccc}
 & 5\,$\sigma$ & area \\ 
 & [mJy] & [deg$^2]$ \\ \hline \hline
FIRST & 1.22 & 10,575 \\
NVSS & 2.5 & 37,216 \\ \hline
IRAS 60\,$\mu$m & 200 & full sky\\
IRAS 100\,$\mu$m & 1000 & full sky\\ \hline
H-ATLAS DR1 100\,$\mu$m & 220 & 161 \\
H-ATLAS DR1 160\,$\mu$m & 245 & 161 \\
H-ATLAS DR1 250\,$\mu$m & 37 & 161 \\
H-ATLAS DR1 350\,$\mu$m & 47 & 161 \\
H-ATLAS DR1 500\,$\mu$m & 51 & 161 \\ \hline
PPSC 100\,$\mu$m & 107$^*$ & $\sim$3,300 \\
PPSC 160\,$\mu$m & 236 & $\sim$3,300 \\
SPSC 250\,$\mu$m & 73 & $\sim$3,700 \\
SPSC 350\,$\mu$m & 73 & $\sim$3,700 \\
SPSC 500\,$\mu$m & 78 & $\sim$3,700 \\ \hline
WISE 22\,$\mu$m & 4.35 & full sky\\
\end{tabular}
\end{table}

\subsection{IR data}
\label{sect::ir_data}

Here we describe the archival data underpinning our IR luminosity measurements via spectral energy distribution fitting (see Sect. \ref{sect::sed_fits}).

\subsubsection{WISE photometry}
The WISE satellite \citep{wright10} carried out all-sky observations in four bands, two of which (12 and 22\,$\mu$m, with resolutions of 6.5 and 12.0\,arcsec, respectively) lie in the 8--1000\,$\mu$m window underpinning our total IR (TIR) luminosity measurement. Rapidly changing SED amplitudes due to polycyclic aromatic hydrocarbon (PAH) features around 12\,$\mu$m make modelling sources with photometric redshifts difficult. To avoid these issues, and to also further minimize the impact of any residual contamination from mid-IR (MIR) torus emission from AGN hosts which were not picked up by our AGN removal criteria in Sect. \ref{sect::agn_sfg_sep}, we only use 22\,$\mu$m flux densities in the following.\\
Exploiting the high-resolution SDSS optical data, \cite{lang16} performed flux extraction with a forced photometry approach on un-blurred, co-added WISE images \citep{lang14} at over 400 million optical source positions, resulting in the unWISE catalogue. The unWISE data are hence naturally linked to our SDSS parent sample. To enter our sample, we require each unWISE 22\,$\mu$m detection to have a signal-to-noise ratio (SNR) of at least 5, despite the availability of lower significance measurements due the forced photometry technique used for unWISE.

\subsubsection{IRAS photometry}
The Infrared Astronomical Satellite \citep[IRAS;][]{neugebauer84} mission covered the full sky at wavelengths 12, 25, 60 and 100\,$\mu$m with an angular resolution varying between $\sim$0.5 arcmin at 12\,$\mu$m and $\sim$2 arcmin at 100\,$\mu$m. We drew 60 and 100\,$\mu$m fluxes from the Revised IRAS Faint Source Redshift Catalog (RIFSCz) of \cite{wang14}, which contains galaxies selected at 60\,$\mu$m with SNR $>$ 5 while covering 60\% of the sky. We discarded IRAS 12 and 25\,$\mu$m fluxes tabulated in the RIFSCz, respectively, due to the difficulty of fitting the PAH features of the SED for sources with photometric redshifts and the availability of better quality WISE photometry at 22\,$\mu$m.\\
\citet{wang14} performed a likelihood ratio matching technique to combine the Faint Source Catalog with the deep WISE 3.4 $\mu$m data. They then cross-matched these sources with SDSS DR10 using the WISE positions and a 3\,arcsec search radius. After reconciliation of SDSS DR10 and DR12 galaxy positions, we find that 17,829 (7,261) of the 60\,$\mu$m (100\,$\mu$m) RIFSCz sources are associated with an entry in our low-$z$ SDSS DR12 parent catalogue.

\subsubsection{Herschel photometry}
During its nearly four years of operation, Herschel produced thousands of maps of varying depth with two cameras: the Photoconductor Array Camera and Spectrometer \citep[PACS;][]{poglitsch10} and the Spectral and Photometric Imaging REceiver \citep[SPIRE;][]{griffin10}, operating at 60--210\,$\mu$m and 200--670\,$\mu$m, respectively, with angular resolutions of 5.6-11.3 arcsec and 17-35 arcsec. This resulted in a large number of data products optimized by numerous science collaborations for different purposes. Here we make use of two data bases that provide Herschel galaxy photometry, namely (i) the Herschel-ATLAS survey \citep[H-ATLAS;][]{eales10, valiante16}, as well as (ii) the PACS Point Source Catalog \citep[PPSC\footnote{\url{https://doi.org/10.5270/esa-rw7rbo7}};][]{marton17} and SPIRE Point Source Catalog (SPSC\footnote{European Space Agency, 2017, Herschel SPIRE Point Source Catalogue, Version 1.0. \url{https://doi.org/10.5270/esa-6gfkpzh}}) for which all archival data, including calibration scans, were mined in a systematic and homogeneous way.

\noindent H-ATLAS Data Release 1 \citep{valiante16} covers the three equatorial fields surveyed by the GAMA \citep[Galaxy and Mass Assembly;][]{driver11} spectroscopic survey. It consists of 120,230 sources detected at 250\,$\mu$m. It contains PACS 100 and 160\,$\mu$m detections at $>$ 3 SNR, and 250, 350, and 500\,$\mu$m photometry at $>$ 4 SNR considering both instrumental and confusion noise. Extended sources were identified and their fluxes extracted using appropriately sized apertures \citep{rigby11}. The DR1 catalogue also provides an optical identification from the SDSS DR7/8 catalogue using the likelihood ratio technique \citep{bourne16}. We cross-correlated the positions of H-ATLAS optical counterparts classified as secure by \citet{bourne16} with our SDSS DR12 parent catalogue using a search radius of 1\,arcsec. This resulted in 8,752 matches (a small fraction of the H-ATLAS DR1 sources due to our redshift cut at $z$\,=\,0.2). We note that for 3 sources we found negative fluxes at 500\,$\mu$m in the H-ATLAS catalogue. These measurements were removed from any subsequent analysis.

\noindent In order to fully exploit the available Herschel coverage in the SDSS footprint, we also adopt PPSC and SPSC fluxes of $>$ 3 SNR, where H-ATLAS photometry is not available. This enables us to increase the subset of galaxies with Herschel photometry in our combined sample (see Sect. \ref{sect::merged_cat}) by a factor of $\sim$8. From the PPSC and SPSC we have removed flagged (edge-flag, blend-flag, warm altitude or solar system map flag for the PPSC and an additional large galaxy flag from \citealt{jarrett2003} for the SPSC) sources, in the process retaining 64\% and 69\% of all 100 and 160\,$\mu$m catalog entries, respectively, and 97\% of the 250 and 350\,$\mu$m sources, as well as 96\% of all 500\,$\mu$m sources. To assign optical counterparts to these remaining PPSC and SPSC entries, we adopted a band-dependent matching radius. The details of this cross-correlation procedure and estimates of spurious match fractions are provided in Appendix \ref{app::contamination}. We find that 5,878 100\,$\mu$m PPSC sources have an optical counterpart in our parent catalog; at 160\,$\mu$m this is the case for 10,149 sources. For the SPSC, we were able to assign 43,665 sources to an optical counterpart at 250\,$\mu$m, 15,614 sources at 350\,$\mu$m, and 2,806 sources at 500\,$\mu$m.

\noindent
Relying on point source measurements extracted with point spread functions ranging from 7 to 35\,arcsec in angular size carries the risk of underestimating fluxes for our low-redshift, $z<0.2$, galaxies. We were able to assess whether our PPSC and SPSC fluxes are subject to any systematic bias by comparing them to H-ATLAS photometry for objects where both types of measurements are available. This comparison reveals an average deficit of $\sim$19 and 10\% for PPSC fluxes relative to H-ATLAS measurements at 100\,$\mu$m and 160\,$\mu$m, while SPIRE fluxes are consistent within 5\% in all bands, suggesting that resolution-related effects only noticeably bias our PPSC photometry. To mitigate these systematics we applied statistical corrections to our PPSC data and Appendix \ref{app::flux_consistency} details how we derived the appropriate scaling factors.\\
In a further test of the overall consistency of our Herschel photometry we also investigated the agreement of flux errors between H-ATLAS and the point source catalogues. For the PPSC flux errors, we adopt either the local RMS or the so-called structure noise\footnote{The structure noise produces statistical estimates on the error of the photometry by measuring the flux of artificial sources injected into the various Herschel fields (for details see the HPPSC Explanatory Supplement and references therein).}, whichever is larger. Both approaches account for the instrumental as well as the confusion noise. We note that in the case of 100\,$\mu$m data, uncertainties are likely underestimated due to fewer detections available for producing noise simulation maps (78 in contrast to 326 at 160\,$\mu$m), resulting in less accurate modelling of the noise in this band, especially for faint sources (based on priv. comm with G\'{a}bor Marton). To consider the uncertainties in the consistency between resolved H-ATLAS and point source PPSC/SPSC photometry, we increased the flux density errors in each band for PPSC/SPSC data as described in Appendix \ref{app::flux_consistency}. However, we emphasize that this was done only to inform our spectral energy distribution modelling (Sect \ref{sect::sed_fits}). For source selection purposes we worked with the errors as tabulated in the catalogues.

\subsection{Radio data}
\label{sect::rad_data}

Our radio fluxes were drawn from two wide-area 1.4\,GHz VLA surveys, FIRST and NVSS. FIRST focused on the SDSS footprint as established in \cite{helfand15}, with higher resolution (5 vs. 45\,arcsec) and sensitivity (1 vs. 2.5 mJy) than NVSS, while NVSS surveyed the entire Northern sky \citep{condon98}. These two catalogues were combined, via positional cross-match, and form the basis of the Unified Radio Catalog \citep{kimball08,kimball14}. An entry in the Unified Radio Catalog is either a FIRST object with an NVSS match, an NVSS object with a FIRST match, or an unmatched object from either survey. The three closest matches within 30 arcsec to a FIRST or NVSS source were also recorded. As a result, several sources appear more than once in the final database. To facilitate flexible and easy sample selection, \citet{kimball08} defined several flags indicating whether FIRST or NVSS was the primary matching source, and the number of objects from the other catalogue within 5, 10, 30 and 120\,arcsec. We made use of these flags in the process of creating our catalogue, as explained in the following.

When both FIRST and NVSS data are available, FIRST does not always clearly provide the best total intrinsic radio flux density of a galaxy, despite its superior resolution and depth. Due to $uv$ coverage from short baselines being absent in the B-array FIRST data, its sensitivity to extended emission is limited. Larger scale flux components thus were potentially captured only by NVSS. As a result, \citet{helfand15} report a 1--5\% statistical flux deficit in the total FIRST sample compared to NVSS. The difference is expected to be larger on average for low redshift galaxies due to their larger angular sizes. On the other hand, multiple galaxies could be blended in the 45 arcsecond NVSS imaging, leading to positive flux biasing. To mitigate the effects of these on our analysis, we selected sources in the Unified Radio Catalog that are:

\begin{enumerate}
    \item NVSS sources with no FIRST counterparts within the 30 arcsec of the NVSS position. Expressed with the flags of the catalogue this selection is \textit{(matchflag\_nvss = -1) \& (matchflag\_first = 0) \& (matchtot\_30 = 0)}.
    \item NVSS sources with only one FIRST detection in their beam. Selection flags for these sources were \textit{(matchflag\_nvss = -1) \& (matchflag\_first = 1) \& (matchtot\_30 = 1)}.
    \item FIRST sources with no NVSS counterparts, i.e.  \textit{(matchflag\_first = -1) \& (matchflag\_nvss = 0) \& (matchtot\_30 = 0)}.
\end{enumerate}

\begin{figure}
\centering
\includegraphics[width=0.45\textwidth]{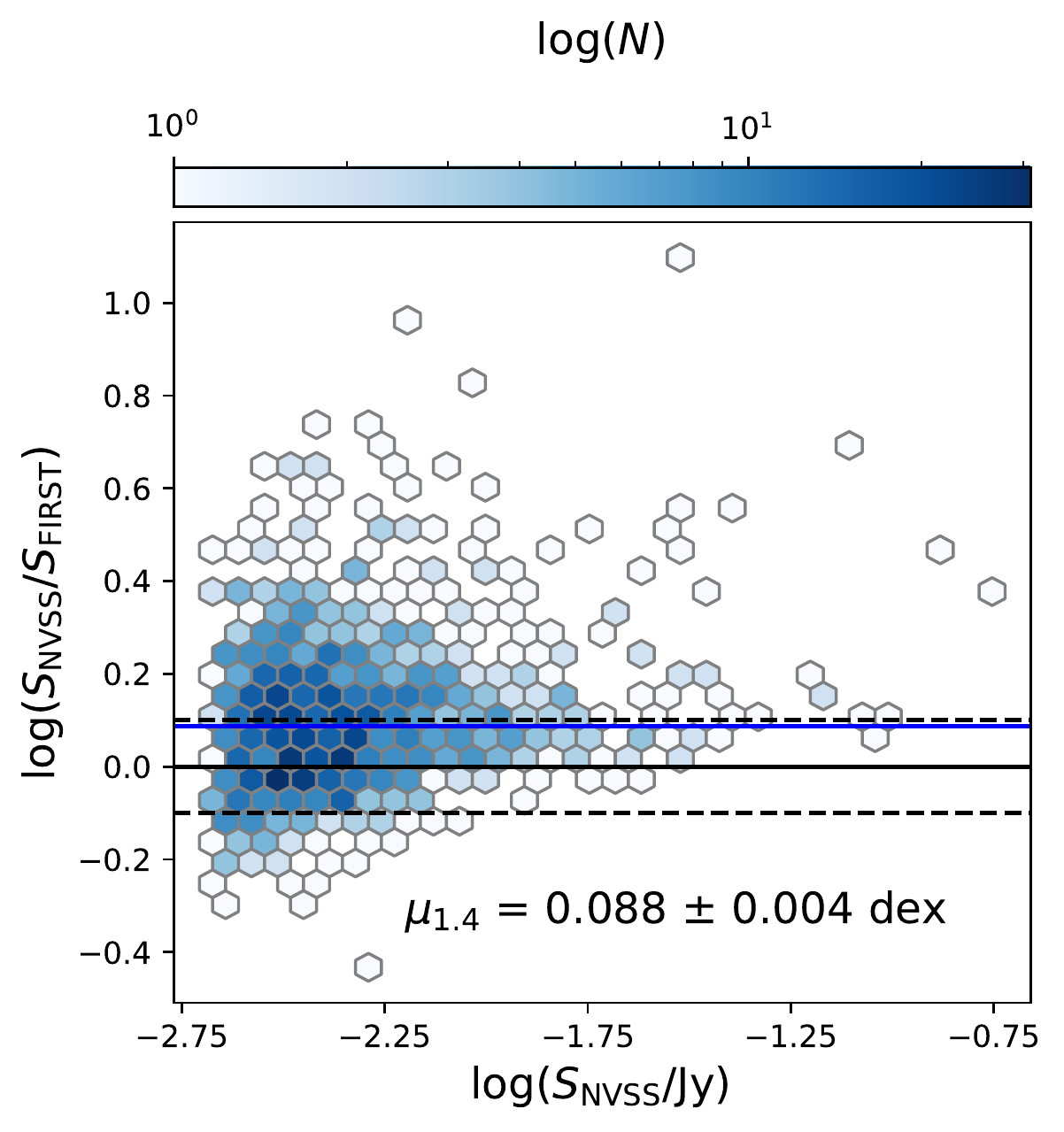}
\caption{Logarithmic ratio of FIRST and NVSS flux densities as a function of the latter using sources in (ii) from Sect \ref{sect::rad_data}. Horizontal black line is drawn at 0, i.e. a flux density ratio of 1, while the dashed lines represent the $\pm$ 0.1 dex offsets. The median ratio, $\mu_{\rm 1.4}$, shown as a horizontal blue line, was used to scale the FIRST flux densities to statistically match the NVSS data.}
\label{fig::nvss_vs_first_fluxes}
\end{figure}

This subset of the Unified Radio Catalog was then spatially matched to our low-$z$ SDSS parent sample. For sources in (i) and (ii) we used NVSS positions and a matching radius of 30 arcsec \citep[equal to the value adopted for the NVSS matching in][]{kimball08,kimball14}, and for sample (iii) FIRST positions with 2.5 arcsec matching radius (half of the 5 arcsec FIRST beam size). Comparing FIRST and NVSS fluxes in Fig. \ref{fig::nvss_vs_first_fluxes} we find a systematic NVSS/FIRST flux ratio of $\sim$1.22. This is largely independent of 1.4 GHz angular size, flux and galaxy type, i.e. SFG or AGN. However, the ratio is closer to 1 ($\sim$1.09) for galaxies selected via (ii) if we do not restrict the selection to low-$z$ sources, supporting the idea that the main reason for the flux offset is missing extended flux in FIRST measurements of nearby sources \citep[see e.g.][]{helfand15}. An alternative explanation for the offset is the possibility that NVSS fluxes are biased high due to blending. We consider this a less likely scenario, because we removed all NVSS sources with more than one FIRST counterpart in order to minimize the effect of blending. Thus, we conclude that for our purposes in this particular sample NVSS provides a more robust measure of the 1.4 GHz flux. As a result we adopted NVSS measurements for (i) and (ii) and applied a statistical correction of 1.22 to all FIRST detections in (iii), in order to make these fluxes consistent with (i) and (ii). This corresponds to a 0.09 dex upward correction in logarithmic 1.4 GHz flux space (i.e. a 0.09 dex downward shift of the IRRC parameter, $q$) for these objects. Table \ref{tab::cat_stats} gives a summary of source counts in (i), (ii) and (iii) in our main sample and its subsamples (defined below).

We note that the final FIRST catalog release \citep{helfand15} was published after the assembly of the \citet{kimball14} Unified Radio Catalog. The \citet{helfand15} catalog contains significantly more robust sidelobe probability estimates, updated flux measurements compared to those in \citep{kimball14} and excludes data from unreliable FIRST pointings. We incorporated these improvements by matching the final FIRST catalogue to the FIRST sources of the Unified Radio Catalog via a simple positional cross-match using a matching radius\footnote{The description of the latest FIRST catalogue -- available at \url{http://sundog.stsci.edu/first/catalogs/readme.html} -- suggests a better than 1\,arcsec positional accuracy at the detection limit of the survey, and 0.5\,arcsec for $\sim$10\,$\sigma$ detections. Accordingly, after the cross-match we find a mean separation of 0.01 arcsec with a scatter of 0.06\,arcsec.} of 1\,arcsec. We also removed FIRST sources with a sidelobe probability greater than 10\% to mitigate contamination by spurious detections prior to the selection steps described above.

\subsection{Combined IR and radio sample}
\label{sect::merged_cat}

\begin{figure}
\includegraphics[width=0.5\textwidth]{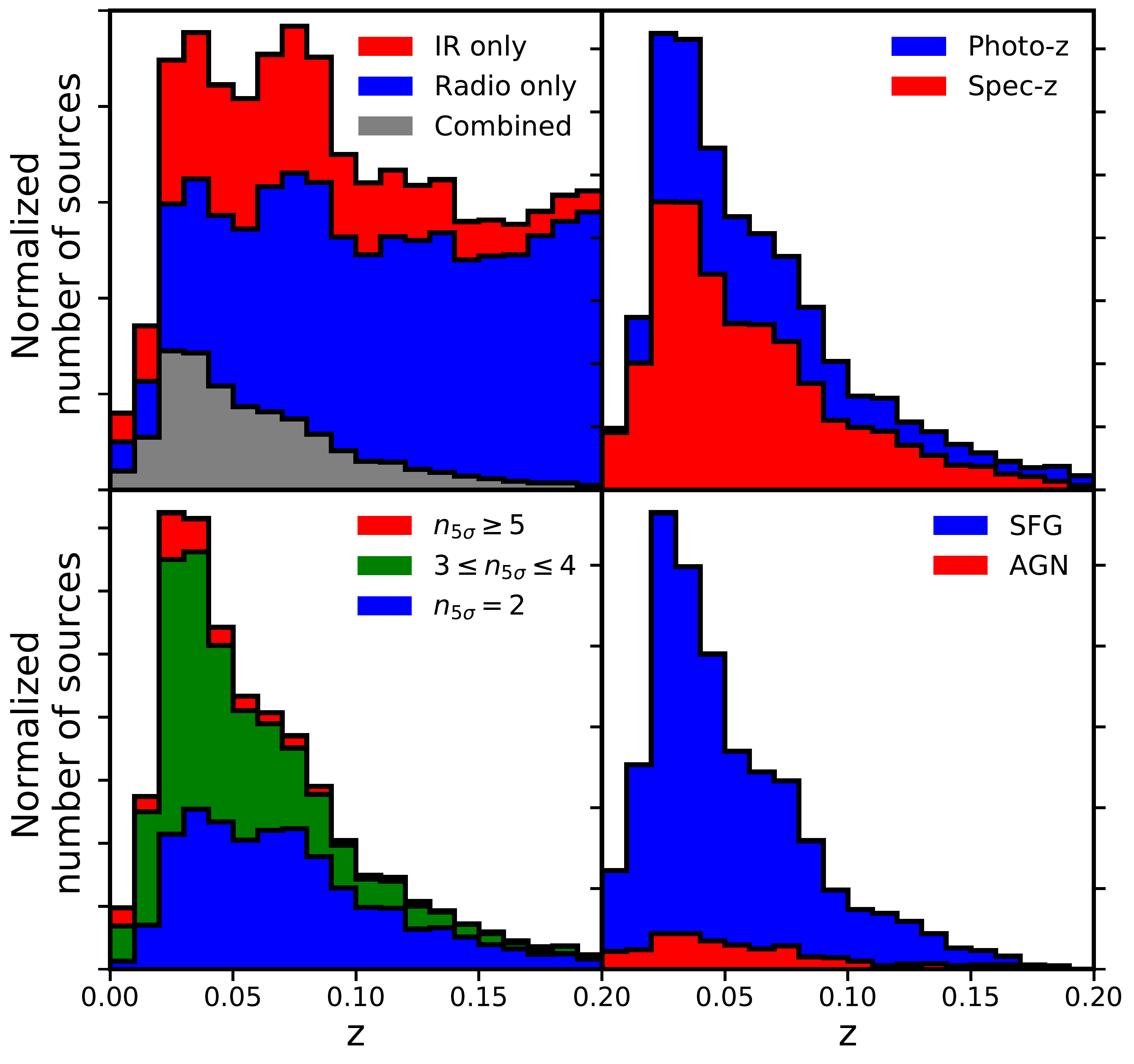}
\caption{Stacked histograms showing the key sample properties. \textbf{(top left)} -- redshift distributions of the only radio- and IR-detected samples, and the combined sample, i.e. sources detected both at IR and radio wavelengths; \textbf{(top right)} -- sources in the combined sample with photometric and spectroscopic redshifts; \textbf{(bottom left)} --  sources in the combined sample with fluxes detected at the $\geq$5\,$\sigma$ level in 2, 3 or 4, and $\geq$5 IR bands; \textbf{(bottom right)} -- galaxies classified as AGN (both optical and WISE selected) or star-forming galaxies in the combined sample.}
\label{fig::joint_z_dist}
\end{figure}

To summarize, we have collected IR and radio flux densities at 8 different wavelengths using 6 archival databases with varying depths and survey areas, as presented in Table \ref{tab::sensitivity}. A source is required to have a SNR $>$ 5 detection in the unWISE catalogue at 22\,$\mu$m and at least one other SNR $>$ 5 measurement in any of the longer wavelength data in Sect \ref{sect::ir_data} to enter our IR-selected sample. Meanwhile, to be considered as a radio-detected object, each source needed an at least 5 SNR 1.4 GHz flux density measurement in the (i), (ii) or (iii) subsamples of the Unified Radio Catalog, as described in Sect \ref{sect::rad_data}. Our compilation thus lead to samples of 25,782 IR- and 51,774 radio-detected galaxies at $z < 0.2$, respectively. Merging these IR- and radio-detected catalogues resulted in a joint catalogue of 67,908 objects. In this joint sample, 9,645 sources are members of both the radio- and IR-detected catalogues and will be referred to as combined sample henceforth, while 16,134 have only IR and 42,126 only radio data. In Fig. \ref{fig::joint_z_dist} we show the redshift distributions of IR- and radio-detected sources and the combined sample; sources with spectroscopic and photometric redshifts in the combined sample; IR-detected sources with different number of IR photometric bands available; and SFGs and AGN. For the latter, Sect. \ref{sect::agn_sfg_sep} details our classification approach.

\begin{table*}
    \caption{Number of galaxies detected in the different IR bands in the combined and the depth-matched samples, as well as for the SFG and AGN subsamples drawn from the latter. Bracketed percentages are the fraction of source detected at $>$5\,$\sigma$ in any given band and sample (e.g. of the 972 galaxies in the combined sample with a measured Herschel/SPIRE 500\,$\mu$m flux, 90\% -- i.e. 875 objects -- have a SNR $>$5\,$\sigma$ detection). The final three rows list the number of sources with only NVSS fluxes, with both NVSS and FIRST fluxes, and with only FIRST detections, respectively.}
    \label{tab::cat_stats}
    \centering
    \begin{tabular}{ccccc}
        & Combined & Depth-matched & Depth-matched SFG & Depth-matched AGN \\ \hline\hline
        unWISE 22\,$\mu$m & 9,645 (100\%) & 6,601 (100\%) & 2,371 (100 \%) &  248 (100 \%)\\ \hline
        IRAS 60\,$\mu$m & 8,720 (90\%) & 6,217 (94\%) & 2,258 (95\%) & 229 (92 \%)\\
        IRAS 100\,$\mu$m & 4,407 (74\%) & 3,519 (77\%) & 1,338 (77\%) & 122 (73 \%)\\ \hline
        Herschel 100\,$\mu$m & 442 (92 \%) & 299 (95\%) & 111 (97\%) & 9 (81 \%)\\
        Herschel 160\,$\mu$m & 415 (91\%) & 263 (95\%) & 98 (96\%) & 11 (100 \%)\\
        Herschel 250\,$\mu$m & 1,710 (99\%) & 1,067 (99\%) & 347 (99\%) & 38 (100 \%)\\
        Herschel 350\,$\mu$m & 1,683 (98\%) & 1,058 (99\%) & 354 (99\%) & 37 (97 \%)\\
        Herschel 500\,$\mu$m & 914 (89\%) & 697 (93\%) & 262 (95\%) & 24 (89 \%)\\ \hline
        NVSS only & 3,117 & 1,280 & 597 & 36 \\
        NVSS+FIRST & 4,442 & 3,891 & 1,289 & 166 \\
        FIRST only & 2,089 & 1,430 & 485 & 46  \\

    \end{tabular}
\end{table*}

\begin{figure*}
\includegraphics[width=0.95\textwidth]{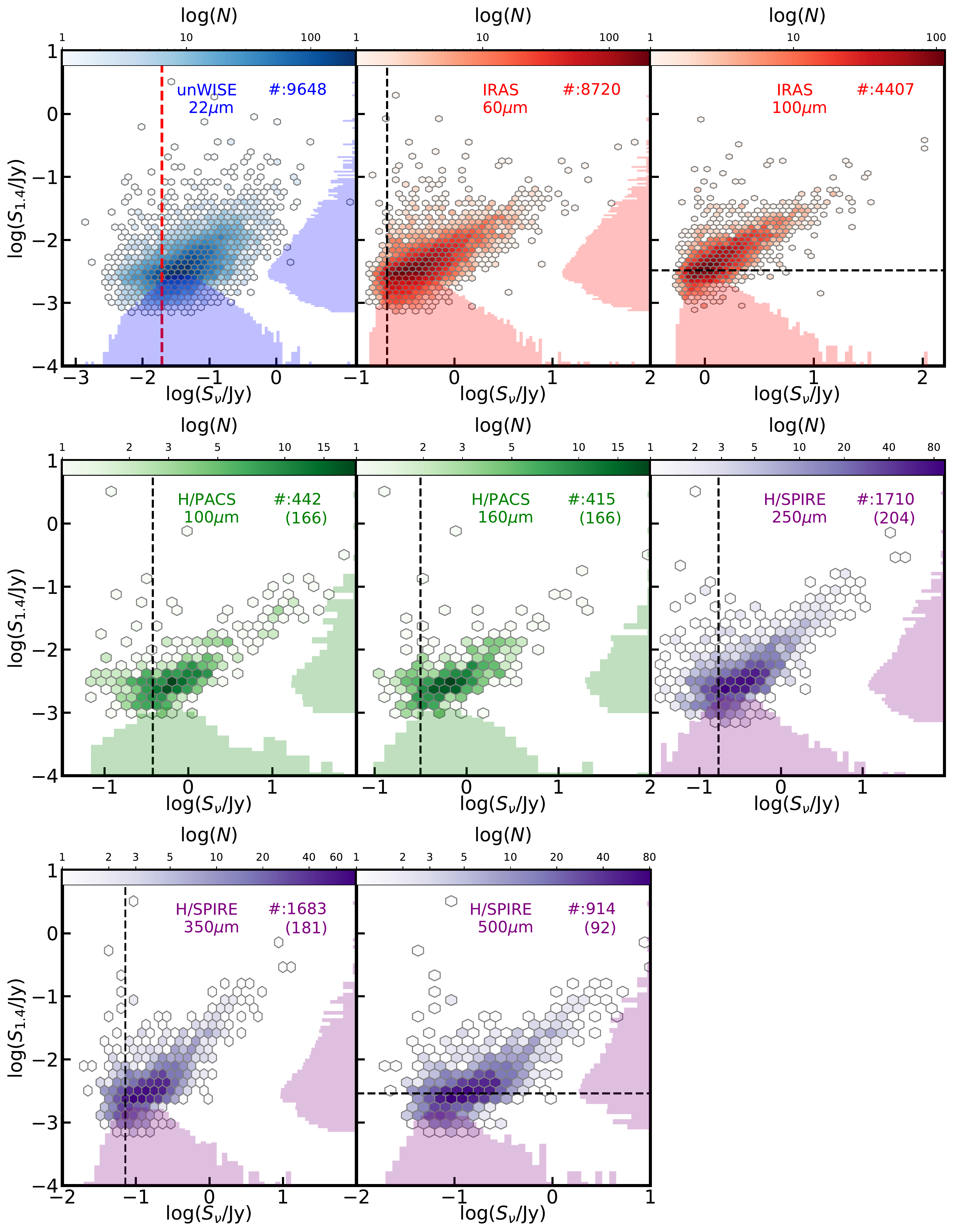}
\caption{Radio versus infrared fluxes from unWISE (22\,$\mu$m), IRAS (60 and 100 \,$\mu$m) and Herschel (100, 160, 250, 350 and 500 \,$\mu$m) in our joint sample (i.e. before flux cuts and removing AGN and ambiguous galaxies). IR and radio flux density distributions are shown on the sides. The dashed lines in each panel illustrate the flux density cuts we applied in order to select our radio-IR depth-matched samples for measuring monochromatic IRRC properties, as described in Sect. \ref{sect::fluxmatch_samp}. Table \ref{tab::fluxlims} contains the corresponding flux density values. We note that only the $\sim$19.5\,mJy flux selection at 22\,$\mu$m was used for the depth-matched sample for investigating the total IRRC. In the upper right corner of each panel we display the number of sources with both radio and IR-detections at a given wavelength. For Herschel bands, values in brackets denote the size of the subset of galaxies with fluxes from the H-ATLAS catalog.}
\label{fig::flux_flux}
\end{figure*}

Table \ref{tab::cat_stats} contains the number of detections in the combined sample, as well as other samples introduced later in Sect. \ref{sect::methods}, per photometric band. Fig. \ref{fig::flux_flux}. shows the IR flux densities in each band against the 1.4 GHz radio flux densities in our combined sample. The IR-radio correlation is apparent already before the conversion to luminosity and removal of contaminating AGN sources. Due to the mismatch between sensitivities of the IR and radio, a selection bias affects the total and the monochromatic IRRC statistics. To mitigate this, we derived flux density cuts that already appear in Fig. \ref{fig::flux_flux}. However, before we detail how these were calculated in Sect. \ref{sect::fluxmatch_samp}, we first describe our luminosity estimation approach in Sect. \ref{sect::calc_lum}.

\section{Methods}
\label{sect::methods}

In this section we describe our approach to calculating IR and radio luminosities for the 9,645 objects in the combined sample (Sect \ref{sect::merged_cat}), and then classifying these as AGN or SFG galaxies. We define a depth-matched subsample of the combined sample in Sect. \ref{sect::fluxmatch_samp} in an effort to mitigate the effect of the sensitivity mismatch between our radio and IR catalogues (Table \ref{tab::sensitivity}) on the IRRC properties studied in Sect. \ref{sect::res}.

\subsection{Infrared and radio luminosity derivations}
\label{sect::calc_lum}

\subsubsection{IR luminosities from SED fitting}
\label{sect::sed_fits}

In order to estimate the total (8 -- 1000\,$\mu$m) infrared luminosity ($L_{\rm TIR}$\footnote{In recent years it has become common practice to denote luminosity in the 8 -- 1000\,$\mu$m range as $L_{\rm IR}$, however, to clearly distinguish between the total infrared and far-infrared radio correlations, we chose this notation throughout the paper.}) of our galaxies, we fitted their IR flux densities with the SED templates of \cite{dale02} \citep[see also][]{dale01}\footnote{We note that, on average, $L_{\mathrm{TIR}}$ values obtained from SED fits using the \cite{chary01} template library are consistent with the ones produced by the \cite{dale02} library within 0.01 dex. We decided to use the latter, because it yielded overall better quality fits and smoother $L_{\mathrm{TIR}}$ posterior distributions.}. This SED library contains IR spectra of different shapes, sorted by their radiation field hardness parameter. We assigned total infrared luminosity values to each template following the relation between $S_{60}/S_{100}$ and $L_{\rm TIR}$ from \citet{marcillac06}. We then sorted the library SED templates according to their IR luminosity, normalised them, and carried out a cubic spline interpolation between the spectra. This allows us to draw not only one of the 119 pre-defined spectra from \cite{dale02}, but transitional shapes between them from the interpolation via a continuous shape parameter, $\gamma$. To relax the assumption made when we ordered the SED library, i.e. that SED shape is intrinsically tied to the IR luminosity of the sources, we fitted $\gamma$ and $L_{\mathrm{TIR}}$ independently of each other by maximizing the logarithmic likelihood function

\begin{equation}
    \mathcal{L} (L_{\mathrm{TIR}},\gamma) = -0.5 \sum_{i=1}^{N} \left( \frac{\log(S_i) - \log(f(\lambda_i, L_{\rm TIR}, \gamma))}{\sigma_i} \right)^{2} + \mathcal{P},
\label{eq::likelihood}
\end{equation}

\noindent where $S_i$ and $\sigma_i$ are the observed flux density and its logarithmic uncertainty\footnote{We approximated logarithmic flux density errors as $\sigma_{i} = 0.434 \cdot \Delta S_{i} / S_{i}$, where $\Delta S_{i}$ is the uncertainty of the flux density in photometric band $i$.} in photometric band $i$. $f(\lambda_i, L_{\rm TIR}, \gamma)$ is the flux density in the same band predicted by the best-fit model with parameters $L_{\rm TIR}$ and $\gamma$ considering the bandpass shape of band $i$, and

\begin{equation}
    \mathcal{P} = 
\begin{cases}
    0,            & \text{if } 10^5 \leq L_{\mathrm{TIR}} \leq 10^{14} \text{ and } 0 \leq \gamma \leq 119 \\
    -\infty,      & \text{otherwise}
\end{cases}
\label{eq::prior}
\end{equation}

\noindent
is the so-called prior function. $\mathcal{P}$ ensures our optimization process probes a physically meaningful parameter space and $\gamma$ is interpolated between the 119 SED templates.

\begin{figure}
\includegraphics[width=\columnwidth]{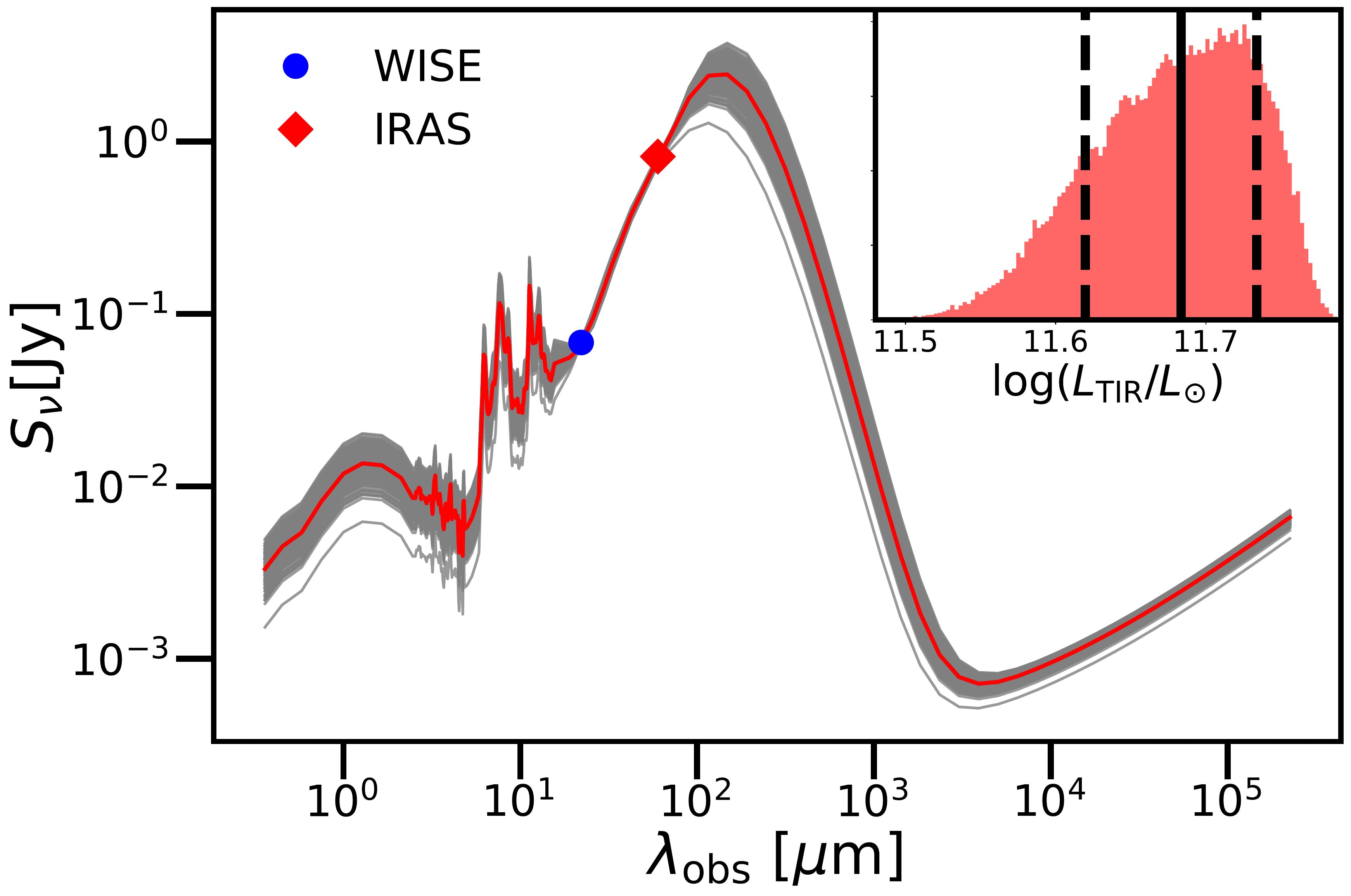}
\includegraphics[width=\columnwidth]{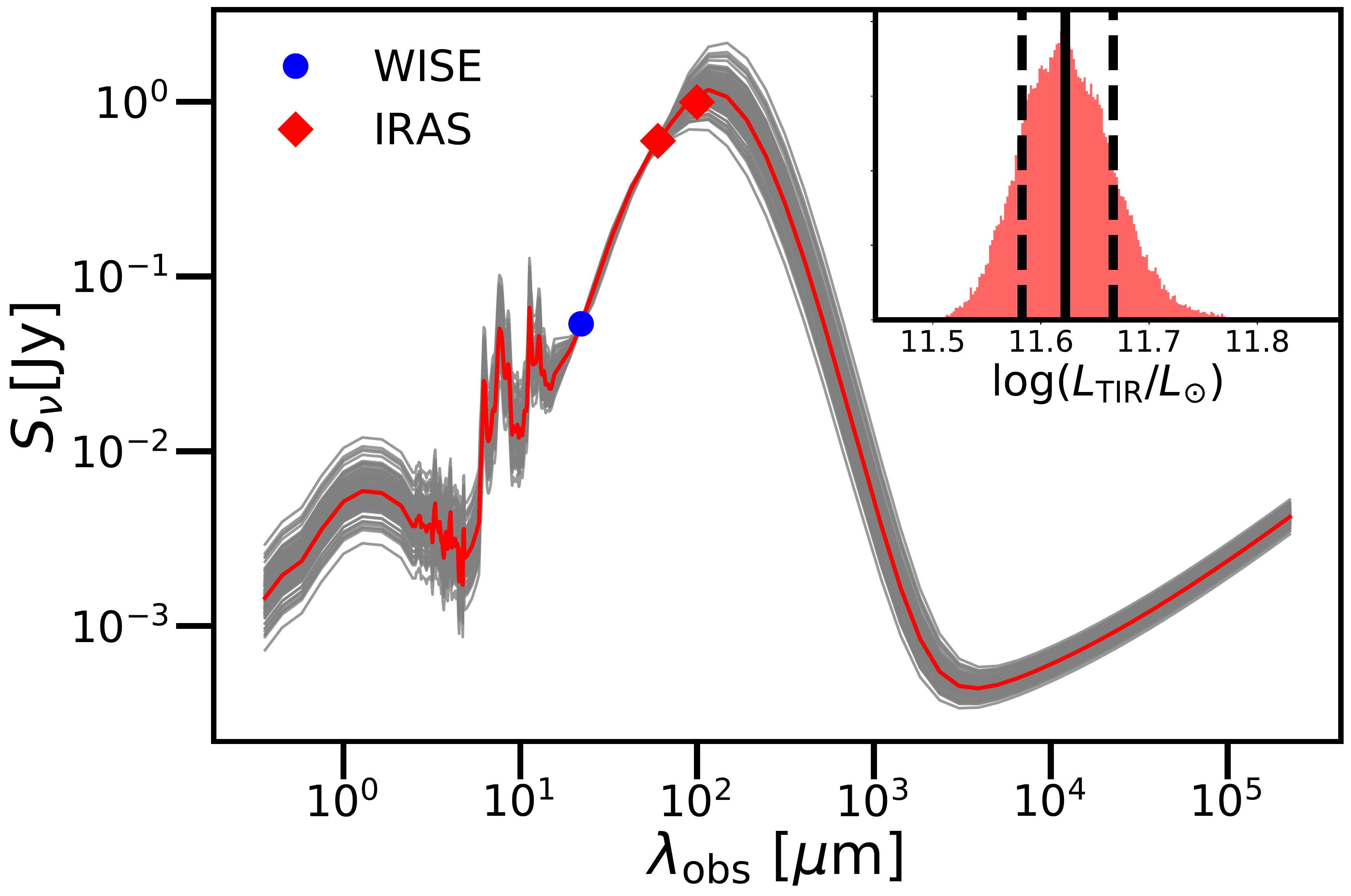}
\includegraphics[width=\columnwidth]{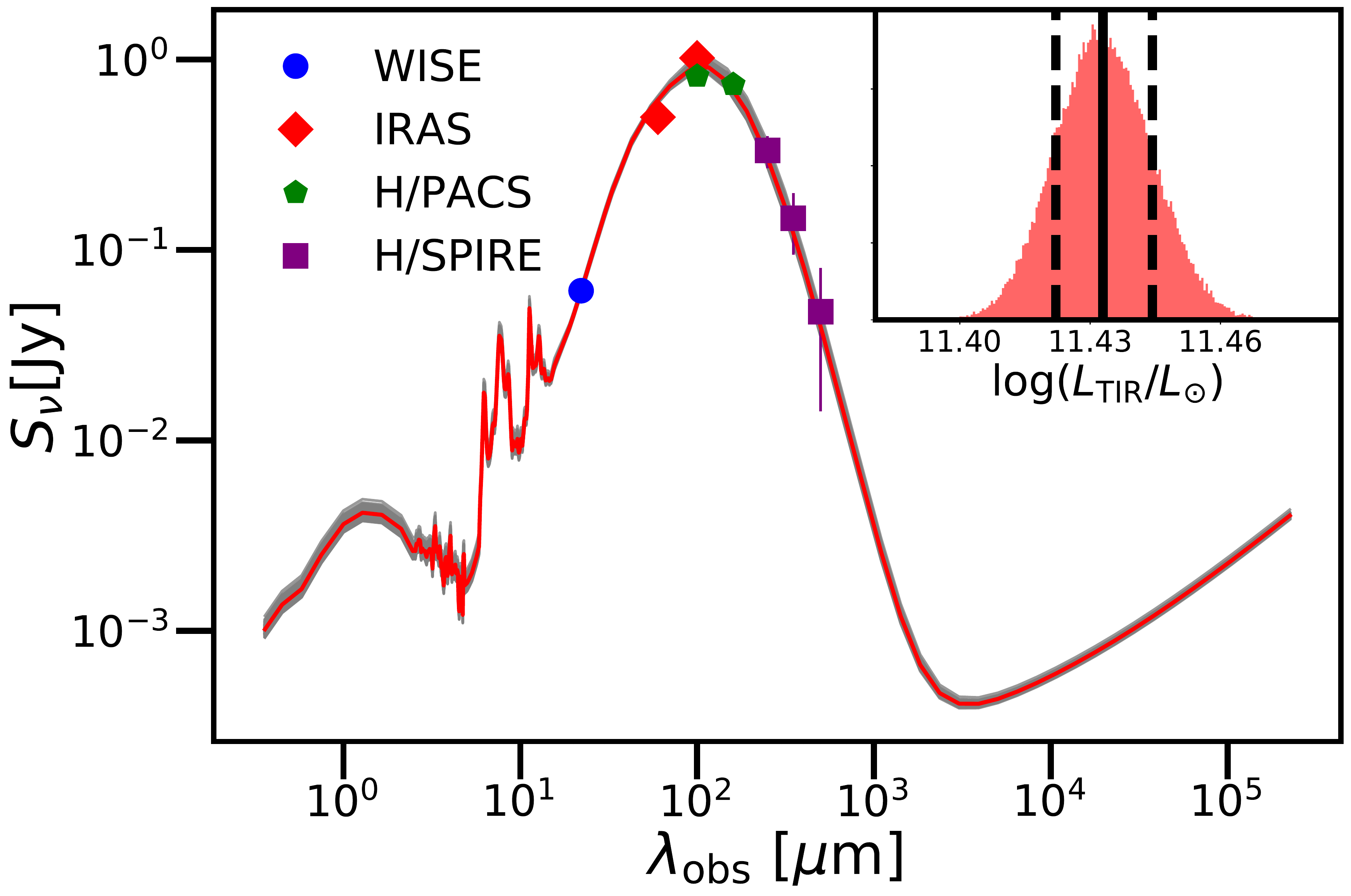}
\caption{Examples of fitted spectral energy distributions with different photometric coverage (i.e. 2, 3 and 6 IR detections of at least 5 SNR from top to bottom), using the templates of \protect\cite{dale02}. Red lines are the best-fit curves, grey ones are 100 randomly selected models from converged MCMC chains, representing the $\sim$1\,$\sigma$ confidence intervals of each fit. The insets show the $L_{\mathrm{TIR}}$ posterior distributions, with the medians (16th/84th percentiles) highlighted by solid (dashed) black lines.}
\label{fig::sed_example}
\end{figure}

To find the best-fit SED model with realistic error estimates on the free parameters, $L_{\rm TIR}$ and $\gamma$, we used the affine invariant Markov chain Monte Carlo (MCMC) sampler \textsc{emcee} \citep{foremanmackey13}, a free, open-source code implemented in \textsc{python}. We initiated 100 walkers with $L_{\mathrm{TIR}}$ values randomly generated according to a uniform distribution between the limits of Eq \ref{eq::prior}, the prior function. In combination with our choice of using $\log(S_{i})$ over $S_{i}$ in the likelihood function (eq. \ref{eq::likelihood}), this setup resulted in the fastest and most robust convergence of the likelihood sampling process. Specifically, we found that the  burn-in period of a typical MCMC chain is $\sim$200 steps. Thus, to achieve sufficient sampling we ran our walkers for 2,200 steps and produced marginalised distributions and statistics after removing the first 200 samples from each, while they were converging on the best-fit parameters. A median acceptance rate\footnote{Acceptance rate or fraction allows for a quick check on MCMC convergence. It is defined as the fraction of proposed steps that are accepted in the chain. An acceptance rate of $\sim$0 indicates that almost all proposed steps are rejected, the chain essentially being stuck and generating very few independent values, such that it does not properly sample the posterior distribution. Vice versa, an acceptance rate of $\sim$1 means that nearly all new steps are accepted, effectively resulting in a random walk, which also does not probe the sought probability density distribution. Depending on the number of free parameters, values between $\sim$0.2 and 0.5 are often considered a sign of a well-sampled posterior distribution.} of 0.4 with a 0.1 standard deviation indicates that the majority of our fits indeed sufficiently converged.

Finally, we assessed the quality of each SED model in order to identify and remove poor SED fits. Even though $\chi^{2}_{\rm red}$ is a widely used metric to judge the goodness of a model fit, it has some potential pitfalls as noted by, e.g., \citet{andrae10}. Chiefly, the number of degrees of freedom for non-linear models (such as our SED templates containing blackbody radiation curves of different temperatures), in general, can be anywhere between 0 and $N - 1$, where $N$ is the number of data points, and may even change during the fit, rendering the use of a single $\chi^{2}_{\rm red}$ value cut to separate poor and robust models inadequate across all our sources which have varying numbers of available photometric bands. To circumvent this issue, we examined the normalised logarithmic residuals, $R_i$, for each source, defined as:

\begin{equation}
    R_i = \frac{\log(S_i) - \log(f(\lambda_i, L_{\rm TIR}, \gamma))}{\sigma_i}.
\end{equation}

Given the data, the true model should produce normalised residuals that follow a standard normal distribution. There are a wide variety of tests to assess whether a set of data is likely to be drawn from such a distribution. However, in order to be sufficiently robust, these require samples larger than the maximum number of 8 measurements an individual galaxy in our analysis can have. Therefore, as a simpler approach, we computed the mean of the normalized residuals, $\mu_0$, for our sources with $N$ photometric datapoints as

\begin{equation}
    \mu_0 = \frac{1}{N} \sum_{i = 0}^{N} R_i.
\end{equation}

\noindent
Since $\mu_0$ can be interpreted as the mean offset between the model and the data in signal-to-noise space, we flagged models as of poor quality, if $\mu_0$ deviated from 0 by more than 1, i.e. on average our model was not consistent with the data within 1\,$\sigma$. We note that inspecting band-by-band normalised residual distributions, we found that on average all of them are consistent with 0 within 1\,$\sigma$ and show no correlation with wavelength, suggesting that there are no statistically significant systematic errors with the flux measurements and that our model library covers the observed IR colour space.

With this method we identified 1,989 sources with fitted models inconsistent with the data, $\sim$21 \% of our overall combined sample. Among star forming galaxies (see Sect. \ref{sect::agn_sfg_sep} for details), 386 (14\%) proved to have poor fits. These were excluded in the subsequent analysis.

Fig. \ref{fig::sed_example} shows typical SEDs with best-fit model and additional randomly drawn models from the MCMC chains representing the 1\,$\sigma$ confidence interval of our fit alongside the marginalised $L_{\mathrm{TIR}}$ distribution from the posterior sampling, which was used to derive $L_{\mathrm{TIR}}$ uncertainties. Typical $L_{\mathrm{TIR}}$ errors in our depth-matched catalogue are 0.12, 0.05 and 0.02\,dex for sources with 2, 3--4, and $>$4 available photometric bands, respectively.

The fitted SEDs were used to derive an empirical K-correction at various wavelengths by taking the ratio of the observed and rest-frame fluxes for each source (see Sect. \ref{sect::lum_lum} for more details). These corrections were applied to the closest adjacent flux measurement when computing the monochromatic IR luminosities presented in Fig. \ref{fig::lum_lum_all}. Finally, for a more direct comparison with e.g. \cite{yun01}, we also calculated the far-IR luminosity, $L_{\mathrm{FIR}}$, for each galaxy by integrating our best-fit SED models between 42.5 \-- 122.5\,$\mu$m.

\subsubsection{1.4\,GHz radio luminosity estimates}

Radio flux densities were converted into 1.4\,GHz rest-frame radio continuum luminosities, $L_{1.4}$, using

\begin{equation}
\label{eq::L_1.4}
\left(\frac{L_{1.4}}{\rm W\,Hz^{-1}}\right) = C_1\,\frac{4\pi}{(1+z)^{(1+\alpha)}}\,\left(\frac{D_L}{\rm Mpc}\right)^2\,\left(\frac{S_\mathrm{1.4}}{\rm mJy}\right)~,
\end{equation}

\noindent
where $C_1 = 9.52\times10^{15}$ is the conversion factor from $\mathrm{Mpc^2\,mJy}$ to $\mathrm{W\,Hz^{-1}}$, $\alpha$ is the radio spectral index\footnote{The radio spectral index is defined as $S_{\nu} \propto \nu^{\alpha}$, where $S_{\nu}$ is the flux density at frequency $\nu$.}, $z$ is redshift, $D_L$ is the luminosity distance and $S_{\mathrm{1.4}}$ is the measured 1.4 GHz flux density. We used the typical $\alpha = -0.7$ assumption \citep[e.g.][]{kimball08}.

\subsection{Identification and removal of AGN host galaxies}
\label{sect::agn_sfg_sep}

It is generally assumed that the IRRC emerges from the correlation of IR and radio flux densities with star formation activity. Since our primary aim is to study this relation, we selected sources in our sample identified as predominantly star-forming. On the other hand, since the IRRC can also be used to detect excess radio emission presumably linked to AGN activity, we also investigated AGN in our sample.

\subsubsection{Classification based on optical emission lines}
\label{sect::bpt}

For galaxies with SDSS DR 8 emission line measurements from the value-added MPA/JHU group \footnote{\url{http://www.sdss3.org/dr10/spectro/galaxy_mpajhu.php}}, we classify galaxies as star-forming, AGN or composite following the method presented in \cite{kewley06} (and first introduced by \citealt{baldwin1981}). The method makes use of the emission line ratios [NII]6584/H$\alpha$ , [SII]6717,6731/H$\alpha$, [OI]6300/H$\alpha$, and [OIII]5007/H$\beta$ which are sensitive to the metallicity and ionization properties of the gas. 
First, we use all the diagnostic line ratios to classify star-forming galaxies using the theoretical ``maximal starburst line" derived by \cite{kewley01}, indicating the theoretical maximum line ratios that could be produced by pure stellar photoionization models alone. 

We then identify galaxies which are classified as star-forming according to their [SII]/H$\alpha$ and [OI]/H$\alpha$ ratios, but fall in the ``composite'' region in the [NII]/H$\alpha$ vs. [OIII]/H$\beta$ diagnostic according to the empirical \cite{kauffmann03} boundary. These are galaxies with a composite spectrum containing a mix of HII region emission and a harder ionizing source, and are thus not included in our SFG sample.

We classify as AGN the sources that lie above the ``maximal starburst line" in the [SII]/H$\alpha$ and [OI]/H$\alpha$ line ratio diagnostics and above the \cite{kauffmann03} line in the [NII]/H$\alpha$ diagnostic diagram.

Sources we label as unclassified lack sufficiently high significance line measurements (we require SNR $>$ 3 in H$\alpha$ following \cite{leslie16}) or observed spectra altogether. Even though it is possible to utilize e.g. a colour-colour selection of SFGs and AGN to increase our sample size, the unclassified population has both radio and IR luminosity distributions coincident with the classified ones, and thus by relying solely on emission lines we retain the ability to probe the entire luminosity regime available in our combined sample.

\subsubsection{Identifying AGN using mid-IR colours}
\label{sect::mir_agn}

\cite{assef18} selected AGN candidates using the WISE 3.4 and 4.6~$\mu$m bands. The selection criteria were calibrated and assessed based on UV- to near-IR spectral energy distribution analysis of AGN in the NOAO Deep Wide-Field Survey Bo\"{o}tes field where deep WISE data are available.
We use their criteria that select AGN candidates with a 90\% confidence. Galaxies meeting the following criteria are flagged as MIR-selected AGN, where W1 and W2 are the WISE 3.4 and 4.6~$\mu$m Vega magnitudes, $\mathrm{SNR}$ is SNR, and ccflags = 0\footnote{As described in the Explanatory Supplement to WISE (\url{https://wise2.ipac.caltech.edu/docs/release/allsky/expsup}), ccflag stands for contamination and confusion flag. It indicates whether a source may be affected by a nearby imaging artifact. A value of $0$ indicates it is not.} for W1 and W2:
 \begin{align*}
 \mathrm{W1}&>8, \\
 \mathrm{W2}&>7,\\
 \mathrm{SNR}_\mathrm{W2}&>5,\\
 \mathrm{W1}-\mathrm{W2} &>\begin{cases} 
      a \exp[b( \mathrm{W2}-c^2)] &  \mathrm{W2}>c  \\
      a & \mathrm{W2}<c,
   \end{cases}
 \end{align*}
where ($a,b,c$) = (0.650, 0.154, 13.86). 
This selection identifies 4,470 MIR-AGN candidates in the joint catalogue (5\%) and these AGN candidates tend to be more luminous than the optically selected AGN candidates. Only 828 (186) objects in the IR-detected (combined IR and radio detected) catalogue satisfy both the MIR and optical AGN selection criteria, indicating the importance of a multiwavelength approach for selecting all types of AGN.

\begin{table}
    \caption{Number of SFG, AGN and composite sources classified by optical emission lines (Sect. \ref{sect::bpt}) in the combined sample and its subset of depth-matched galaxies. Bracketed numbers correspond to sources identified as MIR AGN in each sample (Sect. \ref{sect::mir_agn}). These were excluded from our analysis due to their most likely inaccurate $L_{\rm TIR}$ estimates. Unclassified galaxies lack sufficiently high quality spectra that permits classification.}
    \label{tab::type_stats}
    \centering
    \begin{tabular}{cc|c}
        & Combined & Depth-matched \\ \hline\hline
        SFG & 2,495 (54) & 2,093 (46) \\
        AGN & 313 (80) & 239 (67)\\
        Composite & 1,417 (77) & 1,107 (68)\\
        Unclassified & 3,431 (293) & 1,743 (190) \\ \hline
        Total & 7,656 (504) & 5,182 (371)
    \end{tabular}
\end{table}

As described in Sect. \ref{sect::sed_fits}, the templates used for IR SED fitting assume that the IR emission arises purely from star formation. If a source has a non-negligible AGN-related MIR component, which in most cases enhances only the MIR flux but not the FIR ones, our fitted $L_{\rm TIR}$ values are very likely to be overestimated. We therefore excluded all 504 AGN identified at MIR wavelengths. This leaves 2,441 SFGs and 233 AGN in our sample that show no sign of AGN activity at MIR wavelengths, and have reliable $L_{\rm TIR}$ estimates according to our residual analysis described in Sect. \ref{sect::sed_fits}.

\subsection{Depth homogenization and depth-matched sample}
\label{sect::fluxmatch_samp}

Up to now our selection strategy has produced a sample of 7,656 jointly IR- and radio-detected z $<$ 0.2 galaxies with robust SED fits, of which 2,441 were identified as high-confidence SFGs in the preceding section. We now consider the bias affecting IRRC statistics arising from sensitivity differences between IR and radio \cite[see, e.g.,][]{sargent10a}. To assess the possibility of such an effect on our study, we compared the depths of the data sets used to create our catalogue.

\begin{figure}
\includegraphics[width=\columnwidth]{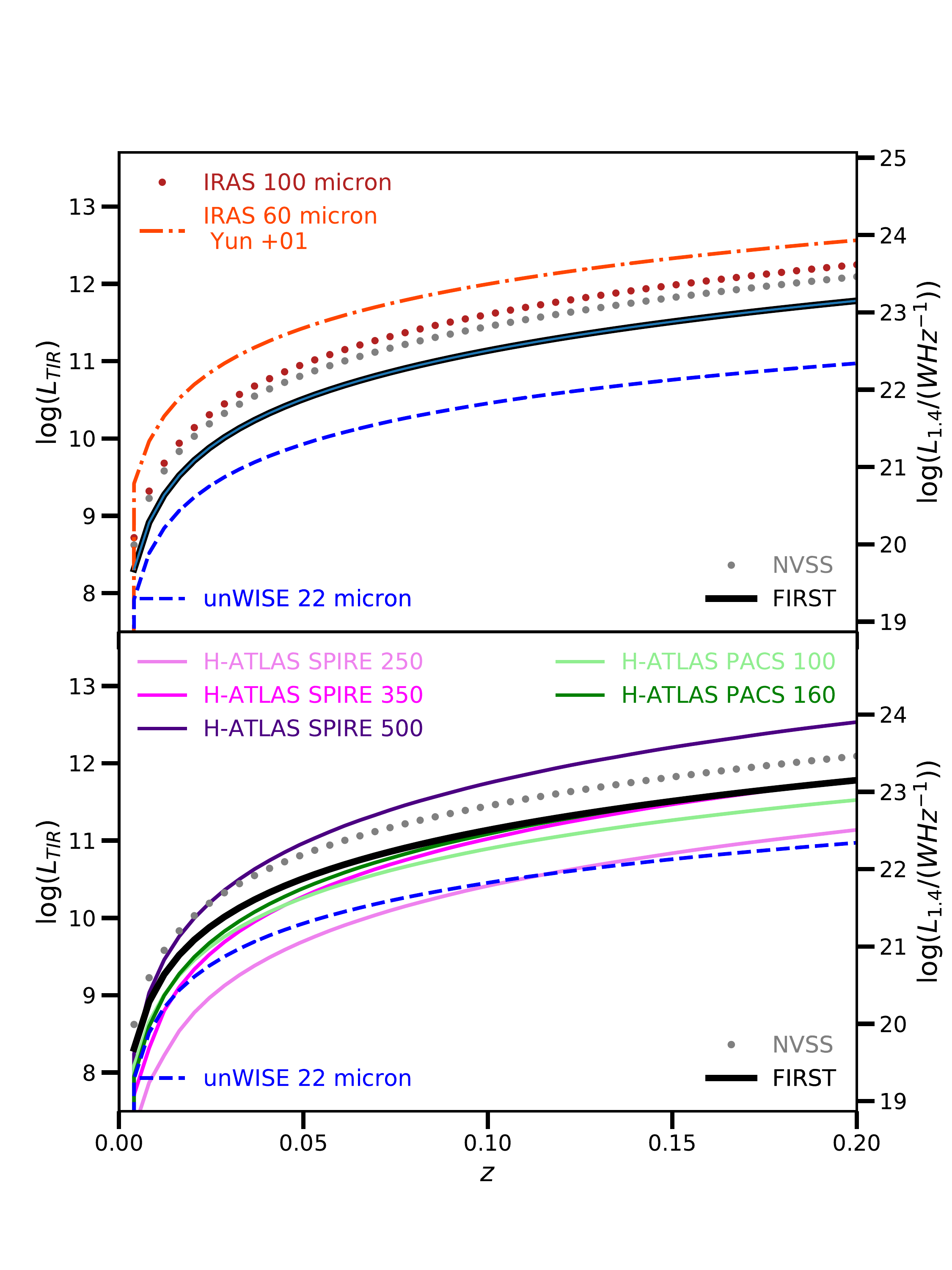}
\caption{Redshift-dependent $L_{\mathrm{TIR}}$ sensitivity for the data sets used in this work. Radio luminosity limits were converted to $L_{\mathrm{TIR}}$ with the canonical $q_{\mathrm{TIR}}\,=\,2.64$ of \protect\cite{bell03}. IRAS (H-ATLAS) sensitivity curves are displayed in the upper (lower) panel. For easier reference, FIRST, NVSS and WISE 22\,$\mu$m limits are shown in both panels.}
\label{fig::tir_lim}
\end{figure}

Fig. \ref{fig::tir_lim} shows the total IR and radio luminosity limits as a function of redshift estimated for the various photometric bands and surveys in our catalogue. For each IR band/survey, at a given redshift, we calculate the predicted flux density value of every SED template in the \cite{dale02} library taking the transmission curves of the specific instrument into account and plot the $L_{\mathrm{TIR}}$ value of the SED template that reproduces the 5\,$\sigma$ flux density limit of the respective catalogue. This approach relies on the assumption that the IR colour -- $L_{\mathrm{TIR}}$ relation in the template library holds. Each individual survey is sensitive to galaxies above its corresponding curve in Fig. \ref{fig::tir_lim}. For radio surveys FIRST and NVSS, we calculate their radio luminosity sensitivity curves by substituting their flux detection limits into Eq. \ref{eq::L_1.4} in our redshift range. These are then subsequently converted into $L_{\mathrm{TIR}}$ by re-scaling them with the currently widely adopted value of the local $q_{\mathrm{TIR}}$ = 2.64 of \cite{bell03}.

Barring the 250\,$\mu$m H-ATLAS coverage, which contributes only $\sim$2\% to the footprint of our sample, unWISE 22\,$\mu$m observations provide the deepest data in our catalogue. The 100, 160 and 350\,$\mu$m H-ATLAS Herschel data are matched quite well in sensitivity with FIRST. Lastly, NVSS, IRAS and H-ATLAS 500\,$\mu$m provide shallower measurements relative to FIRST. NVSS and IRAS encompass a larger area and probe similar luminosity regimes in $z\,<$\,0.2 galaxies. However, the 2 Jy IRAS 60 $\mu$m flux cut imposed by \cite{yun01} results in significantly shallower IR coverage than the radio data from NVSS in their sample. The consequences of such a mismatch between IR and radio measurements are explored in Sect. \ref{sect::bias}.

Following \cite{sargent10a}, to avoid selection effects  biasing our IRRC measurement and ensure that both our radio and IR data have comparable depths, we applied a flux cut of 19.5\,mJy to our unWISE 22\,$\mu$m detections. This flux limit was calculated using the ratio between unWISE and FIRST luminosity limits (Fig. \ref{fig::tir_lim}) at $z$\,=\,0.046, the median redshift of our combined sample. However, this simple flux cut leaves 1,612 galaxies in the catalogue that are outside of the FIRST footprint, and are only covered by the shallower NVSS. These sources would require a $\sim$0.4 dex higher 22\,$\mu$m flux cut in order to match IR and radio sensitivity in this region. In principle, we could define a second sensitivity tier in our catalogue and include these objects, however, firstly, only $\sim$5\% of them have high quality spectroscopy enabling an SFG/AGN classification based on spectra, and secondly, their overall IR and radio luminosity distribution closely match the luminosity range of the rest of our sample, and thus their inclusion would not improve our ability to probe the IRRC. Therefore, we simply discard all sources that were not covered by the FIRST survey when defining our final, depth-matched sample which includes 6,611 galaxies. The effects of unmatched radio and IR luminosity limits are further investigated in Sect. \ref{sect::bias}. The number of sources available in each photometric band across all wavelengths in the combined and the depth-matched catalogues is shown in Table \ref{tab::cat_stats}.

Finally, to measure the unbiased monochromatic IRRC in each of our available IR bands, we introduce band-by-band flux cuts to apply to either IR or radio fluxes, depending on which is deeper in relative terms. This was done via the same method as outlined above for the 22\,$\mu$m data, using the median redshift of sources jointly detected in the radio and in a given IR band. Due to the heterogeneous nature of the fields used for the PPSC and SPSC catalogues it is not straightforward to quantify their overall sensitivity. We thus limit ourselves to H-ATLAS data when constructing the monochromatic IRRCs for the different Herschel bands (i.e. we do not consider galaxies which only have PPSC or SPSC flux measurements). We report all adopted band-by-band flux cuts in Table \ref{tab::fluxlims}, and show them in Fig. \ref{fig::flux_flux}. We note that IRAS 100\,$\mu$m and Herschel 500\,$\mu$m data, unlike all other IR bands, are less sensitive than the radio coverage, such that we have to apply a cut to the 1.4\,GHz flux distribution, rather than to the 100 or 500\,$\mu$m fluxes. These monochromatic depth-matched samples were only used when examining monochromatic IRRCs. For all other aspects of our analysis, when referring to the ``depth-matched sample", we mean the subset of the combined sample which was selected via a single flux cut in the selection band at 22$\mu$m.

\begin{table}
    \caption{Selection cuts in each IR band calculated at the median redshift, $\overline{z}$, of galaxies detected in the appropriate band. The WISE 22\,$\mu$m flux cut was used when defining our depth-matched sample. All other cuts are instead applied exclusively when fitting monochromatic IRRCs. Both the IRAS 100\,$\mu$m and the H-ATLAS 500\,$\mu$m data are IR-limited, i.e. the radio data are deeper than the IR, and thus require a flux density cut at 1.4\,GHz, as indicated by the third column of the table.}
    \label{tab::fluxlims}
    \centering
    \begin{tabular}{c|c|c|c}
        IR band & flux limit & $\overline{z}$ & flux cut\\
        & [mJy] & & applied to\\ \hline\hline
        WISE 22\,$\mu$m & 19.5 & 0.046 & IR \\ \hline
        IRAS 60\,$\mu$m & 233.6 & 0.044 & IR \\
        IRAS 100\,$\mu$m & 3.5 & 0.035 & radio \\
        H-ATLAS 100\,$\mu$m & 419.9 & 0.050 & IR \\
        H-ATLAS 160\,$\mu$m & 313.6 & 0.049 & IR \\
        H-ATLAS 250\,$\mu$m & 165.5 & 0.051 & IR \\
        H-ATLAS 350\,$\mu$m & 41.6 & 0.049 & IR \\
        H-ATLAS 500\,$\mu$m & 4.3 & 0.032 & radio \\
    \end{tabular}
\end{table}

\section{Results}
\label{sect::res}

\begin{figure*}

\includegraphics[width=0.92\textwidth]{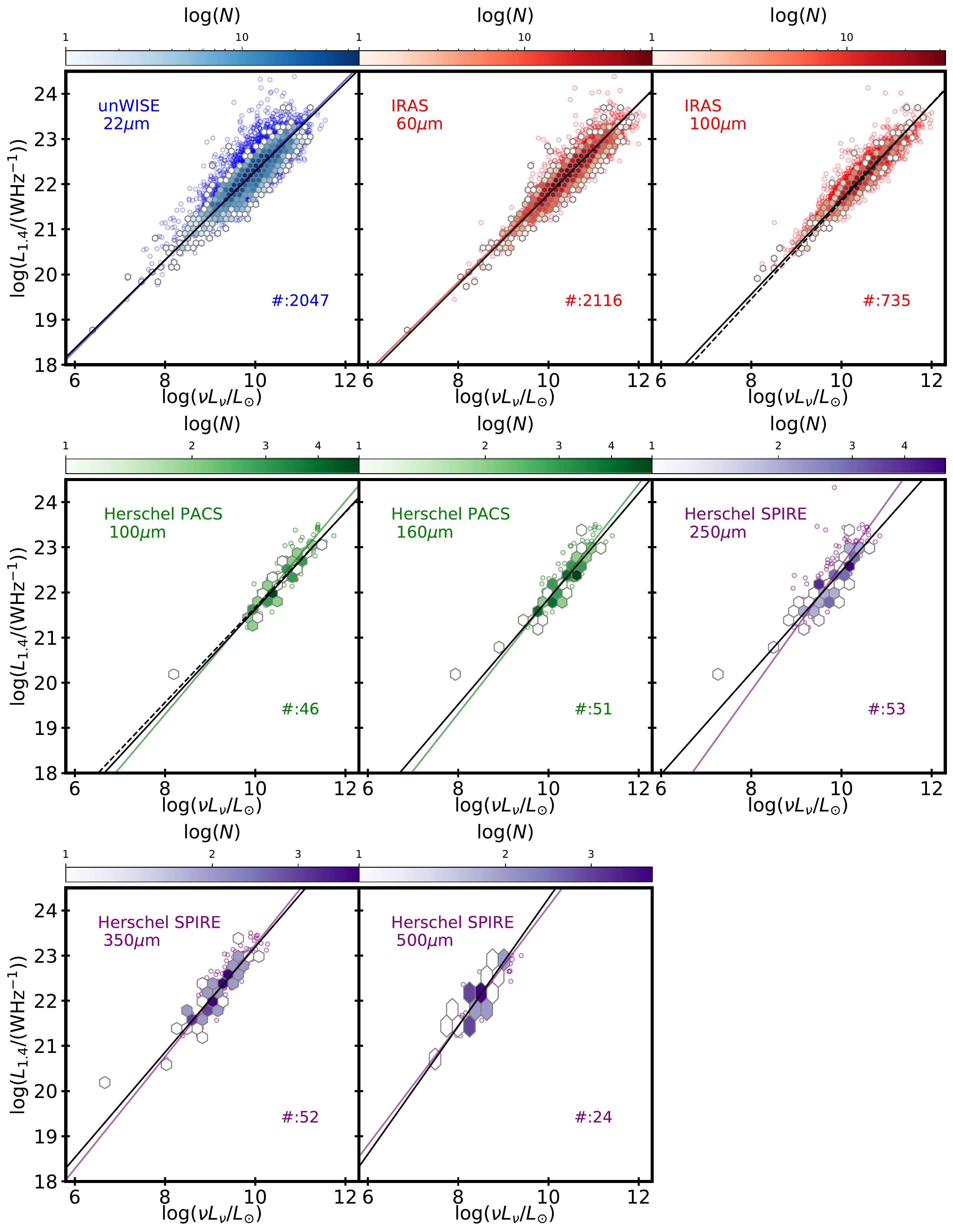}
\caption{Radio luminosity at 1.4 GHz as a function of different monochromatic IR luminosities. Coloured hexagonal bins (empty circles) represent sources in the depth-matched SFG sample (full combined sample) with a $>$5 SNR detection in the IR band of the respective panel (see Table \ref{tab::fluxlims} for the selection criteria in each band). Black lines are the best-fit linear relations using depth-matched SFGs, while coloured lines are fits to the full combined sample. The dashed lines in the two 100\,$\mu$m panels show the best fit, respectively, to the measurements of the other 100\,$\mu$m data set. The number of sources used in the fit is specified in the bottom right quarter of each panel. Table \protect\ref{tab::lum_lum} reports all best-fit slope and dispersion values for the depth-matched SFGs, and additionally also for the combined sample and the entire depth-matched sample.}
\label{fig::lum_lum_all}
\end{figure*}

\subsection{Monochromatic infrared-radio correlations}
\label{sect::lum_lum}

For the $>$5\,$\sigma$ detections in each IR band we calculate monochromatic luminosities, $\nu L_{\nu}$, at the rest-frame frequency $\nu$ (with $\nu$ corresponding to the characteristic wavelength/frequency of each band). Measured flux densities were converted to rest-frame monochromatic luminosities via:

\begin{equation}
    \left(\frac{\nu L_{\nu}}{L_{\odot}}\right) = C_2 \, 4\pi \left(\frac{D_L}{\rm Mpc}\right)^2\, K\, \left(\frac{S_{\nu,\mathrm{obs}}}{\rm Jy}\right)\,\left(\frac{\nu}{\mathrm{Hz}}\right)~,
\end{equation}

\noindent
where $C_2 = 3.64\cdot10^{-7}$ is the conversion factor from Mpc$^2$\,Jy\,Hz to solar luminosity and $K$ is the K-correction factor containing both colour correction (computed as the ratio of $S_{\nu}$ and $S_{\nu,\mathrm{obs}}$, i.e. flux density at the observed wavelength) and bandpass compression terms. Uncertainties on the adopted $K$-correction values were calculated by re-sampling SEDs from the posterior distributions of $\gamma$. Due to the low redshift of our sample, these proved to be negligible compared to the uncertainty of the fluxes, which thus dominate the error budget of the $\nu L_{\nu}$ measurements.

As Fig. \ref{fig::lum_lum_all} shows, all observed monochromatic IR luminosities correlate with 1.4\,GHz radio continuum luminosity. Each panel contains only SFG sources from samples depth-matched on a band-by-band basis (see Table \ref{tab::fluxlims} for the selection criteria). Arguably, it is not appropriate to treat either the radio or the IR luminosities as the independent variable. Correspondingly, \cite{bell03} carried out a bisector fit to determine the slope of the relation. However, as \cite{hogg10} pointed out, it is preferable to adopt other approaches for linear regression. Therefore, we inferred the best-fit model

\begin{equation}
\label{eq::irrc_fit}
     \log\left(\frac{L_{1.4}}{\rm W\,Hz^{-1}}\right) = m \cdot \log\left(\frac{\nu L_{\nu}}{L_{\odot}}\right) + b
\end{equation}

\noindent
with the bivariate correlated errors and intrinsic scatter \citep[BCES;][]{akritas96,nemmen12} method, in particular by minimizing the squared orthogonal distances to the modelled relation. We measure the dispersion as the standard deviation of the orthogonal offset distribution of the data relative to the best-fit model. 

Table \ref{tab::lum_lum} contains the slopes ($m_{\mathrm{all}}$, $m_{\mathrm{dm}}$ and $m_{\mathrm{dmSFG}}$), intercepts ($b_{\mathrm{all}}$, $b_{\mathrm{dm}}$ and $b_{\mathrm{dmSFG}}$) and dispersions ($\sigma_{\mathrm{all}}$, $\sigma_{\mathrm{dm}}$ and $\sigma_{\mathrm{dmSFG}}$) of all monochromatic IRRCs for the full combined sample, the depth-matched sample (see Sect. \ref{sect::bias}) and its subset of depth-matched SFGs, only using galaxies with at least 5\,$\sigma$ flux density measurements in a given IR band. Removing AGN from the samples reduces the dispersion of all correlations, in some cases by almost 50\%, and in particular for Herschel data it systematically brings the IRRC slope closer to unity. This is due to the radio-loud AGNs that tend to be high radio luminosity outliers and are typically IR luminous as well, thus simultaneously steepening the IRRC slope and adding to its dispersion. We include the fits to the full combined sample in order to illustrate the effect our flux-matching approach has on the derived monochromatic IRRC parameters. The most striking example of this bias occurs when we compare the 100\,$\mu$m IRRCs based on IRAS and H-ATLAS data. Initially, with no flux cut applied, their slopes are inconsistent at the $\sim$2.5\,$\sigma$ level, but fitting their depth-matched SFG subsamples we find IRRC parameters consistent within 1\,$\sigma$.

\begin{figure}
\includegraphics[width=\columnwidth]{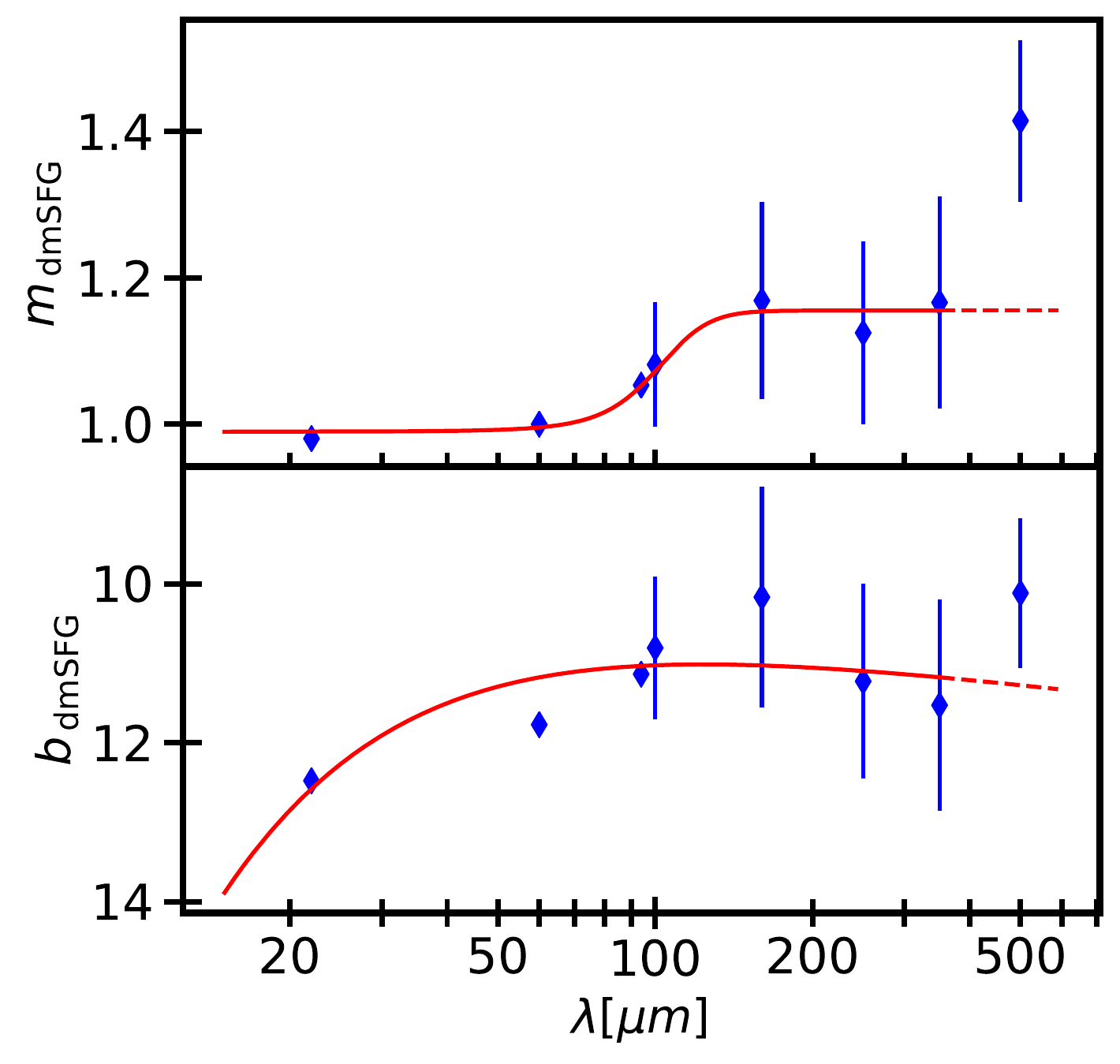}
\caption{Measured monochromatic IRRC slopes and intercepts from Table \protect\ref{tab::lum_lum}/Fig. \ref{fig::lum_lum_all} as a function of their wavelengths. Empirical approximations (red lines) give a calibration for any $\nu L_{\nu}$ -- $L_{1.4}$ relation between 22 and 500\,$\mu$m. We plot the IRAS 100\,$\mu$m data points at 94\,$\mu$m to visually demonstrate the difference in uncertainties between IRRC parameters from Herschel and IRAS at 100\,$\mu$m. Our choice to fit $L_{\rm 1.4}$ as a function of $\nu L_{\nu}$ mirrored the grey body-like trend in the bottom panel, and thus for visualization purposes we inverted the y-axis.}
\label{fig::slope_wl}
\end{figure}

Overall, as shown in Fig. \ref{fig::slope_wl}, we find slopes ($m_{\lambda}$) near unity below 100\,$\mu$m transitioning to slopes of $\sim$1.2 in the 160 -- 350\,$\mu$m regime. This change of IRRC monochromatic slope values likely evidences the transition from bands probing warmer dust emission (which more closely correlates with on-going star-formation activity) to a regime sampling colder dust components in the interstellar medium below and above 100\,$\mu$m, respectively. At the same time, we see a decrease in best-fit intercepts ($b_{\lambda}$) which reaches a minimum around 100-160\,$\mu$m, and then again rises up to 350\,$\mu$m (see lower panel of Fig. \ref{fig::slope_wl}). Considering the broadly similar slope values at all wavelengths (within 20\%), to zeroth order we expect the intercepts to generally reflect the changing SED amplitude at the respective wavelengths, with the addition that our choice to fit $L_{\rm 1.4}$ as a function of $\nu L_{\nu}$ mirrored this trend, and thus for visualization purposes we inverted the y-axis of in the bottom panel of Fig. \ref{fig::slope_wl}.

With this physical picture in mind we empirically approximated the wavelength ($\lambda$) dependence of $m_{\lambda}$ as:

\begin{equation}
\label{eq::mlambd}
     m_{\lambda} = 0.08 \cdot \tanh \left[ 0.04 \cdot \left( \frac{\lambda}{\mu m} - 100 \right) \right] + 1.07,
\end{equation}

\noindent
and the $\lambda$ vs. $b_{\lambda}$ data with a grey body-like model:

\begin{equation}
\label{eq::blambd}
    b_{\lambda} = \log(4{\times}10^{6}) - \log\left( \frac{\lambda^{-1.9}}{(e^{175/\lambda}-1)}\right).
\end{equation}

\noindent 
The uncertainties on our best-fit monochromatic IRRC measurements are strongly dominated by sample sizes, and thus for optimizing the models that describe their $\lambda$ dependence, we weighted each point equally. As a result, our trends should be considered tentative exploratory models rather than fully-realised fits. Furthermore, as seen in Fig. \ref{fig::slope_wl}, the 500\,$\mu$m measurement is an outlier to the general behaviour of other bands. According to our simple physical interpretation outlined above, it should have a slope similar to the 250 and 350\,$\mu$m measurements, and correspondingly a smaller intercept than measured at 350\,$\mu$m. We note, however, that this band is by far the most sensitive to the depth matching approach we employ due to the small numbers of detections involved. Indeed, a small change of $\sim$0.1\,dex in the flux density cut applied to the 1.4 GHz detections alters the best-fit $m_{500}$ and $b_{500}$ values such that they become consistent with the estimates in the other SPIRE bands, and such that the 500\,$\mu$m measurement would conform much better to our simple physical interpretation of the slope and intercept variations of monochromatic IRRCs.

Ideally, an analysis of the monochromatic IRRC parameters as a function of wavelength would be based on datasets of similar sizes, or even the same galaxies altogether at all wavelengths. Nevertheless, our Eqs \ref{eq::mlambd} and \ref{eq::blambd} provide a tool to estimate the properties of monochromatic IRRCs across a wide IR wavelength range. In particular, this will be useful when studying high-z galaxies. At sub-millimetre wavelengths instruments such as the Atacama Large Millimeter Array or the James Clerk Maxwell Telescope sample the peak of IR SED of redshift of 2 -- 3 galaxies. Meanwhile sub-1\,GHz observations from the Giant Metrewave Radio Telescope, the Low-Frequency Array or MeerKAT are capable of detecting 1.4\,GHz radio emission from the same sources. One could envisage even more combinations of monochromatic IR and radio observations across a wide redshift range, which will be compatible with each other through empirical formulae such as Eqs \ref{eq::mlambd} and \ref{eq::blambd}, and as a result, will help constraining the evolution of the IRRC already from suitably chosen single-band observations.

Finally, to facilitate comparison with other studies of the monochromatic IRRC, we computed the more commonly used $q_{\nu} = \log(L_{\nu} / L_{\mathrm{1.4}})$ values in each band. In Table \ref{tab::mono_q} we report our median $q_{\nu}$ measurements derived in two different ways. On the one hand, we used our best-fit monochromatic IRRC models (parametrised by $m_{\mathrm{flSFG}}$ and $b_{\mathrm{flSFG}}$ from Table \ref{tab::lum_lum}) and substituted the median $\log(L_{\nu})$ value into Eq. \ref{eq::irrc_fit} to calculate the expected median $\log(L_{\mathrm{1.4}})$ and consequently median $q_{\nu}$ in each band (denoted as $\overline{q}_{\nu,\mathrm{fit}}$ in Table \ref{tab::mono_q}). On the other hand, we also calculated individual $q_{\nu}$ values for galaxies shown in Fig. \ref{fig::lum_lum_all} and measured the median of their resulting distribution ($\overline{q}_{\nu}$ in Table \ref{tab::mono_q}). Due to the small sample size of the depth-matched H-ATLAS subsets, the statistical error on the median $\log(L_{\nu})$ is quite large, leading to large uncertainties on the $\overline{q}_{\nu,\mathrm{fit}}$ values for Herschel bands. Nevertheless, our $\overline{q}_{250} = 2.03 \pm 0.03$, calculated as the median of all $q_{250}$ in the depth-matched 250\,$\mu$m sample matches well with the $\overline{q}_{250} = 2.01 \pm 0.04$ of \cite{jarvis10}. The $q_{250} = 1.95 \pm 0.2$ measurement of \citet{gurkan18} is also compatible with our results. We note that their measured 250\,$\mu$m IRRC slope of 0.96 $\pm$ 0.01 is shallower than our 1.12 $\pm$ 0.12, but still consistent within $\sim$ 1.5 $\sigma$. Meanwhile, \citet{read18} reported a higher $q_{250} = 2.30 \pm 0.04$ using the same catalogue as \citet{gurkan18}, and the  $q_{250} \approx 2.61$ of \citet{smith14} is even more offset from our median $q_{250}$. However, the latter study was based on a sample selected at 250\,$\mu$m, and thus it is likely biased towards high $q_{250}$ values. As this example shows, and for reasons explained in the following Sect. and in Sect. \ref{sect::bias} and \ref{sect::q_vs_lum}, we caution against comparing median IR-radio ratios without considering the IR and radio luminosity coverage of particular samples.

\begin{figure*}
\includegraphics[width=0.48\textwidth]{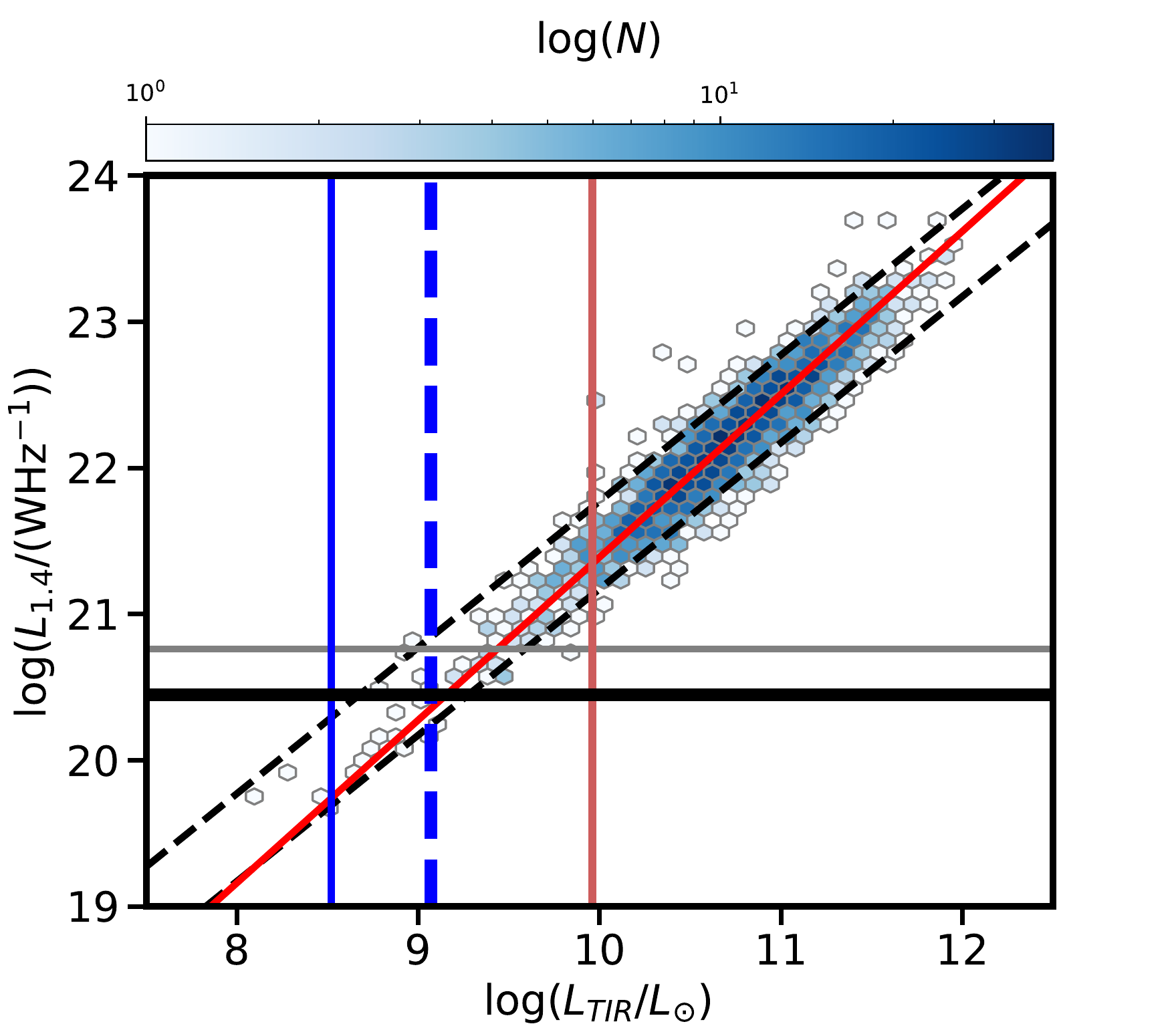}
\includegraphics[width=0.455\textwidth]{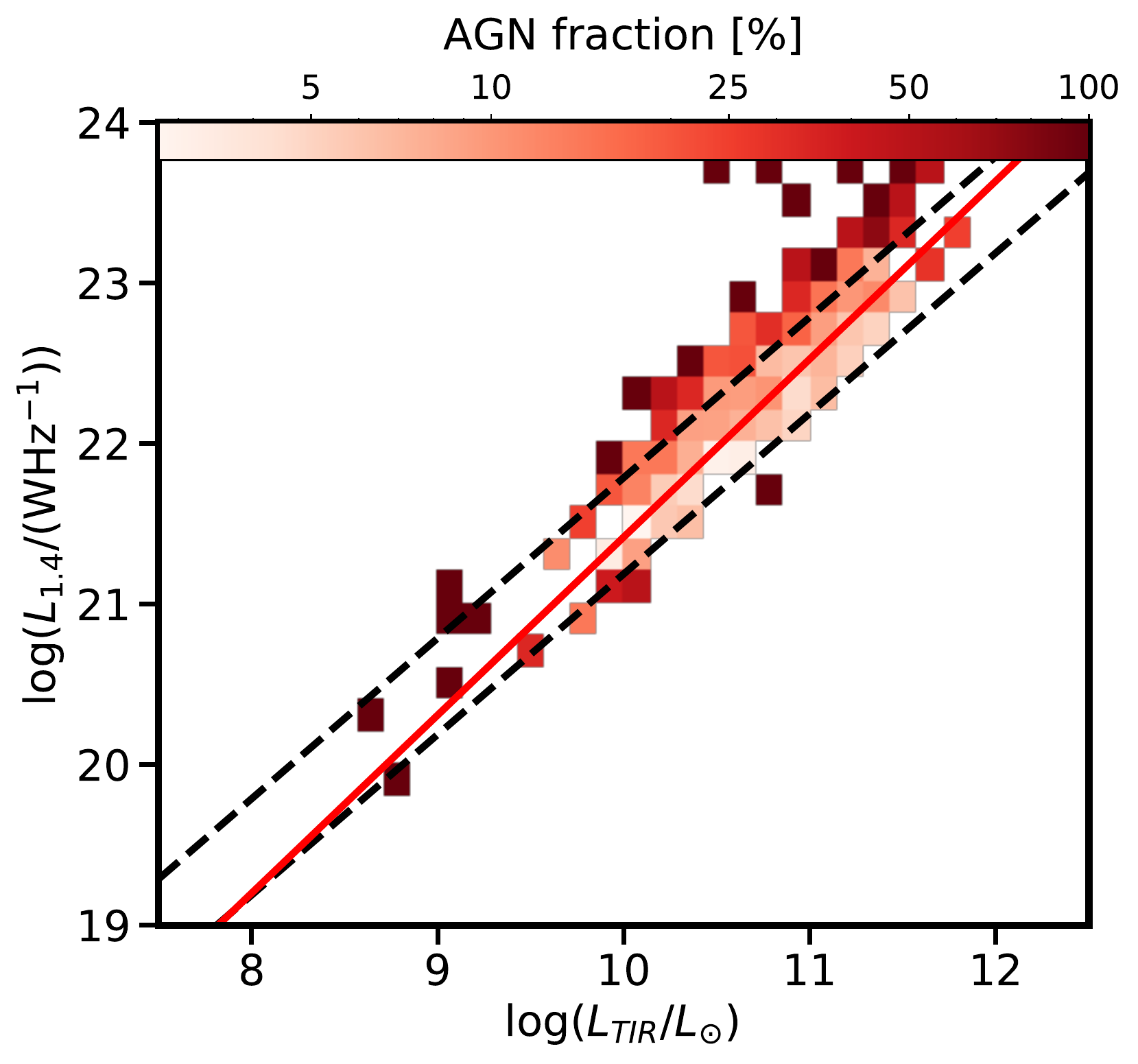}

\caption{\textbf{Left:} 1.4\,GHz luminosity as a function of the total infrared luminosity in the depth-matched SFG sample. The red line is our best-fit linear model, while dashed black lines are drawn at constant $\mathrm{q}_{\rm TIR}$ values of $\mathrm{\overline{q}}_{\rm TIR}{\pm}$0.3\,dex. Table \protect\ref{tab::lum_lum} lists the slope and scatter of the best-fit relation for the depth-matched SFGs, and additionally also for the combined sample and the entire depth-matched sample. The 2\,$\sigma$ confidence band has a similar width as the best-fit line. Vertical and horizontal lines illustrate the depth of various surveys at $z$\,=\,0.01: the blue solid line is the $\log(L_{\rm TIR} / L_{\odot})$ $\sim$ 8.7 limit of unWISE, the red vertical one is drawn at $\log(L_{\rm TIR} / L_{\odot}) = 9.54$ for IRAS 100, and the horizontal black and grey lines are the FIRST and NVSS limits of $\log(L_{1.4} / (WHz^{-1})) =$ 20.35 and 20.75, respectively. The dashed blue line is the luminosity limit at z = 0.01 in our depth-matched sample.\newline
\textbf{Right:} Fraction of AGN hosts on the $L_{\rm TIR}$ -- $L_{1.4}$ plane in our depth matched sample after the removal of MIR AGN. The largest AGN fractions occur in the radio-loud regime (at high $L_{1.4}$, above the locus of the IRRC), and -- due to the larger dispersion of their $q_{\rm TIR}$ distribution (Fig. \protect\ref{fig::q_sfg_agn}) -- also on the opposite, radio-quiet, side of the IRRC. The average AGN fraction of the combined sample is 9\%.}
\label{fig::irrc_main}
\end{figure*}

\begin{table*}
\scriptsize

\caption{Slope measurements ($m_{\mathrm{all}}$, $m_{\mathrm{dm}}$ and $m_{\mathrm{dmSFG}}$), intercepts ($b_{\mathrm{all}}$, $b_{\mathrm{dm}}$ and $b_{\mathrm{dmSFG}}$) and scatters ($\sigma_{\mathrm{all}}$, $\sigma_{\mathrm{dm}}$ and $\sigma_{\mathrm{dmSFG}}$) of each monochromatic IR -- 1.4 GHz radio luminosity correlation, and the total IRRC for the combined sample, the full depth-matched sample and for the subset of depth-matched SFGs.}
\label{tab::lum_lum}
\begin{tabular}{ccccccccc|c}
& $L_{22\mu m}$ & $L_{60\mu m}$ & $L_{100\mu m, \mathrm{IRAS}}$ & $L_{100\mu m, \mathrm{H-ATLAS}}$ & $L_{160\mu m}$ & $L_{250\mu m}$ & $L_{350\mu m}$ & $L_{500\mu m}$ & $L_{\rm TIR}$  \\ \hline \hline

$m_{\mathrm{all}}$ & 1.01 $\pm$ 0.01 & 0.99 $\pm$ 0.00 & 1.00 $\pm$ 0.06 & 1.18 $\pm$ 0.07 & 1.26 $\pm$ 0.08 & 1.36 $\pm$ 0.13 & 1.23 $\pm$ 0.08 & 1.30 $\pm$ 0.07 & 1.170 $\pm$ 0.009 \\
$b_{\mathrm{all}}$ & 12.30 $\pm$ 0.09 & 11.90 $\pm$ 0.05 & 11.63 $\pm$ 0.60 & 9.83 $\pm$ 0.70 & 9.32 $\pm$ 0.83 & 9.05 $\pm$ 1.31 & 10.98 $\pm$ 0.76 & 11.08 $\pm$ 0.61 & 9.7 $\pm$ 0.1 \\
$\sigma_{\mathrm{all}}$ & 0.22 & 0.14 & 0.17 & 0.15 & 0.15 & 0.23 & 0.17 & 0.15 & 0.18 \\ \hline
$m_{\mathrm{dm}}$ & 0.97 $\pm$ 0.01 & 1.00 $\pm$ 0.01 & 0.99 $\pm$ 0.07 & 1.08 $\pm$ 0.06 & 1.22 $\pm$ 0.08 & 1.19 $\pm$ 0.08 & 1.22 $\pm$ 0.08 & 1.61 $\pm$ 0.20 & 1.15 $\pm$ 0.01 \\
$b_{\mathrm{dm}}$ & 12.54 $\pm$ 0.09 & 11.73 $\pm$ 0.06 & 11.77 $\pm$ 0.75 & 10.83 $\pm$ 0.60 & 9.68 $\pm$ 0.82 & 10.61 $\pm$ 0.81 & 11.04 $\pm$ 0.74 & 8.54 $\pm$ 1.72 & 9.9 $\pm$ 0.1 \\
$\sigma_{\mathrm{dm}}$ & 0.17 & 0.17 & 0.24 & 0.12 & 0.14 & 0.15 & 0.16 & 0.15 & 0.14 \\ \hline
$m_{\mathrm{dmSFG}}$ & 0.98 $\pm$ 0.01 & 1.01 $\pm$ 0.01 & 1.06 $\pm$ 0.01 & 1.05 $\pm$ 0.09 & 1.17 $\pm$ 0.13 & 1.12 $\pm$ 0.12 & 1.21 $\pm$ 0.14 & 1.63 $\pm$ 0.21 & 1.114 $\pm$ 0.009 \\
$b_{\mathrm{dmSFG}}$ & 12.48 $\pm$ 0.09 & 11.69 $\pm$ 0.08 & 11.09 $\pm$ 0.12 & 11.11 $\pm$ 0.92 & 10.13 $\pm$ 1.33 & 11.27 $\pm$ 1.18 & 11.14 $\pm$ 1.30 & 8.31 $\pm$ 1.78 & 10.2 $\pm$ 0.1 \\
$\sigma_{\mathrm{dmSFG}}$ & 0.16 & 0.12 & 0.09 & 0.12 & 0.14 & 0.16 & 0.18 & 0.10 & 0.12

\end{tabular}

\end{table*}

\begin{table}
    \centering
    \begin{tabular}{c c c c}
         IR survey & $\log(\overline{\nu L_{\nu}} / L_{\odot})$  & $\overline{q}_{\nu,\mathrm{fit}}$ & $\overline{q}_{\nu}$ \\ \hline
         WISE 22 $\mu$m & 9.90 $\pm$ 0.01 & 1.2 $\pm$ 0.1 & 1.140 $\pm$ 0.006 \\
         IRAS 60 $\mu$m & 10.39 $\pm$ 0.01 & 2.11 $\pm$ 0.08 & 2.098$\pm$ 0.005 \\
         IRAS 100 $\mu$m & 10.43 $\pm$ 0.02 & 2.4 $\pm$ 0.1 & 2.414 $\pm$ 0.006 \\
         H-ATLAS 100 $\mu$m & 10.4 $\pm$ 0.1 & 2.5 $\pm$ 0.9 & 2.45 $\pm$ 0.04 \\
         H-ATLAS 160 $\mu$m & 10.30 $\pm$ 0.09 & 2.4 $\pm$ 1.4 & 2.42 $\pm$ 0.04 \\
         H-ATLAS 250 $\mu$m & 9.76 $\pm$ 0.09 & 2.0 $\pm$ 1.2 & 2.03 $\pm$ 0.04 \\
         H-ATLAS 350 $\mu$m & 9.10 $\pm$ 0.09 & 1.6 $\pm$ 1.3 & 1.60 $\pm$ 0.04 \\
         H-ATLAS 500 $\mu$m & 8.41 $\pm$ 0.09 & 1.2 $\pm$ 0.9 & 1.21 $\pm$ 0.06 \\
    \end{tabular}
    \caption{Median monochromatic  luminosities ($\log(\overline{\nu L_{\nu}} / L_{\odot})$), and median monochromatic IR-radio ratios calculated (a) using the best-fit relations (Eq \ref{eq::irrc_fit}) to the monochromatic IRRC in Fig. \ref{fig::lum_lum_all} ($\overline{q}_{\nu,\mathrm{fit}}$) and (b) by taking the median of each $q_{\nu}$ distribution ($\overline{q}_{\nu}$). Errors on (a) are computed by propagating the uncertainties on the best-fit slopes and intercepts (Table \ref{tab::lum_lum}, final three rows) and the uncertainty on the median monochromatic luminosity, while for (b) we measured the standard error on the median of the $q_{\nu}$ distributions.}
    \label{tab::mono_q}
\end{table}

\subsection{The total infrared-radio correlation}
\label{sect::q}

A general parametrization of the IRRC is possible via fitting the equation

\begin{equation}
    \log\left(\frac{L_{1.4}}{\rm W\,Hz^{-1}}\right) = M \cdot \log\left(\frac{L_{\rm TIR}}{L_{\odot}}\right) + B.
\end{equation}

\noindent
With the use of same BCES fitting methodology as employed for characterising the monochromatic IRRCs (Sect. \ref{sect::lum_lum}), and considering the 2,047 SFGs with reliable SED models in our depth-matched sample we obtain $M = 1.114 \pm 0.009$ (see left-hand panel of Fig. \ref{fig::irrc_main}), consistent with the $1.10 \pm 0.04$ measurement of \citet{bell03}. This greater than unity slope implies a non-linear IRRC, which has consequences for the most widely used metric of the IRRC, $q_{\mathrm{TIR}}$, defined as the logarithmic ratio, of the total infrared and 1.4\,GHz radio ($L_{1.4}$) luminosities:

\begin{equation}
\label{eq::q_tir}
q_{\mathrm{TIR}} \equiv \log\left( \frac{L_{TIR}}{3.75{\times}10^{12}\,\mathrm{W}} \right) - \log\left( \frac{L_{1.4}}{\mathrm{W\,Hz^{-1}}} \right).
\end{equation}

Median or mean $q_{\mathrm{TIR}}$ values are often used to characterize the IRRC properties of a sample of galaxies throughout the literature. Based on Eq. \ref{eq::q_tir}, constant $q_{\mathrm{TIR}}$ values lie alongside lines with slopes of unity in the $L_{\rm TIR}$ -- ${L_{\rm 1.4}}$ parameter space. Our estimated slope of $\sim$ 1.11 therefore suggests that the average $q_{\mathrm{TIR}}$ is decreasing towards higher luminosities, as seen in Fig. \ref{fig::irrc_main} and found by e.g. \citet{jarvis10,moric10} and \citet{ivison10b}. Our best-fit model shown in red indeed connects two dashed lines of constant $q_{\mathrm{TIR}}$ values with a difference of 0.6\,dex across the $\sim$4 dex luminosity range of our data. Therefore, median $q_{\mathrm{TIR}}$ values are dependent on the luminosity range of a given galaxy sample. In Sect. \ref{sect::q_vs_lum} we explore the implications of this in more detail.

Another effect on the $q_{\mathrm{TIR}}$ statistics demonstrated in Fig. \ref{fig::irrc_main} is related to the IR and radio sensitivity of a particular dataset, as mentioned in Sect. \ref{sect::fluxmatch_samp} and described in \citet{sargent10a}. For example, the FIRST survey at $z$\,=\,0.01 is typically sensitive to galaxies that lie above the black horizontal line in Fig. \ref{fig::irrc_main}, while IRAS 100\,$\mu$m is likely to detect sources to the right of the red vertical line. If we require a detection by both surveys, the resulting sample will likely miss several galaxies on the low $q_{\mathrm{TIR}}$ region above our best-fit model, and therefore have a median $q_{\mathrm{TIR}}$ biased towards higher values. The magnitude of this bias is dependent on the mismatch between the depth of the IR and radio data. To obtain our depth-matched sample, we applied a 22\,$\mu$m flux density cut of 19.5 mJy in order for a source to enter our IR-detected sample. This shifted the nominal $L_{\mathrm{TIR}}$ sensitivity of the unWISE catalog at $z$\,=\,0.01 (shown as solid blue line) to a higher $L_{\mathrm{TIR}}$ (marked by the vertical dashed blue line) and matches the FIRST sensitivity well, in that the intersection of the lines of limiting $L_{\mathrm{TIR}}$ and ${L_{\rm 1.4}}$ lies almost on top of the best-fit total IRRC model. In Sect. \ref{sect::bias} we further discuss the quantitative impact of relative IR and radio survey depths on the median $q_{\mathrm{TIR}}$.

Regardless of these potential issues affecting $q_{\mathrm{TIR}}$ measurements, which also complicate comparisons between results from different datasets, they remain the basis of many widely used $L_{1.4}$ -- SFR conversions \citep[e.g.][]{yun01,murphy11,delhaize17}. Specifically, with a typical/representative $q_{\rm TIR}$ value one can estimate the SFR of a galaxy via

\begin{equation}
\label{eq::sfr_q}
    SFR \propto 10^{q_{\rm TIR}} L_{1.4}.
\end{equation}

\noindent
under the assumption that $L_{\rm TIR}$ is a good proxy of the total galaxy SFR (see further discussion of this in Sect. \ref{sect::sfr_radio}). Thus the dependencies of $q_{\mathrm{TIR}}$ on various galaxy properties and redshift are crucial for improving the accuracy of these radio-based SFR estimates.

\begin{figure}
\includegraphics[width=0.5\textwidth]{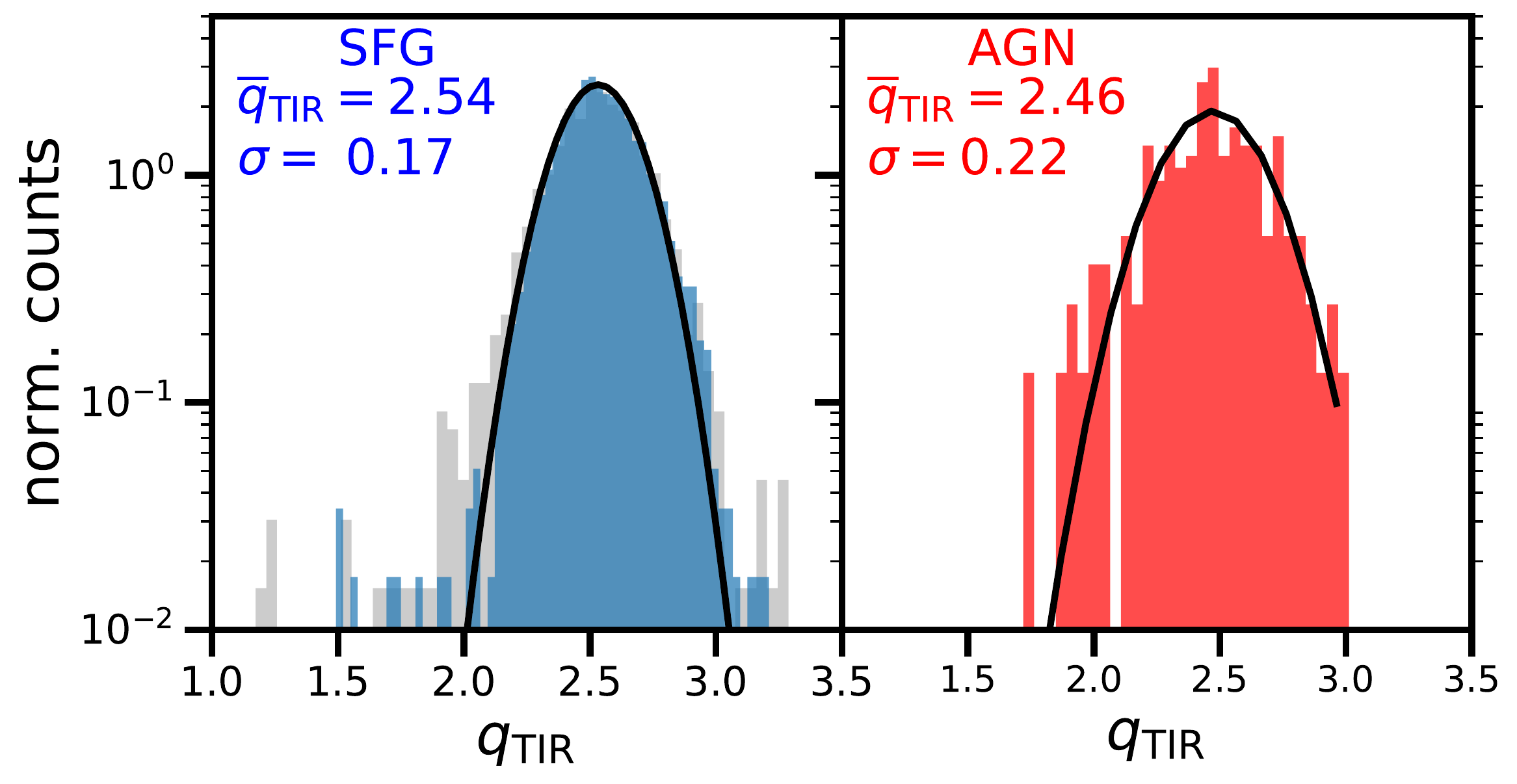}
\caption{Distribution of $q_{\mathrm{TIR}}$ values for SFGs (\textbf{left}) and AGN (\textbf{right}) in the depth-matched, unbiased samples. In the left panel we show the  $q_{\mathrm{TIR}}$ distribution of unclassified sources in the depth-matched sample as a grey histogram in the background. The ordinate axis is set to logarithmic scale. AGN show lower $q_{\mathrm{TIR}}$ on average, and larger scatter.}
\label{fig::q_sfg_agn}
\end{figure}

In our combined sample, we measure a median value of $\overline{q}_{\mathrm{TIR}} =$ 2.47$\pm$0.01 with a scatter of 0.27\,dex.  In our depth-matched sample including AGN, SFGs, and unclassified objects, we find $\overline{q}_{\mathrm{TIR}} =$ 2.51$\pm$0.01, while the scatter is 0.22\,dex. This 0.04\,dex higher value is broadly consistent with predictions from Sect. \ref{sect::bias} considering the $\sim$0.65\,dex sensitivity offset between unWISE 22\,$\mu$m fluxes, our deepest and widest IR photometry, and the FIRST flux limit, as seen in Fig. \ref{fig::tir_lim}. If we consider only SFGs in the depth-matched sample and thus remove most radio-loud objects, we find $\overline{q}_{\mathrm{TIR}} =$ 2.54$\pm$0.01 and a scatter of 0.17\,dex\footnote{We note that the formal error on the median (estimated as $1.253 \sigma / \sqrt{n}$, where $\sigma$ is the standard deviation of the sample and $n$ is the number of sources in the sample), is one order of magnitude smaller than the quoted value of 0.01 due to the large number of $q_{\mathrm{TIR}}$ measurements. However, systematic errors, such as the choice of SED template, do not permit a more precise determination of the median value. When the sample size is small enough for the formal error on the median to exceed this 0.01\,dex threshold from systematics, we will quote a different error. \newline A (small) source of systematic error is the median $q_{\rm TIR}$ value assumed to match the sensitivity curves in Fig. \ref{fig::tir_lim} when defining our depth-matched sample. We have set our present flux limit assuming $\overline{q}_{\mathrm{TIR}} =$ 2.64 following \citet{bell03}. However, if we re-calculate the flux cut according to our $\overline{q}_{\mathrm{TIR}} =$ 2.54, we only lower our $\overline{q}_{\mathrm{TIR}}$ in the SFG sample by 0.01 dex. This is consistent with the prediction of Fig. \protect\ref{fig::q_bias}.}. This median measurement is $\sim$0.1\,dex lower than that by \cite{bell03} and \cite{yun01}. However, it is an excellent match to the average IR-radio ratio of $q_{\rm TIR} = 2.52 \pm 0.03$ measured by \citet{jarvis10} using H-ATLAS and FIRST data and considering lower limits. The  $q_{\mathrm{TIR}}$ distribution of SFGs in our depth-matched sample is shown in Fig. \ref{fig::q_sfg_agn}. A consequence of the non-linear IRRC is that the scatter of $q_{\rm TIR}$ in any given sample is systematically larger than the dispersion relative to the best-fit IRRC models of the same sample -- in the case of the depth-matched SFGs these values are 0.17 dex and 0.12 dex, respectively.

Radio emission not related to the process of star-formation in AGN host galaxies leads, on average, to IR-radio ratios of AGN being lower than for pure SFG samples \citep[see e.g.][]{ibar08,moric10,delhaize17}. Indeed, the AGN fraction on the $L_{\mathrm{TIR}}$ -- $L_{1.4}$ plane, shown in the right panel of Fig. \ref{fig::irrc_main}, is found to be higher in the radio excess region (above/to the left of the best-fit line), than on the main locus of the IRRC.

We measure a $\overline{q}_{\mathrm{TIR}} =$ 2.46$\pm$0.02 for AGNs in our depth-matched sample, with a scatter of 0.2\,dex. This larger scatter compared to SFGs is also seen in Fig. \ref{fig::irrc_main}, and is in qualitative agreement with the findings of e.g. \cite{moric10}.

Composite sources have $\overline{q}_{\mathrm{TIR}} =$ 2.54$\pm$0.01 with a scatter of 0.2\,dex. Since these sources likely harbour a complex mix of AGN, shock, and SF activity, we exclude them from further analysis and will concentrate on ``pure'' SFGs and AGN for the rest of this study.

\subsection{The effect of flux limits on IRRC statistics}
\label{sect::bias}

\begin{figure}
\includegraphics[width=\columnwidth]{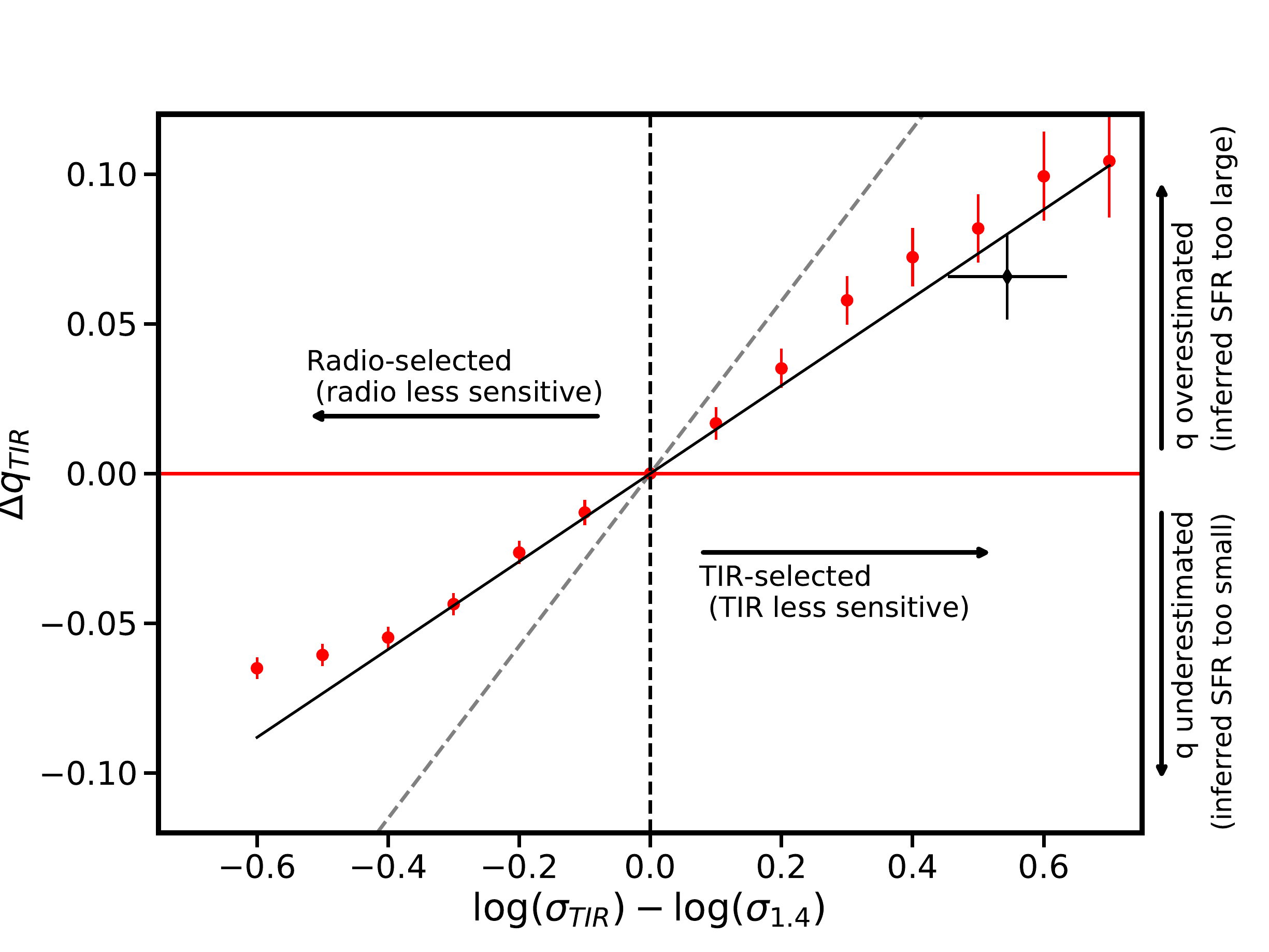}
\caption{Median $q_{\mathrm{TIR}}$ difference to the median $q_{\mathrm{TIR}} = 2.54$ of our depth-matched SFG sample as a function of offset between the sensitivity of TIR and radio data. The black point represents the measured offset of $\overline{q}_{\mathrm{TIR}}$ value for IRAS 60\,$\mu$m $>$ 2 Jy sources \protect\citep[i.e. the selection used in][]{yun01}. It is broadly consistent with the bias predicted from our analysis. The dashed grey line is the $\Delta\overline{q}_{\mathrm{TIR}}$ vs. $\left ( \log(\overline{\sigma}_{\mathrm{IR}}) - \log(\overline{\sigma}_{1.4}) \right )$ trend for a data set with a 40\% larger scatter than our sample.}
\label{fig::q_bias}
\end{figure}

\cite{sargent10a} discuss how a mismatch in the sensitivities of IR and radio data can bias the median $q_{\mathrm{TIR}}$ value ($\overline{q}_{\mathrm{TIR}}$) of a sample. In order to demonstrate the impact of such a difference in the flux limits, we created various flux-limited subsamples from our combined sample. Each of these subsamples was selected by applying different flux cuts either at 22\,$\mu$m or at 1.4\,GHz.
We quantified the mismatch between the IR and radio sensitivities of these samples, $\sigma_{\mathrm{IR}}$ and $\sigma_{1.4}$ respectively, as the mean difference between the $L_{\textrm{TIR}}$ curves calculated for their given 22\,$\mu$m and radio flux limits (see Fig. \ref{fig::tir_lim}). We then measured $\overline{q}_{\mathrm{TIR}}$ in each of these subsamples and computed the $q_{\mathrm{TIR}}$ bias, $\Delta\overline{q}_{\mathrm{TIR}}$, as the difference between the measured $\overline{q}_{\mathrm{TIR}}$ in the sample and $q_{\mathrm{TIR}} = 2.54$, i.e. the median IR-radio ratio of our depth-matched SFG sample.

Fig. \ref{fig::q_bias} shows $\Delta\overline{q}_{\mathrm{TIR}}$ as a function of the logarithmic ratio of the IR and 1.4 GHz luminosity limits. At the origin of the figure is our depth-matched SFG sample with matched IR and radio sensitivies. To the left of it, represented by negative $\log(\overline{\sigma}_{\mathrm{TIR}}) - \log(\overline{\sigma}_{1.4})$ values, there are subsamples where we applied increasingly higher radio flux density cuts, and thus obtained $q_{\mathrm{TIR}} < 2.54$ values. Conversely, in the right hand side of the figure, towards positive $\log(\overline{\sigma}_{\mathrm{IR}}) - \log(\overline{\sigma}_{1.4})$ values, we measure larger than $2.54$ median IR-radio ratios.

\noindent
To this trend we fitted a linear model,

\begin{equation}
\label{eq::q_bias}
\Delta\overline{q}_{\mathrm{TIR}} = m_{\rm bias}\big \{ \log(\overline{\sigma}_{\mathrm{TIR}}) - \log(\overline{\sigma}_{1.4}) \big \},
\end{equation}

\noindent
with the slope $m_{\rm bias}$, that has a best-fit value of 0.147 $\pm$ 0.007. As discussed above, and in line with the analysis of \cite{sargent10a}, we find that IR-limited (i.e. when radio data are more sensitive than IR) $\overline{q}_{\mathrm{TIR}}$ measurements are positively biased, whereas radio-limited samples lead to lower $\overline{q}_{\mathrm{TIR}}$ values. The magnitude of this bias is proportional to the mismatch between the radio and IR data. We note that this bias can be mitigated with either a selection based on a third, uncorrelated selection criterion, e.g. by studying a mass-selected sample, or techniques that allow probing IR and radio flux densities below their nominal limits, such as stacking or survival analysis, at least in the regime of not too strongly mismatched depths.

Our fit in Fig. \ref{fig::q_bias} can in principle be used to quantitatively estimate and compare IRRC selection biases across different studies in the literature \citep{sargent10a}. However, in practice reconciling the median $q_{\rm TIR}$ values of different samples is not that straightforward. To see why, consider the following analytical formula for the difference between the average IR/radio ratio of IR- and radio-detected samples (\citealp{sargent10a}; see also analogous expressions in a variety of contexts in \citealp{kellermann64}; \citealp{condon84}; \citealp{francis93}; \citealp{lauer07}):

\begin{equation}
\label{eq:dq_mark}
    \Delta q = \ln{(10)} (\beta - 1)\, \sigma_{q}^2.
\end{equation}

\noindent
Here $\beta$ is the flux-dependent power-law index of the number counts (which, for the sake of simplicity, are assumed to have the same $\beta$ in the IR and radio band when dealing with a pure SFG sample) and $\sigma_{q}$ is the observed scatter of the IRRC. The  power law indices of IR and radio number counts thus directly influence the value of $\Delta \overline{q}_{\rm TIR}$, and since they are naturally a function of survey depth, Eq. \ref{eq:dq_mark} should not be viewed as producing a single, universal offset estimate, but instead has some dependence on luminosity. Similarly, changes in the observed scatter -- to which both the intrinsic dispersion and measurement errors contribute -- will change the slope of Eq. \ref{eq::q_bias}. To illustrate this, we show in Fig. \ref{fig::q_bias} with a dashed grey line the $\Delta\overline{q}_{\mathrm{TIR}}$ vs. $\left ( \log(\overline{\sigma}_{\mathrm{IR}}) - \log(\overline{\sigma}_{1.4}) \right )$ trend for a data set with a 40\% larger scatter than our sample, resulting in a slope that is twice as steep. Nevertheless, if we bear in mind these different factors, Eq. \ref{eq::q_bias} can still be used to reconcile apparently inconsistent results, and identify residual disagreement beyond the bias caused by this selection effect if present. 

As an example, we consider the sample of \cite{yun01}. Fig. \ref{fig::q_bias} predicts that due to its unmatched selection criteria (for an illustration of this see Fig. \ref{fig::irrc_main}), the median $q_{\mathrm{FIR}}$ of \cite{yun01} is biased high. In order to find evidence for this in our catalogue, we examined the $\overline{q}_{\mathrm{FIR}}$ values reported in \citet{yun01} ($\overline{q}_{\mathrm{FIR, Yun}} =$ 2.34 $\pm$ 0.01) and the median $q_{\mathrm{FIR}}$ in a sample suitably depth-matched between the IR and the radio at the level of the NVSS sensitivity curve in Fig. \ref{fig::tir_lim} ($\overline{q}_{\mathrm{FIR, dm}} =$ 2.26 $\pm$ 0.01). For a fair comparison, we re-scaled the difference of these values, $\Delta q$\,=\,0.08\,dex, with the ratio of their scatters squared $(\sigma_{\rm q,dm,SFG}/\sigma_{\rm q,Yun})^2$ following Eq. \ref{eq:dq_mark} above. Substituting the dispersion of 0.26\,dex reported in \citet{yun01}, and our 0.23\,dex scatter of $q_{\mathrm{FIR}}$ distribution (measured similar to \citet{yun01}, i.e. not as the standard deviation of orthogonal distances which would result in a lower dispersion value) results in $\Delta q = 0.07$. The difference between average sensitivities in the \cite{yun01} catalogue was computed from the 2 Jy IRAS 60\,$\mu$m and NVSS curves in Fig. \ref{fig::tir_lim}, and its uncertainty was taken as the standard deviation of the differences in the plotted redshift range. The $\overline{q}_{\mathrm{FIR}}$ difference agrees within $\sim$1.5\,$\sigma$ with our expectation based on Eq. \ref{eq::q_bias}. This suggests that the $\overline{q}_{\mathrm{FIR}}$\,=\,2.34 reported in \cite{yun01} is biased high and that the low-z $\overline{q}_{\mathrm{FIR}}$ value is instead nearer 2.26 in the relevant luminosity range.

Finally, we note that due to the different sample selection philosophy of \citet{bell03} -- who aim to maximise wavelength coverage from the far-ultraviolet to the radio, rather than basing sample selection on (a) tiered survey(s) -- their data set does not lend itself to the same kind of systematic comparison we carried out above for the \citet{yun01} analysis. However, a more qualitative comparison is possible by considering the median luminosity $L_{\rm TIR} = 10^{9.68}\,L_{\odot}$ of the \citet{bell03} sample. At this luminosity, our IRRC best-fit parameters in Eq. \ref{eq::q_vs_ltir} below translate to an IR-to-radio ratio of $q_{\rm TIR}$\,=\,2.63, which closely matches the median $q_{\rm TIR}$\,=\,2.64$\pm$0.02 of \citet{bell03} and suggests that their sample -- while situated in a lower luminosity regime than that of \citet{yun01}, see Fig. \ref{fig::yun_bell_us} -- is not subject to strong selection biases.

Fig. \ref{fig::q_bias} also provides clues as to the potential issues with future studies seeking to investigate the IRRC using upcoming radio surveys if they are matched to already existing IR data. For example, from an IRRC perspective the targeted 1\,$\mu$Jy sensitivity of MIGHTEE \citep[][]{jarvis16} at z $<$ 0.2 will be $\sim$ 2.4, 3.5 and 2.3 dex deeper than the unWISE 22\,$\mu$m, IRAS 100\,$\mu$m and Herschel 250\,$\mu$m data we use in this paper, respectively. If the fitted trend in Fig. \ref{fig::q_bias} is taken at face value, calculating $\overline{q}_{\mathrm{TIR}}$ based on a cross match between MeerKAT detections and these IR data without considering the different survey depths could result in $\overline{q}_{\mathrm{TIR}}$ estimates biased high by $\sim$ 0.25 -- 0.4 dex due to the significantly deeper radio observations. The 10\,$\mu$Jy detection limit of the EMU survey \citep{norris11} with ASKAP will likely lead to a qualitatively similar bias, if not mitigated using an appropriate flux cut. With this newly arising large gap between IR and radio surveys, other calibration methods will become more important, e.g. the utilization of utilizing shorter wavelength or combined SFR tracers when exploiting deep radio data \citep[see e.g.][]{hodge08,brown17,davies17,gurkan18,read18,duncan20} as opposed to IR measurements. Arguably, for such low-luminosity sources, the IR emission may not do very well at capturing the bulk of the SFR anyway since typically their SFR-budget is dominated by the unobscured component.

\section{Discussion}
\label{sect::disc}

\subsection{The non-linearity of the IRRC}
\label{sect::q_vs_lum}

\begin{figure*}
\includegraphics[width=\textwidth]{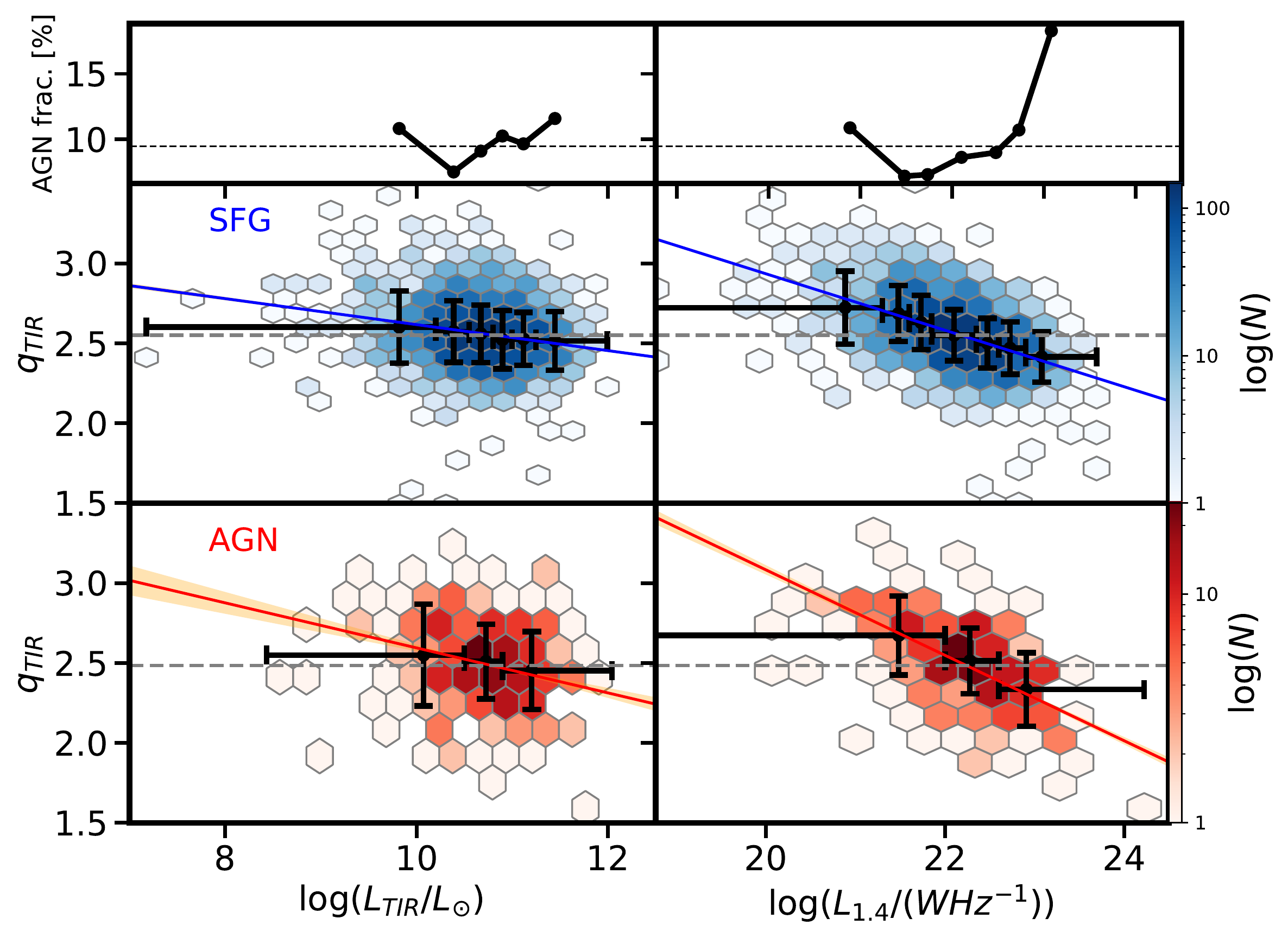}
\caption{Median $q_{\mathrm{TIR}}$ of SFGs (middle) and AGN (bottom) as a function of IR (left) and radio (right) luminosity in the depth-matched sample. Black points are median $q_{\mathrm{TIR}}$ values of luminosity bins, vertical and horizontal error bars represent the measured scatter of these binned $q_{\mathrm{TIR}}$ distributions (for details see Figs. \ref{fig::q_sfg_ir}, \ref{fig::q_sfg_rad} and \ref{fig::q_agn_irrad}) and the range of each luminosity bin, respectively. The dashed horizontal grey lines in these panels highlight the median $q_{\mathrm{TIR}}$ value found in the corresponding galaxy sample. Lines and shaded regions are the best fit and the 2-$\sigma$ uncertainty bands from our BCES fits, respectively. The upper row shows the variation of the AGN fraction with $L_{\mathrm{TIR}}$ and $L_{\mathrm{1.4}}$. Fine dashed lines are drawn at the $\sim$9\% average AGN fraction of the depth-matched sample.}
\label{fig::q_vs_lum}
\end{figure*}

In this section we investigate the luminosity dependence of the IR/radio ratio, $q_{\mathrm{TIR}}$. Fig. \ref{fig::q_vs_lum} presents $q_{\mathrm{TIR}}$ values for depth-matched SFGs and AGN host galaxies as a function of their radio and total IR luminosities. We fitted a linear model of the form of

\begin{equation}
\label{eq::q_vs_lum}
    q_{\rm TIR} = s\,\log({L}) + i~,
\end{equation}

\noindent
where best-fit parameters $s$ and $i$ were found using the BCES method with orthogonal distance minimization. As the uncertainties on both IR and radio luminosities and $q_{\mathrm{TIR}}$ correlate strongly, we included their covariance in the BCES fits.

We find that $q_{\mathrm{TIR}}$ is only weakly dependent on $L_{\mathrm{TIR}}$ for both SFGs and AGN. The best-fit linear relation using depth-matched SFGs is 

\begin{equation}
    q_{\mathrm{TIR, SFG}} = (-0.08\pm0.01) \cdot \,\log \left( \frac{L_{\rm TIR}}{\rm W} \right) + (3.4\pm0.1)~,
\label{eq::q_vs_ltir}
\end{equation}{}

\noindent while for depth-matched AGN

\begin{equation}
   q_{\mathrm{TIR, AGN}} = (-0.14\pm0.05) \cdot \,\log \left( \frac{L_{\rm TIR}}{\rm W} \right) + (4\pm0.5)~.
\end{equation}{}

\noindent Conversely, radio-bright, $\log{(L_{\mathrm{1.4}} / (W Hz^{-1}))} \geq$ 22.5, sources tend to have lower than average $q_{\mathrm{TIR}}$ values, while galaxies that are faint in radio have higher $q_{\mathrm{TIR}}$. The best-fit relation between $q_{\mathrm{TIR}}$ and $\log{(L_{\mathrm{1.4}})}$ in the depth-matched SFG sample is

\begin{equation}
\label{eq::l_vs_q}
    q_{\mathrm{TIR, SFG}} = (-0.177\pm0.009) \cdot \,\log \left( \frac{L_{\rm 1.4}}{\rm W\,Hz^{-1}} \right) + (6.5\pm0.2)~,
\end{equation}{}

\noindent while in depth-matched AGN sample it is

\begin{equation}
   q_{\mathrm{TIR, AGN}} = (-0.27\pm0.03) \cdot \,\log \left( \frac{L_{\rm 1.4}}{\rm W\,Hz^{-1}} \right) + (8.4\pm0.7)~.
\end{equation}{}

\noindent The fact that $q_{\rm TIR}$ appears to vary more with radio luminosity than with IR luminosity is consistent with the findings of \cite{moric10}, \cite{jarvis10} and \cite{ivison10b}. We note that even though AGN hosts and SFGs show qualitatively a similar behaviour, AGN have systematically steeper relations. This supports the scenario that the radio emission in AGN arises from different processes than for SFGs. 

\begin{figure*}
\includegraphics[width=\textwidth]{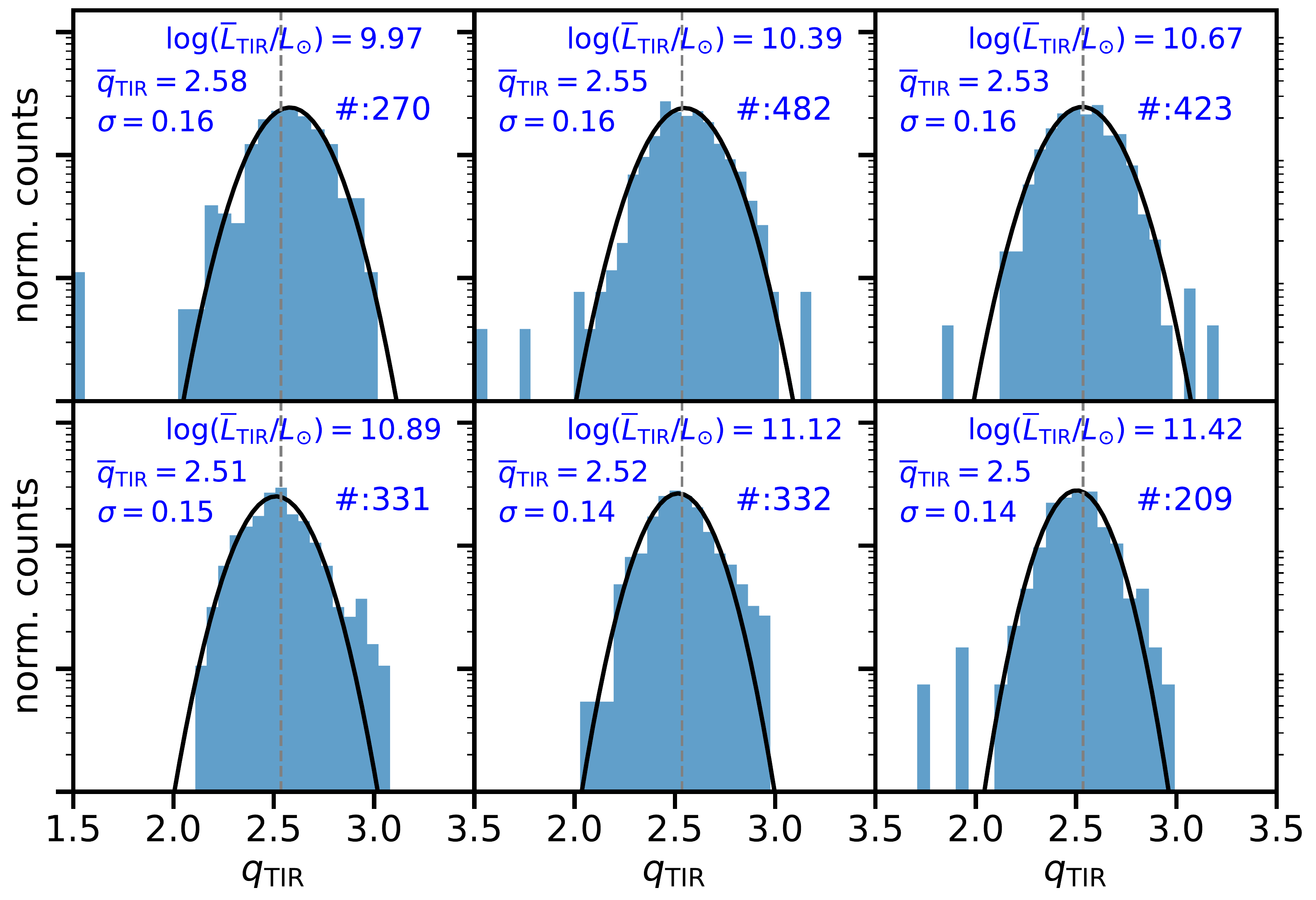}
\caption{Distribution of $q_{\mathrm{TIR}}$ values for SFGs in the depth-matched sample in bins of increasing  $L_{\mathrm{TIR}}$ (from the upper left to the lower right corner). Median IR luminosity values ($\overline{L}_{\mathrm{TIR}}$), fitted median $q_{\mathrm{TIR}}$ ($\overline{q}_{\mathrm{TIR}}$), scatter ($\sigma$) and number of sources in a given bin (\#) are displayed in each panel, while bin widths are represented as horizontal errorbars in Fig. \ref{fig::q_vs_lum}. Vertical, dashed grey lines represent the $\overline{q}_{\mathrm{TIR}}$ value of the entire depth-matched SFG sample for reference. The black curves are best-fit Gaussians to each distribution. The ordinate axis is set to logarithmic scale.}
\label{fig::q_sfg_ir}
\end{figure*}

\begin{figure*}
\includegraphics[width=\textwidth]{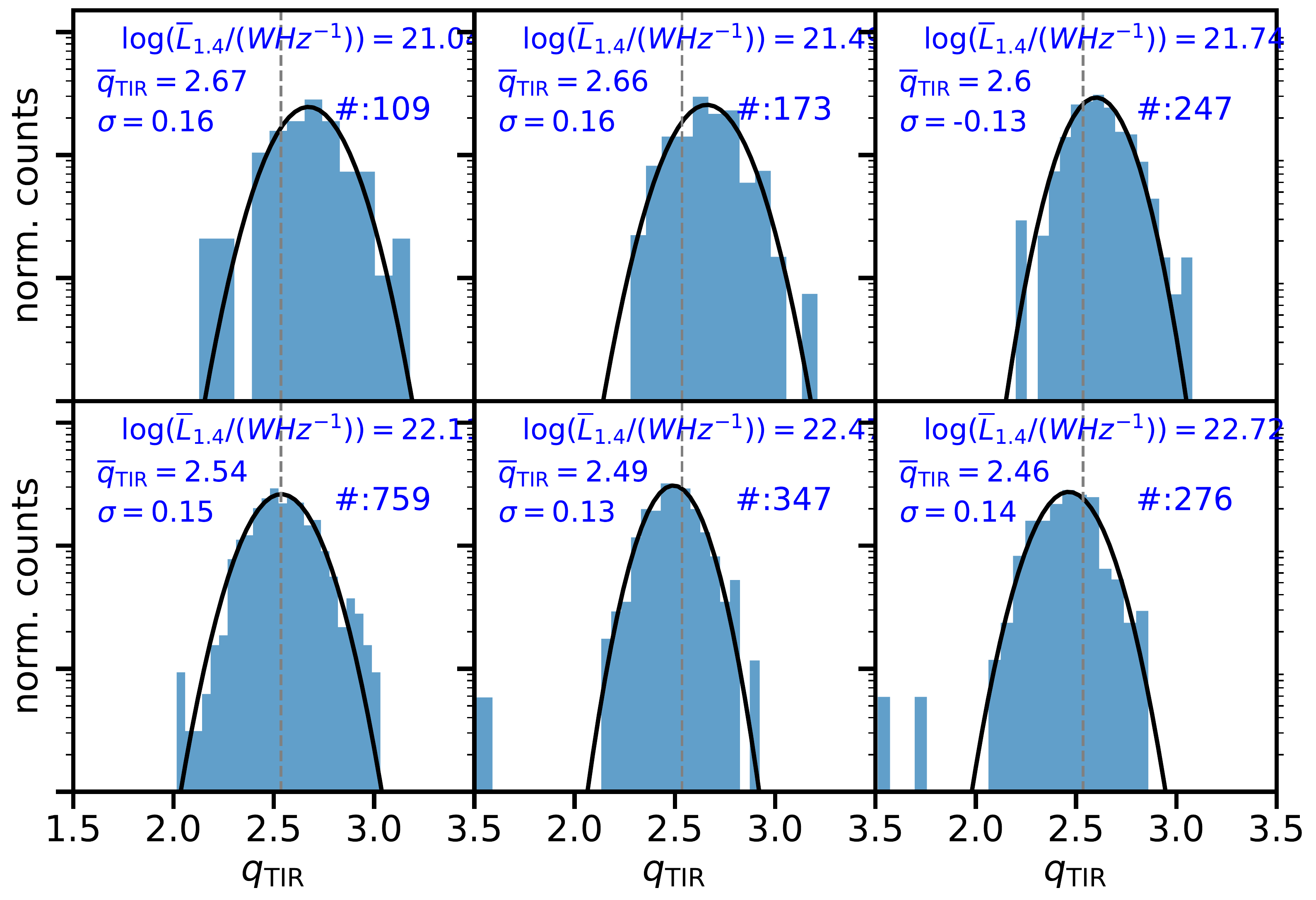}
\caption{Same as Fig. \protect\ref{fig::q_sfg_ir}, but with $L_{\mathrm{1.4}}$.}
\label{fig::q_sfg_rad}
\end{figure*}

\begin{figure*}
\includegraphics[width=\textwidth]{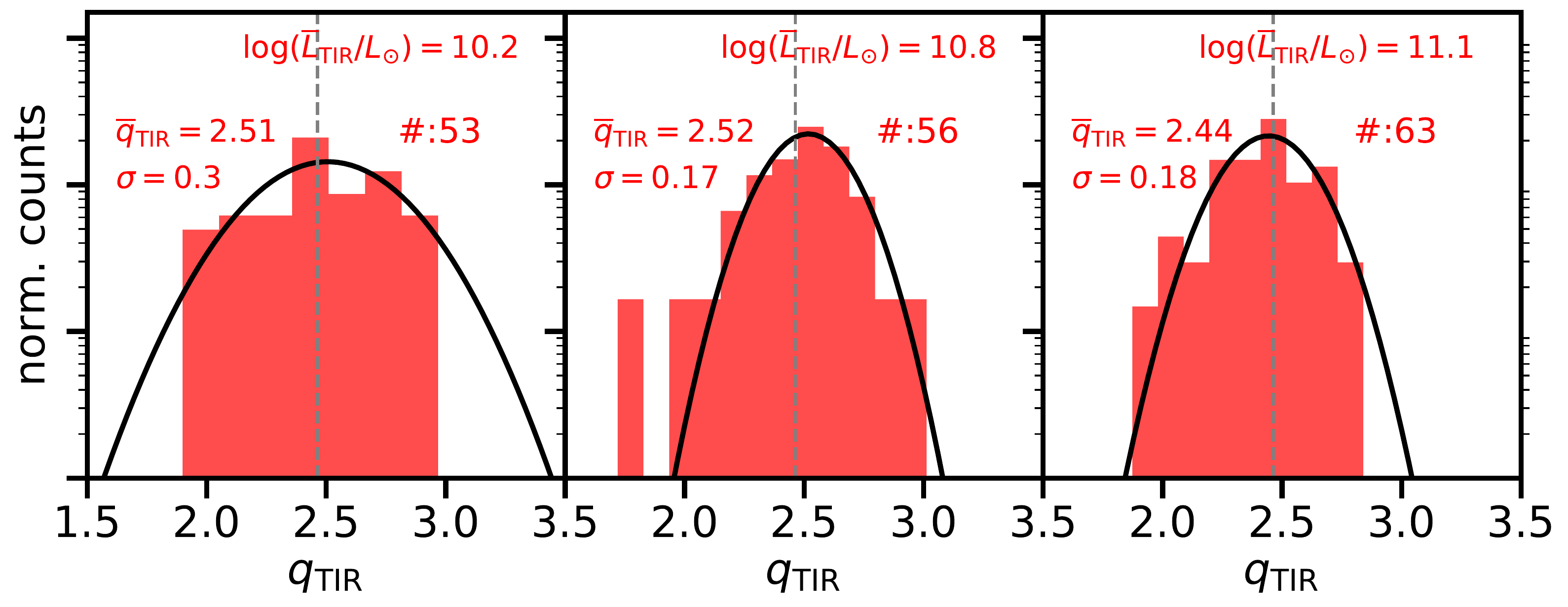}
\includegraphics[width=\textwidth]{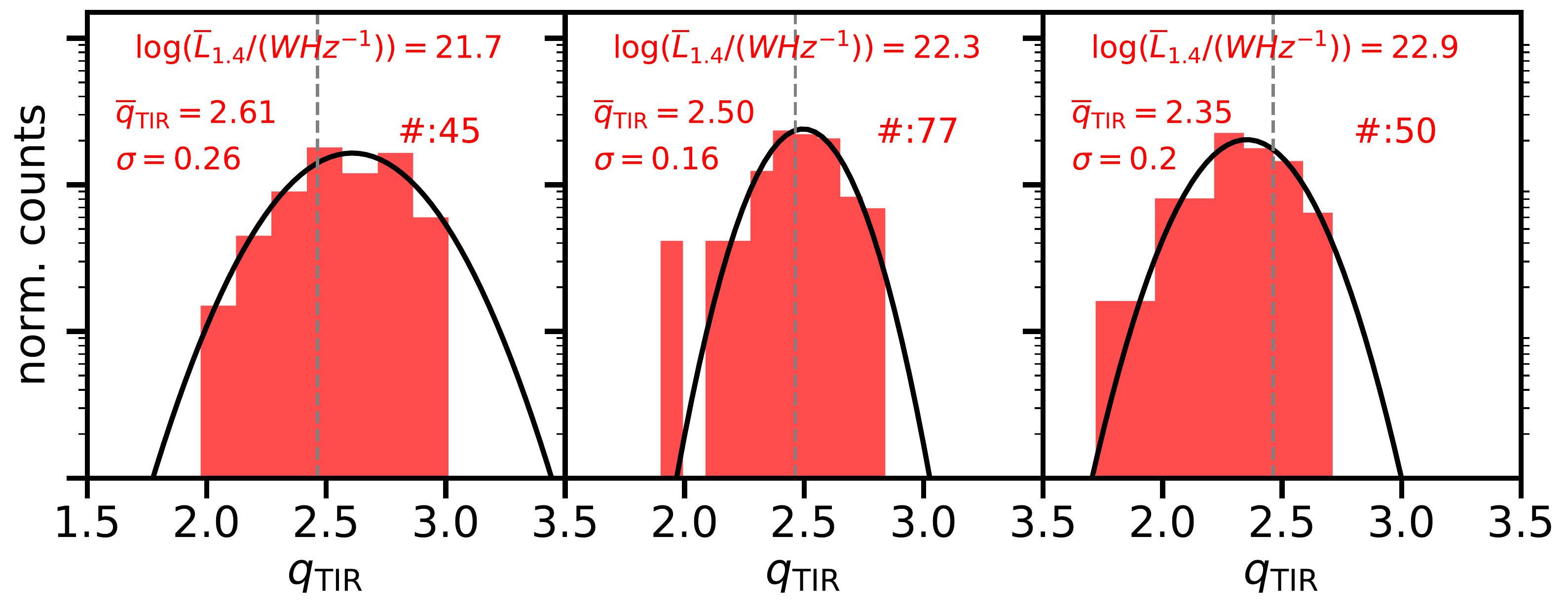}
\caption{Same as Fig. \protect\ref{fig::q_sfg_ir} and \protect\ref{fig::q_sfg_rad}, but for AGN host galaxies. The upper (lower) row shows $q_{\rm TIR}$ distributions for AGN hosts at different $L_{\rm TIR}$ ($L_{1.4}$), respectively.}
\label{fig::q_agn_irrad}
\end{figure*}

In Figs. \ref{fig::q_sfg_ir} and \ref{fig::q_sfg_rad} we present the $q_{\mathrm{TIR}}$ distributions of SFGs in the radio and IR luminosity bins defined in the 2nd row of Fig. \ref{fig::q_vs_lum}. Fig. \ref{fig::q_agn_irrad} shows the same information for AGN hosts (with bins defined as in the lower row of Fig. \ref{fig::q_vs_lum}). These figures demonstrate a consistent Gaussian $q_{\mathrm{TIR}}$ profile around the fitted lines in the entire $\sim$4\,dex luminosity range. We observe a slightly decreasing scatter for SFGs both with increasing $L_{\mathrm{TIR}}$ and $L_{\mathrm{1.4}}$. This is seemingly at odds with the findings of e.g. \cite{yun01}, who measure a higher scatter at high radio and IR luminosities. However, the sharply increasing AGN fraction towards this luminosity range both at IR and radio wavelengths (see Fig. \ref{fig::q_vs_lum} and \ref{fig::irrc_main}) suggest that studies that do not separate these populations may find an artificially increased IRRC scatter especially at high luminosities, due to the on-average lower values and higher spread of AGN IR/radio ratios. Conversely, the increasing scatter towards low radio and IR luminosities may, at least to some extent, be caused by the expected break-down of the IRRC due to UV and optical photons not being fully reprocessed by dust \citep[see e.g.][]{bell03,lacki10a}.

As \cite{moric10} discussed, in a given sample, where ${q}_{\mathrm{TIR}}$ has no $L_{\mathrm{TIR}}$ dependence, and a non-zero dispersion, a declining trend with $L_{\mathrm{1.4}}$ must be present due to the definition of ${q}_{\mathrm{TIR}}$ \citep[for details on why this is the case see e.g.][]{condon84,moric10}. To test whether the measured slope of the ${q}_{\mathrm{TIR}}$ -- $L_{\mathrm{1.4}}$ relation can be fully explained by this mathematical interdependence, \citet{moric10} generated mock data by sampling the observed ${q}_{\mathrm{TIR}}$ distribution and calculated $L_{\mathrm{1.4}}$ by combining these randomized ${q}_{\mathrm{TIR}}$ values with the real $L_{\mathrm{TIR}}$ measurements. Finally the bootstrapped ${q}_{\mathrm{TIR}}$ -- $L_{\mathrm{1.4}}$ relation was fitted. Indeed, having carried out this exercise using our ${q}_{\mathrm{TIR}}$ distribution in Fig. \ref{fig::q_sfg_agn} we find a consistent ${q}_{\mathrm{TIR}}$ -- $L_{\mathrm{1.4}}$ trend with our data, with a slope of -0.14 $\pm$ 0.01.

The luminosity dependence of average $q_{\mathrm{TIR}}$ values is linked to the non-linearity of the IRRC. By definition (see Eq. \ref{eq::q_tir}), lines of constant, luminosity-independent $q_{\mathrm{TIR}}$ values have slopes of unity in the $L_{\rm TIR}$ -- $L_{1.4}$ plane (see Fig. \ref{fig::irrc_main}). However, if, as we see in our data, at low radio luminosities the average IR-radio ratio is high, while at high radio luminosities $q_{\mathrm{TIR}}$ tends to be low, a fit across the whole range has to deviate from a slope of unity to connect these regions. Thus, rather than adopting a single, constant $q_{\mathrm{TIR}}$ value, the dependence on 1.4 GHz luminosity of $q_{\mathrm{TIR}}(L_{1.4})$ should be incorporated into radio continuum based SFR estimates. Substituting Eq. \ref{eq::q_vs_lum} into Eq. \ref{eq::sfr_q} results in

\begin{equation}
\label{eq::sfr_q_lrad_lin}
    \mathrm{SFR} \propto 10^{s\,\log (L_{1.4})+i}\,L_{1.4},
\end{equation}

\noindent
or, in log-log space

\begin{equation}
    \log(\mathrm{SFR}) = (s+1) \log(L_{1.4}) + [i+C]
\end{equation}{}

\noindent
and with units and our best-fit parameters substituted

\begin{multline}
\label{eq::sfr_q_lrad}
    \log \left( \frac{\mathrm{SFR}}{\rm M_{\odot}yr^{-1}}  \right) = (0.823 \pm 0.009) \cdot \\ \cdot \log \left( \frac{L_{1.4}}{\rm W Hz^{-1}} \right) - (17.5 \pm 0.2),
\end{multline}

\noindent
assuming a $L_{\rm TIR}$ -- SFR scaling factor\footnote{We adopted the $4.5 \cdot 10^{-44} \, M_{\odot} \mathrm{yr}^{-1} \, \mathrm{erg}^{-1} s$ of \citealt{kennicutt98} (found in their Eq 4) and multiplied it by 0.61 \citep[see e.g.][]{madau14} to account for the difference between the \citet{salpeter55} IMF assumed by \citet{kennicutt98} and the \citet{chabrier03} IMF we adopt.} of $10^{-10} M_{\odot} \mathrm{yr}^{-1} \, L_{\odot}^{-1}$.

Since Eq. \ref{eq::sfr_q_lrad_lin} assumes $\log(\rm SFR) \propto \log(L_{\rm TIR})$, we can see that its slope is related to the slope of the IRRC, where $\log(\rm SFR)$ is essentially re-scaled to $\log(L_{\rm TIR})$. A comparison of Eq. \ref{eq::sfr_q_lrad_lin} and Eq. \ref{eq::irrc_fit} shows that the IRRC slope should therefore be roughly the inverse of the slope found for the $\log(\rm SFR)$ -- $\log(L_{1.4})$ calibration. Indeed the latter is $(s+1) \approx 0.85$, while the IRRC slope is $1.11$ (reported in Table \ref{tab::lum_lum}). Finally we note that as opposed to calibrations assuming a constant $q_{\rm TIR}$ value our $\log(SFR)$ -- $\log (L_{1.4})$ relation, and by extension the IRRC, is a power law. This contradicts the previously suggested conspiracy of $L_{1.4}$ and $L_{\rm TIR}$ to equally underestimate SFR in low luminosity galaxies \citep{bell03,lacki10a}.

\subsection{1.4 GHz radio emission as a star-formation rate tracer}
\label{sect::sfr_radio}

\begin{figure}
\centering
\includegraphics[width=0.45\textwidth]{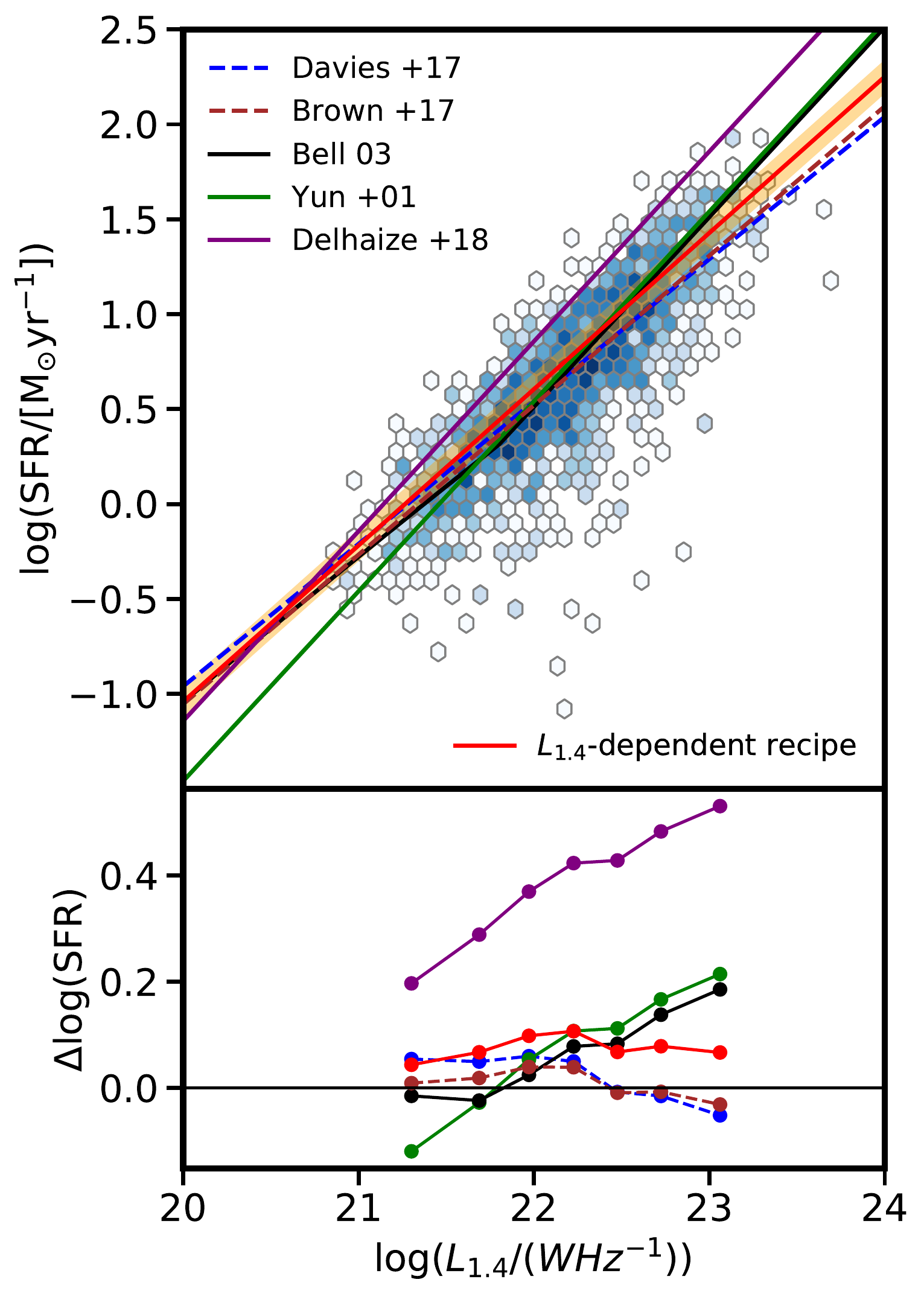}
\caption{Top: The star formation rate -- $L_{1.4}$ correlation using our depth-matched SFG sample. SFR values are taken from the GSWLC catalogue of \protect\citet{salim16}. Various models are presented as well as our best-fit to the data, and the SFR-conversion using our $L_{1.4}$-dependent IR-radio ratio. The 1\,$\sigma$ significance band around the latter was calculated from the correlated uncertainties on our $q_{\rm TIR}$ -- $\log({L_{1.4}})$ fit's parameters. Bottom: Mean logarithmic ratio of the various SFR recipes and the reference SFR estimate of \protect\citet{salim16} in bins of $L_{1.4}$.}
\label{fig::L20_sfr}
\end{figure}

We tested the validity of an $L_{1.4}$-dependent SFR calibration by comparing our recipe to radio-independent SFR estimates from the GALEX-SDSS-WISE Legacy Catalog \citep[GSWLC;][]{salim16}. GSWLC SFRs were obtained via UV/optical SED fitting with CIGALE \citep{noll09}, independently of both TIR or radio luminosity. 1,740 of our depth-matched SFGs have SFR estimates in the GSWLC catalogue. In Fig. \ref{fig::L20_sfr} we plot these against our $L_{1.4}$ measurements, and add several commonly used radio based SFR recipes (after conversion to a \citealt{chabrier03} IMF where necessary, see e.g. \citet{madau14} for conversion factors). They have either been calibrated through the IRRC at low \citep[][however, the latter applies a correction at $L_{1.4} < 6.4 \cdot 10^{21}\, \mathrm{W\,Hz^{-1}}$]{yun01,bell03} or high redshift \citep[][we use their Eq. 4 with the median redshift, $z$\,=\,0.04, of the galaxies shown in Fig. \protect\ref{fig::L20_sfr}]{delhaize17}, or calibrated against non-IR tracers \citep{brown17,davies17}\footnote{We used Eq 3 from \protect\citet{davies17}, i.e. 1.4 GHz radio luminosity calibrated against SED-fit derived SFRs.}. We also show our $L_{1.4}$-dependent IRRC SFR calibration (Eq. \ref{eq::sfr_q_lrad}). The lower panel of Fig. \ref{fig::L20_sfr} shows the mean offsets of SFRs in the GSWLC catalogue and SFRs estimated using the aforementioned $L_{1.4}$ -- SFR calibrations.

In the $\log(L_{1.4} / W\,Hz^{-1}) <$ 21.5 regime our conversion is consistent with the \cite{bell03}, \cite{brown17} and \cite{davies17} formulae as well as the SED-derived SFRs at the $\sim$10\% level. On the other hand, the \cite{yun01} recipe predicts $\sim$25\% lower SFRs compared to the reference SFRs. Meanwhile, as a result of their  $L_{1.4}$-independent $q_{\rm TIR}$ values, \citet{yun01} and \citet{bell03} predict systematically higher SFR values in the range $\log(L_{1.4}/W\,Hz^{-1}) >$ 22, reaching an $\sim$0.2\,dex excess at $\log(L_{1.4}/W\,Hz^{-1}) \approx $ 23. In comparison, our conversion is consistent within  $\sim$15\% with the GSWLC values, while \citet{brown17} and \citet{davies17} stay below 5 -- 10\%. Using the \citet{delhaize17} calibration on the other hand would yield 0.2 -- 0.4\,dex higher SFR estimate across the full luminosity range in this low-$z$ sample, due to their median $q_{\rm TIR}$ being $\sim$0.2\,dex higher at $z\,{\sim}$\,0.04 relative to the measurements in \citet{yun01} and \citet{bell03}, and $\sim$0.3\,dex higher than our $\overline{q}_{\rm TIR} = 2.54$.

In conclusion, if we assume the SFRs based on the UV/optical SED fitting in \citet{salim16} are a robust benchmark (see their Sects. 7 and 8 for a comparison to other widely used SFR measurement techniques and catalogues, respectively), IRRC-based  $L_{1.4}$ -- SFR conversions with a constant $q_{\rm TIR}$ that is  independent of $L_{1.4}$ systematically underestimate SFRs in low-luminosity sources and overestimate them at high luminosities. The \citet{bell03} recipe sought to resolve this issue by providing a modified prescription below $\log(L_{\rm 1.4} / W\,Hz^{-1}) \approx 21.8$, but remains less accurate than more recent calibrations at higher luminosities. Having dropped the assumption of a fixed $q_{\rm TIR}$, our calibration --  which is more akin to the approach of \citealt{hodge08} -- achieves a significantly better agreement with studies that do not solely use IR emission to infer SFRs.\\
It is important to note that, while our calibration performs well in the luminosity regime we probe (SFR\,$\gtrsim$\,0.5\,$M_{\odot}$/yr), a purely IRRC-based approach to calibrating SFRs becomes less and less tenable as one pushes to systems with lower mass and luminosity, where a much larger fraction of the star formation activity is not obscured by dust. Both these astrophysical reasons, as well as the pragmatic desire not to discard large numbers of faint sources due to dissimilar IR and radio survey depths (see discussion at the end of Sect. \ref{sect::bias}), imply that SFR measurements from multi-wavelength photometry or nebular emission lines will play an important role for the calibration of radio SFRs in deep radio surveys with SKA and its precursors \citep[see][for examples of such studies that have already pursued this approach using current radio data]{hodge08,davies17,brown17,gurkan18,duncan20}. Nevertheless, we still expect that there will continue to be applications where a purely IRRC-based calibration, and in particular a depth-matching approach as we discuss in this paper, remain useful. For example, this could be the case where a study focuses on a measurement of the scatter of the IRRC (which is not easily recoverable by stacking) in the high-luminosity regime in order to learn about the underlying physical processes, or when dealing with the rare population of highly dust-obscured starbursts, of which larger numbers will be picked up out to higher redshifts thanks to the higher survey speeds of the new generation of radio telescope arrays.

\subsection{The redshift dependence of the infrared-radio correlation}
\label{sect::q_vs_z}

In the past decade it has been intensely debated whether the IRRC -- and hence the relation between radio luminosity and SFR -- evolves with redshift. Statements on the (lack of) evolution of the IRRC have almost exclusively been based on measurements of representative $q$-values for galaxy populations across different redshifts. A number of recent studies \citep{ivison10a,magnelli15,calistro-rivera17,delhaize17} have found evidence for a declining radio-IR ratio across cosmic time \citep[but see also][]{garrett02,appleton04,garn09,jarvis10,sargent10a,sargent10b,mao11,smith14, pannella15}. Similar rates of modest, but statistically significant evolution have been reported at different observed frequencies, e.g, $(1+z)^{-(0.19\pm0.01)}$ at 3 GHz in \citet{delhaize17}, $(1+z)^{-(0.12\pm0.04)}$ at 1.4\,GHz in \citet{magnelli15}, and in low-frequency LOFAR \citep{harleem13} data \citet{calistro-rivera17} measured a consistent redshift dependency of $(1+z)^{-(0.15\pm0.03)}$ for SFGs. Since we have access to the according data, in the following we will discuss the evolutionary trend of \citet{delhaize17} in the COSMOS field in more detail.

\begin{figure}
\includegraphics[width=0.45\textwidth]{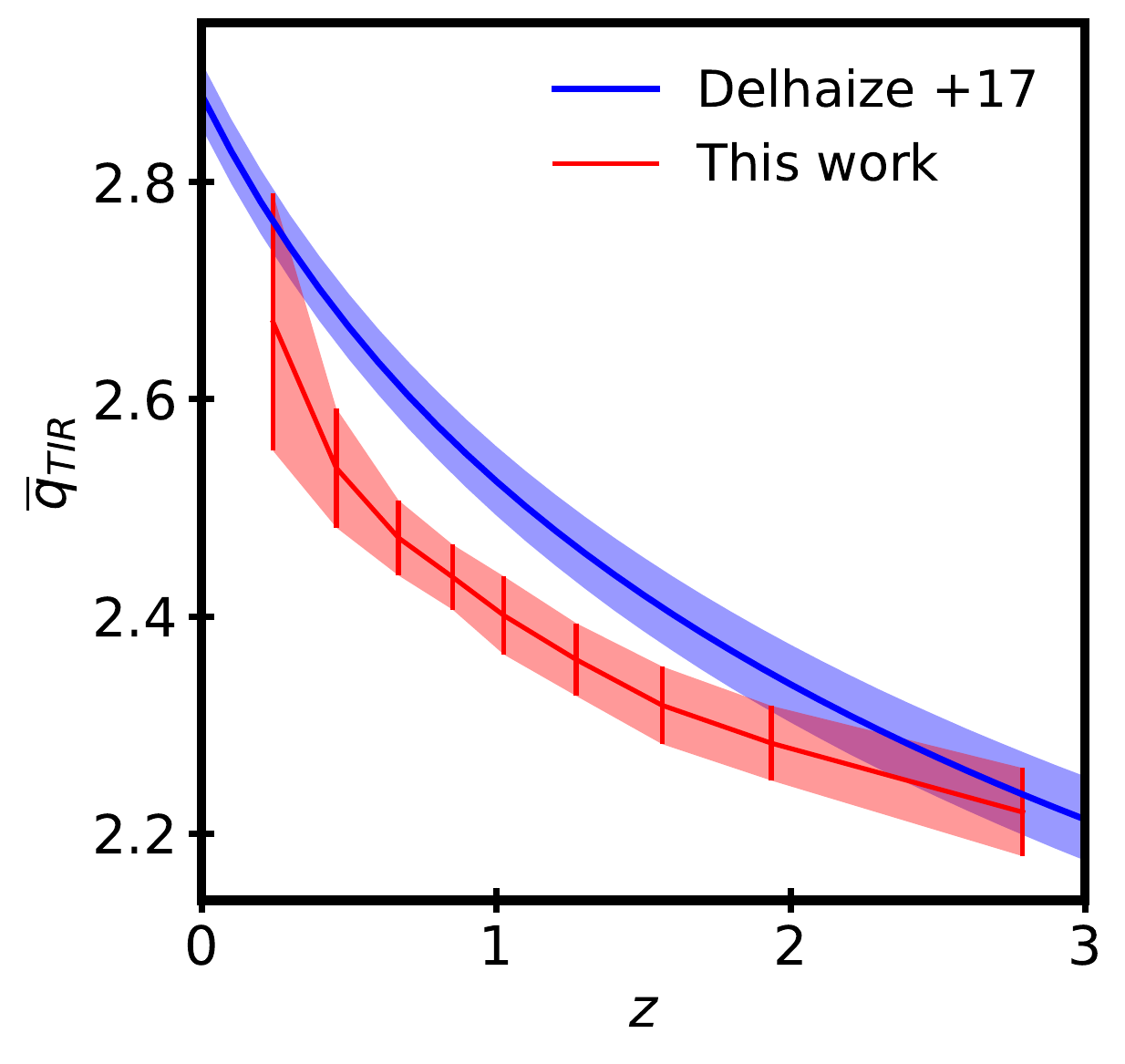}
\caption{Redshift dependence of the IRRC. Blue symbols represent median $\overline{q}_{\rm TIR}$ values in the COSMOS field from \protect\citet{delhaize17}. Blue line with the shaded are is the best-fit from \protect\citet{delhaize17} with 1$\sigma$ confidence interval. Red line shows our predicted $\overline{q}_{\rm TIR}$ as a function of redshift based on the median $L_{1.4}$ luminosity of COSMOS SFGs in each $z$ bin using Eq \protect\ref{eq::l_vs_q}. The red band shows the 1$\sigma$ confidence interval of our prediction based on the uncertainties and covariance of our fit parameters and the median $L_{1.4}$ observed in COSMOS data.}
\label{fig::q_z_mod}
\end{figure}

\citet{delhaize17} found no physical explanation for the trend of decreasing IR-to-radio ratios. Moreover, as discussed in Sect. \ref{sect::sfr_radio}, the extrapolation from the COSMOS sample in particular overestimates the $z = 0$ IR-radio ratios most commonly cited in the literature \citep{yun01,bell03}. Using our $q_{\rm TIR}$ -- $L_{1.4}$ relation we revisited the results of \citet{delhaize17}. We computed median $L_{1.4}$ values in equal number redshift bins in the \cite{delhaize17} SFG sample using single-sided survival analysis\footnote{For a given a set of data containing both limits and direct measurements, survival analysis estimates the cumulative distribution function (CDF) of the underlying distribution they were drawn from. If, as in our particular case, besides direct detections, only either upper or lower limits occur (i.e. the data are singly censored), the CDF can be constrained analytically with the Kaplan--Meier product limit estimator \citep{kaplan58}.}, and with the best-fit Eq. \ref{eq::l_vs_q} we predicted the expected median $q_{\rm TIR}$ in each bin. Fig. \ref{fig::q_z_mod} shows the median $q_{\rm TIR}$ values reported by \citet{delhaize17}, and our predicted values. The 1\,$\sigma$ confidence interval was calculated from the correlated slope and intercept uncertainties of our best-fit $q_{\rm TIR}$ -- $L_{1.4}$ line combined with uncertainties on the median $L_{1.4}$ values in each redshift bin. Our empirical model qualitatively recovers the observed declining $q_{\rm TIR}$ -- $z$ trend, albeit with a $\sim$0.1 dex lower normalization and slightly shallower slope. In particular, compared to the $(1+z)^{-(0.19\pm0.01)}$ of \citet{delhaize17}, our predicted $q_{\rm TIR}$ -- $z$ fit for this COSMOS sample follows a $(1+z)^{-(0.16\pm0.01)}$ curve. Nevertheless, our model is typically in $\sim$1.5\,$\sigma$ agreement with their measurements. Furthermore, it is an even better match to their radio-excess cleaned sample (shown in their Fig. 16 in cyan), which follows a $(1+z)^{-(0.15\pm0.01)}$ redshift evolution.

The preceding calculations suggest that the apparent redshift evolution of $q_{\rm TIR}$ is primarily a selection effect. On the one hand there is the {\it physical} effect that, in the early Universe, galaxy star formation activity -- and hence radio luminosity -- was higher, which via Eq. \ref{eq::l_vs_q} implies a lower $\overline{q}_{\rm TIR}$ for high-$z$ populations. On the other hand, {\it observational} selection effects cause galaxies fainter than the detection limit both in the radio and IR bands not to enter high-$z$ samples, again skewing the measurement of IR-to-radio ratios towards high-luminosity objects with lower $q$ values. To a varying extent, this Malmquist bias is present regardless of the details of the analysis/selection method, e.g. when considering only detected sources, but also when including upper flux limits for undetected sources. The combination of both effects produces a redshift-dependent sampling of an underlying non-linear relation, leading to declining IR-to-radio ratio measurements at higher redshifts.

Consistent with this interpretation of the redshift evolution of the IRRC reported in recent literature, the bivariate analysis carried out by \citet{delvecchio21} revealed that IR-to-radio ratios depend mostly on stellar mass and much less on redshift. As the SFR and stellar mass of SFG are correlated \citep[e.g.][]{whitaker12, schreiber15, leslie20}, the preferential sampling of higher-luminosity, high-mass galaxies in the early Universe will lead to a qualitatively similar behaviour as outlined above. We also find a good quantitative agreement with the work of \citet{delvecchio21}. For $M_{\star}\,{=}\,10^{10}\,M_{\odot}$, the characteristic mass scale in their bivariate fitting formula \citep[see Eq. (6) in][]{delvecchio21} and at $z\,{\sim}$\,0 these authors predicts $q_{\rm TIR}$\,=\,2.65. Given normalisation and slope of $z\,{\sim}$\,0 literature star-forming main sequence fits \citep[e.g.][]{brinchmann04, renzinipeng15}, $M_{\star}\,{=}\,10^{10}\,M_{\odot}$ translates to an SFR of $\sim1\,M_{\odot}$/yr or $L_{\rm TIR}\,{\sim}\,10^{10}\,L_{\odot}$. Using our IRRC best-fit parameters in Eq. \ref{eq::l_vs_q}, this luminosity implies an IR-to-radio ratio of $q_{\rm TIR}$\,=\,2.67, in excellent agreement with \citet{delvecchio21}. In this context, the high $z\,{=}\,0$ $q_{\rm TIR}$ values in COSMOS in \citet{delhaize17} are thus likely due to the small volume probed at low redshifts, which causes few bright galaxies to enter the sample of \citet{delhaize17}. As a result, the average $L_{1.4}$ in low-$z$ bins is lower than that of wider surveys, with the consequence that the non-linear IRRC is probed in a regime with a higher effective $q_{\rm TIR}$ (see Fig. \ref{fig::irrc_main}). By the same logic, the varying slopes of different $q_{\rm TIR}$ - $z$ fits in previously mentioned works are potentially related to their different observed radio luminosity distributions at each redshift.

In conclusion, we suggest that the various proposed $q_{\rm TIR}$ -- $z$ calibrations in the literature so far are in general accurate for the data sets they were derived from. For instance, \citet{delhaize17} suggested a redshift-dependent $L_{1.4}$ -- SFR calibration based on their declining $q_{\rm TIR}$ measurements. \citet{novak17} then applied this calibration to infer the cosmic SFR density (SFRD) out to $z\,{\sim}$\,5 from 3\,GHz radio data in COSMOS, finding a SFRD evolution that is broadly consistent with previous measurements. This is due to the fact that the underlying $q$-measurements by-and-large produce the correct relation between radio synchrotron luminosity and SFR for the $L^*$ population which contributes most to the SFRD. A more universally consistent approach, however, would be the use of Eq. \ref{eq::sfr_q_lrad} or a similar formula \citep[e.g.][]{davies17,brown17,gurkan18,duncan20}, until physically motivated models fitting observational data emerge.

\section{Summary}

In order to provide a new, comprehensive study of the IRRC, and ultimately improve the $L_{1.4}$ -- SFR calibration (crucial for the upcoming new generation of radio surveys with MeerKAT, ASKAP, and eventually SKA and ngVLA), we assemble an SDSS-based catalogue of 9,645 IR- and radio-detected galaxies in the nearby ($z<0.2$) Universe. Thanks to our large initial pool of galaxies, even with stringent selection criteria excluding AGN or low-quality measurements, we retain $\sim$ 2,400 SFGs in our final sample. To improve on previous similar works, we (i) utilize more recent IR surveys to achieve a better IR wavelength coverage and measure more accurate IR luminosities via SED-fitting, (ii) add deeper FIRST radio data to our catalogue, (iii) select pure star forming galaxies, (iv) consider the bias in the median IR-radio ratio arising from non-matching IR and radio survey depths and (v) employ a fitting technique to model the IRRC that is shown to be more robust than the typical least square fit or bisector approaches.

With galaxy emission line ratios we separated pure star-forming galaxies from optically-selected AGN (Sect \ref{sect::agn_sfg_sep}). Since our IR-SED fitting employs star-forming galaxy template SEDs, and some AGN can have substantial non-SF related IR emission, we also removed sources identified as AGN by their MIR colours. With the SFG sample we investigate selection effects biasing the median IR/radio ratio q$_\mathrm{TIR}$, and which arise from poorly matched sensitivity levels in the radio and IR data. We were able to quantify the level of bias present in the widely referenced \citet{yun01} study and mitigated it through a "matched-depth" approach, i.e. by applying a 22\,$\mu$m flux cut (see Sect \ref{sect::bias}). More generally, the details of any such bias corrections will depend on the flux/luminosity regime, but we demonstrate that they are important to consider. In particular, we expect that deep radio surveys such as MIGHTEE and EMU with MeerKAT and ASKAP (and ultimately SKA surveys), combined with existing IR data (e.g. from WISE, IRAS or Herschel) in the low-z Universe will obtain median q$_\mathrm{TIR}$ values biased high by $\sim$ 0.2 -- 0.4 dex, if the difference in sensitivies between IR and radio data is not taken into account. 

In Sects \ref{sect::lum_lum} and \ref{sect::q} we characterise the IRRC properties of our depth-matched SFG sample using both monochromatic and total IR luminosities. We find that, in general, 22 -- 100 $\mu$m monochromatic correlations have close to unity IRRC slopes, which becomes steeper towards longer wavelengths. The total IR-based correlation has a slope of 1.11 with a dispersion of 0.12 dex. This significant decrease in dispersion relative to previous literature is, in part, due to the difference in the dispersion measurement itself. In order to carry out a fairer comparison, we also fitted our data considering $L_{\rm TIR}$ as the independent variable, and estimated the scatter as the standard deviation of the offsets from this line in the y-direction. This resulted in a $\sim$0.18\,dex scatter, suggesting a genuine improvement compared to the previously reported 0.26 dex scatter of \cite{yun01}, regardless of the fitting approach.

Recipes for deriving SFRs from $L_{1.4}$ often involve the median IR-radio ratio which we also measure for the galaxies in our sample. In the total depth-matched sample we find $\overline{q}_{\mathrm{TIR}}= 2.51$ with a scatter of 0.22 dex, for depth-matched SFGs we obtained $\overline{q}_{\mathrm{TIR}}=2.54\pm 0.01$ and scatter of 0.19 dex, while depth-matched AGN have $\overline{q}_{\mathrm{TIR}} = 2.46\pm 0.02$ and a scatter of 0.27 dex. These scatters are systematically higher than those found for the IRRC itself, due to the non-unity slope of the correlation, which in turn is the result of q$_\mathrm{TIR}$ values (anti-)correlating with radio luminosity. Thus, instead of using a fixed q$_\mathrm{TIR}$ value, we propose an $L_{1.4}$-dependent $\overline{q}_\mathrm{TIR}$, and consequently, SFR calibration (Eq \ref{eq::sfr_q_lrad}). In Sect \ref{sect::sfr_radio} with IR- and radio-independent SFR estimates available for our SDSS sources, we confirm that such a recipe is compatible with other existing $L_{1.4}$ -- SFR calibrations in the local Universe. More importantly, it goes a long way to empirically explain the apparent evolution of $\overline{q}_\mathrm{TIR}$ found by recent studies as the consequence of a selection effect whereby different parts of the non-linear IRRC are sampled depending on redshift and sample depth (Sect \ref{sect::q_vs_z}). Hence, it provides a robust $L_{1.4}$ -- SFR recipe for both low and high-redshift 1.4 GHz radio observations.

\section*{Acknowledgements}

We thank the anonymous reviewer for the thorough and insightful reports allowing us to improve this work. We thank G\'{a}bor Marton, Paolo Serra, Filippo Maccagni, Mpati Ramatsoku, Dane Kleiner, Scott Clay and Beno\^{i}t Fournier for all the useful discussions. DCM acknowledges support from the Science and Technology Facilities Council (grant number ST/M503836/1).
SL acknowledges funding from Deutsche Forschungsgemeinschaft (DFG) Grant BE 1837 / 13-1 r. MTS acknowledges support from a Royal Society Leverhulme Trust Senior Research Fellowship (LT150041). ES acknowledges funding from the European Research Council (ERC) under the European Union's Horizon 2020 research and innovation programme (grant agreement No. 694343). B.M. acknowledges support from the Collaborative Research Centre 956, sub-project A1, funded by the Deutsche Forschungsgemeinschaft (DFG) -- project ID 184018867. JD acknowledges financial assistance from the South African Radio Astronomy Observatory (SARAO; www.ska.ac.za).

\section*{Data Availability}

The data underlying this article are available in its online supplementary material. The same data are also available in Zenodo, at \url{https://zenodo.org/badge/latestdoi/344211798}.

%%%%%%%%%%%%%%%%%%%%%%%%%%%%%%%%%%%%%%%%%%%%%%%%%%%%%%%%%%%%%%%%%%%%%%%

%%%%%%%%%%%%%%%%%%%% REFERENCES %%%%%%%%%%%%%%%%%%

\bibliographystyle{mnras}
\bibliography{ref}

\begin{thebibliography}{}
\makeatletter
\relax
\def\mn@urlcharsother{\let\do\@makeother \do\$\do\&\do\#\do\^\do\_\do\%\do\~}
\def\mn@doi{\begingroup\mn@urlcharsother \@ifnextchar [ {\mn@doi@}
  {\mn@doi@[]}}
\def\mn@doi@[#1]#2{\def\@tempa{#1}\ifx\@tempa\@empty \href
  {http://dx.doi.org/#2} {doi:#2}\else \href {http://dx.doi.org/#2} {#1}\fi
  \endgroup}
\def\mn@eprint#1#2{\mn@eprint@#1:#2::\@nil}
\def\mn@eprint@arXiv#1{\href {http://arxiv.org/abs/#1} {{\tt arXiv:#1}}}
\def\mn@eprint@dblp#1{\href {http://dblp.uni-trier.de/rec/bibtex/#1.xml}
  {dblp:#1}}
\def\mn@eprint@#1:#2:#3:#4\@nil{\def\@tempa {#1}\def\@tempb {#2}\def\@tempc
  {#3}\ifx \@tempc \@empty \let \@tempc \@tempb \let \@tempb \@tempa \fi \ifx
  \@tempb \@empty \def\@tempb {arXiv}\fi \@ifundefined
  {mn@eprint@\@tempb}{\@tempb:\@tempc}{\expandafter \expandafter \csname
  mn@eprint@\@tempb\endcsname \expandafter{\@tempc}}}

\bibitem[\protect\citeauthoryear{{Akritas} \& {Bershady}}{{Akritas} \&
  {Bershady}}{1996}]{akritas96}
{Akritas} M.~G.,  {Bershady} M.~A.,  1996, \mn@doi [\apj] {10.1086/177901},
  \href {https://ui.adsabs.harvard.edu/\#abs/1996ApJ...470..706A} {470, 706}

\bibitem[\protect\citeauthoryear{{Alam} et~al.,}{{Alam} et~al.}{2015}]{alam15}
{Alam} S.,  et~al., 2015, \mn@doi [\apjs] {10.1088/0067-0049/219/1/12}, \href
  {http://adsabs.harvard.edu/abs/2015ApJS..219...12A} {219, 12}

\bibitem[\protect\citeauthoryear{{Andrae}, {Schulze-Hartung}  \&
  {Melchior}}{{Andrae} et~al.}{2010}]{andrae10}
{Andrae} R.,  {Schulze-Hartung} T.,   {Melchior} P.,  2010, arXiv e-prints,
  \href {https://ui.adsabs.harvard.edu/abs/2010arXiv1012.3754A} {p.
  arXiv:1012.3754}

\bibitem[\protect\citeauthoryear{{Appleton} et~al.,}{{Appleton}
  et~al.}{2004}]{appleton04}
{Appleton} P.~N.,  et~al., 2004, \mn@doi [\apjs] {10.1086/422425}, \href
  {http://adsabs.harvard.edu/abs/2004ApJS..154..147A} {154, 147}

\bibitem[\protect\citeauthoryear{{Assef}, {Stern}, {Noirot}, {Jun}, {Cutri}  \&
  {Eisenhardt}}{{Assef} et~al.}{2018}]{assef18}
{Assef} R.~J.,  {Stern} D.,  {Noirot} G.,  {Jun} H.~D.,  {Cutri} R.~M.,
  {Eisenhardt} P.~R.~M.,  2018, \mn@doi [\apjs] {10.3847/1538-4365/aaa00a},
  \href {https://ui.adsabs.harvard.edu/abs/2018ApJS..234...23A} {234, 23}

\bibitem[\protect\citeauthoryear{{Baldwin}, {Phillips}  \&
  {Terlevich}}{{Baldwin} et~al.}{1981}]{baldwin1981}
{Baldwin} J.~A.,  {Phillips} M.~M.,   {Terlevich} R.,  1981, \mn@doi [\pasp]
  {10.1086/130766}, \href
  {https://ui.adsabs.harvard.edu/abs/1981PASP...93....5B} {93, 5}

\bibitem[\protect\citeauthoryear{{Becker}, {White}  \& {Helfand}}{{Becker}
  et~al.}{1995}]{becker95}
{Becker} R.~H.,  {White} R.~L.,   {Helfand} D.~J.,  1995, \mn@doi [\apj]
  {10.1086/176166}, \href {http://adsabs.harvard.edu/abs/1995ApJ...450..559B}
  {450, 559}

\bibitem[\protect\citeauthoryear{{Bell}}{{Bell}}{2003}]{bell03}
{Bell} E.~F.,  2003, \mn@doi [\apj] {10.1086/367829}, \href
  {http://adsabs.harvard.edu/abs/2003ApJ...586..794B} {586, 794}

\bibitem[\protect\citeauthoryear{{Booth}, {de Blok}, {Jonas}  \&
  {Fanaroff}}{{Booth} et~al.}{2009}]{booth09}
{Booth} R.~S.,  {de Blok} W.~J.~G.,  {Jonas} J.~L.,   {Fanaroff} B.,  2009,
  arXiv e-prints, \href {https://ui.adsabs.harvard.edu/abs/2009arXiv0910.2935B}
  {p. arXiv:0910.2935}

\bibitem[\protect\citeauthoryear{{Bourne} et~al.,}{{Bourne}
  et~al.}{2016}]{bourne16}
{Bourne} N.,  et~al., 2016, \mn@doi [\mnras] {10.1093/mnras/stw1654}, \href
  {http://adsabs.harvard.edu/abs/2016MNRAS.462.1714B} {462, 1714}

\bibitem[\protect\citeauthoryear{{Brinchmann}, {Charlot}, {White}, {Tremonti},
  {Kauffmann}, {Heckman}  \& {Brinkmann}}{{Brinchmann}
  et~al.}{2004}]{brinchmann04}
{Brinchmann} J.,  {Charlot} S.,  {White} S.~D.~M.,  {Tremonti} C.,  {Kauffmann}
  G.,  {Heckman} T.,   {Brinkmann} J.,  2004, \mn@doi [\mnras]
  {10.1111/j.1365-2966.2004.07881.x}, \href
  {https://ui.adsabs.harvard.edu/abs/2004MNRAS.351.1151B} {351, 1151}

\bibitem[\protect\citeauthoryear{{Brown} et~al.,}{{Brown}
  et~al.}{2017}]{brown17}
{Brown} M. J.~I.,  et~al., 2017, \mn@doi [\apj] {10.3847/1538-4357/aa8ad2},
  \href {https://ui.adsabs.harvard.edu/abs/2017ApJ...847..136B} {847, 136}

\bibitem[\protect\citeauthoryear{{Calistro Rivera} et~al.,}{{Calistro Rivera}
  et~al.}{2017}]{calistro-rivera17}
{Calistro Rivera} G.,  et~al., 2017, \mn@doi [\mnras] {10.1093/mnras/stx1040},
  \href {http://adsabs.harvard.edu/abs/2017MNRAS.469.3468C} {469, 3468}

\bibitem[\protect\citeauthoryear{{Chabrier}}{{Chabrier}}{2003}]{chabrier03}
{Chabrier} G.,  2003, \mn@doi [\pasp] {10.1086/376392}, \href
  {http://adsabs.harvard.edu/abs/2003PASP..115..763C} {115, 763}

\bibitem[\protect\citeauthoryear{{Charlot} \& {Fall}}{{Charlot} \&
  {Fall}}{2000}]{charlot2000}
{Charlot} S.,  {Fall} S.~M.,  2000, \mn@doi [\apj] {10.1086/309250}, \href
  {http://adsabs.harvard.edu/abs/2000ApJ...539..718C} {539, 718}

\bibitem[\protect\citeauthoryear{{Chary} \& {Elbaz}}{{Chary} \&
  {Elbaz}}{2001}]{chary01}
{Chary} R.,  {Elbaz} D.,  2001, \mn@doi [\apj] {10.1086/321609}, \href
  {http://adsabs.harvard.edu/abs/2001ApJ...556..562C} {556, 562}

\bibitem[\protect\citeauthoryear{{Condon}}{{Condon}}{1984}]{condon84}
{Condon} J.~J.,  1984, \mn@doi [\apj] {10.1086/162705}, \href
  {https://ui.adsabs.harvard.edu/abs/1984ApJ...287..461C} {287, 461}

\bibitem[\protect\citeauthoryear{{Condon}}{{Condon}}{1992}]{condon92}
{Condon} J.~J.,  1992, \mn@doi [\araa] {10.1146/annurev.aa.30.090192.003043},
  \href {http://adsabs.harvard.edu/abs/1992ARA%26A..30..575C} {30, 575}

\bibitem[\protect\citeauthoryear{{Condon}, {Cotton}, {Greisen}, {Yin},
  {Perley}, {Taylor}  \& {Broderick}}{{Condon} et~al.}{1998}]{condon98}
{Condon} J.~J.,  {Cotton} W.~D.,  {Greisen} E.~W.,  {Yin} Q.~F.,  {Perley}
  R.~A.,  {Taylor} G.~B.,   {Broderick} J.~J.,  1998, \mn@doi [\aj]
  {10.1086/300337}, \href {http://adsabs.harvard.edu/abs/1998AJ....115.1693C}
  {115, 1693}

\bibitem[\protect\citeauthoryear{{Dale} \& {Helou}}{{Dale} \&
  {Helou}}{2002}]{dale02}
{Dale} D.~A.,  {Helou} G.,  2002, \mn@doi [\apj] {10.1086/341632}, \href
  {http://adsabs.harvard.edu/abs/2002ApJ...576..159D} {576, 159}

\bibitem[\protect\citeauthoryear{{Dale}, {Helou}, {Contursi}, {Silbermann}  \&
  {Kolhatkar}}{{Dale} et~al.}{2001}]{dale01}
{Dale} D.~A.,  {Helou} G.,  {Contursi} A.,  {Silbermann} N.~A.,   {Kolhatkar}
  S.,  2001, \mn@doi [\apj] {10.1086/319077}, \href
  {http://adsabs.harvard.edu/abs/2001ApJ...549..215D} {549, 215}

\bibitem[\protect\citeauthoryear{{Davies} et~al.,}{{Davies}
  et~al.}{2017}]{davies17}
{Davies} L.~J.~M.,  et~al., 2017, \mn@doi [\mnras] {10.1093/mnras/stw3080},
  \href {http://adsabs.harvard.edu/abs/2017MNRAS.466.2312D} {466, 2312}

\bibitem[\protect\citeauthoryear{{DeBoer} et~al.,}{{DeBoer}
  et~al.}{2009}]{deboer09}
{DeBoer} D.~R.,  et~al., 2009, \mn@doi [IEEE Proceedings]
  {10.1109/JPROC.2009.2016516}, \href
  {https://ui.adsabs.harvard.edu/abs/2009IEEEP..97.1507D} {97, 1507}

\bibitem[\protect\citeauthoryear{{Delhaize} et~al.,}{{Delhaize}
  et~al.}{2017}]{delhaize17}
{Delhaize} J.,  et~al., 2017, \mn@doi [\aap] {10.1051/0004-6361/201629430},
  \href {http://adsabs.harvard.edu/abs/2017A%26A...602A...4D} {602, A4}

\bibitem[\protect\citeauthoryear{{Delvecchio} et~al.,}{{Delvecchio}
  et~al.}{2020}]{delvecchio21}
{Delvecchio} I.,  et~al., 2020, arXiv e-prints, \href
  {https://ui.adsabs.harvard.edu/abs/2020arXiv201005510D} {p. arXiv:2010.05510}

\bibitem[\protect\citeauthoryear{{Driver} et~al.,}{{Driver}
  et~al.}{2011}]{driver11}
{Driver} S.~P.,  et~al., 2011, \mn@doi [\mnras]
  {10.1111/j.1365-2966.2010.18188.x}, \href
  {http://adsabs.harvard.edu/abs/2011MNRAS.413..971D} {413, 971}

\bibitem[\protect\citeauthoryear{{Duncan}, {Shivaei}, {Shapley}, {Reddy},
  {Mobasher}, {Coil}, {Kriek}  \& {Siana}}{{Duncan} et~al.}{2020}]{duncan20}
{Duncan} K.~J.,  {Shivaei} I.,  {Shapley} A.~E.,  {Reddy} N.~A.,  {Mobasher}
  B.,  {Coil} A.~L.,  {Kriek} M.,   {Siana} B.,  2020, \mn@doi [\mnras]
  {10.1093/mnras/staa2561}, \href
  {https://ui.adsabs.harvard.edu/abs/2020MNRAS.498.3648D} {498, 3648}

\bibitem[\protect\citeauthoryear{{Eales} et~al.,}{{Eales}
  et~al.}{2010}]{eales10}
{Eales} S.,  et~al., 2010, \mn@doi [\pasp] {10.1086/653086}, \href
  {http://adsabs.harvard.edu/abs/2010PASP..122..499E} {122, 499}

\bibitem[\protect\citeauthoryear{{Foreman-Mackey}, {Hogg}, {Lang}  \&
  {Goodman}}{{Foreman-Mackey} et~al.}{2013}]{foremanmackey13}
{Foreman-Mackey} D.,  {Hogg} D.~W.,  {Lang} D.,   {Goodman} J.,  2013, \mn@doi
  [\pasp] {10.1086/670067}, \href
  {http://adsabs.harvard.edu/abs/2013PASP..125..306F} {125, 306}

\bibitem[\protect\citeauthoryear{{Francis}}{{Francis}}{1993}]{francis93}
{Francis} P.~J.,  1993, \mn@doi [\apj] {10.1086/172533}, \href
  {https://ui.adsabs.harvard.edu/abs/1993ApJ...407..519F} {407, 519}

\bibitem[\protect\citeauthoryear{{Garn}, {Green}, {Riley}  \&
  {Alexander}}{{Garn} et~al.}{2009}]{garn09}
{Garn} T.,  {Green} D.~A.,  {Riley} J.~M.,   {Alexander} P.,  2009, \mn@doi
  [\mnras] {10.1111/j.1365-2966.2009.15073.x}, \href
  {http://adsabs.harvard.edu/abs/2009MNRAS.397.1101G} {397, 1101}

\bibitem[\protect\citeauthoryear{{Garrett}}{{Garrett}}{2002}]{garrett02}
{Garrett} M.~A.,  2002, \mn@doi [\aap] {10.1051/0004-6361:20020169}, \href
  {http://adsabs.harvard.edu/abs/2002A%26A...384L..19G} {384, L19}

\bibitem[\protect\citeauthoryear{{Griffin} et~al.,}{{Griffin}
  et~al.}{2010}]{griffin10}
{Griffin} M.~J.,  et~al., 2010, \mn@doi [\aap] {10.1051/0004-6361/201014519},
  \href {http://adsabs.harvard.edu/abs/2010A%26A...518L...3G} {518, L3}

\bibitem[\protect\citeauthoryear{{G{\"u}rkan} et~al.,}{{G{\"u}rkan}
  et~al.}{2018}]{gurkan18}
{G{\"u}rkan} G.,  et~al., 2018, \mn@doi [\mnras] {10.1093/mnras/sty016}, \href
  {https://ui.adsabs.harvard.edu/abs/2018MNRAS.475.3010G} {475, 3010}

\bibitem[\protect\citeauthoryear{{Helfand}, {White}  \& {Becker}}{{Helfand}
  et~al.}{2015}]{helfand15}
{Helfand} D.~J.,  {White} R.~L.,   {Becker} R.~H.,  2015, \mn@doi [\apj]
  {10.1088/0004-637X/801/1/26}, \href
  {http://adsabs.harvard.edu/abs/2015ApJ...801...26H} {801, 26}

\bibitem[\protect\citeauthoryear{{Helou} \& {Bicay}}{{Helou} \&
  {Bicay}}{1993}]{helou93}
{Helou} G.,  {Bicay} M.~D.,  1993, \mn@doi [\apj] {10.1086/173146}, \href
  {http://adsabs.harvard.edu/abs/1993ApJ...415...93H} {415, 93}

\bibitem[\protect\citeauthoryear{{Helou}, {Soifer}  \&
  {Rowan-Robinson}}{{Helou} et~al.}{1985}]{Helou85}
{Helou} G.,  {Soifer} B.~T.,   {Rowan-Robinson} M.,  1985, \mn@doi [\apjl]
  {10.1086/184556}, \href {http://adsabs.harvard.edu/abs/1985ApJ...298L...7H}
  {298, L7}

\bibitem[\protect\citeauthoryear{{Hodge}, {Becker}, {White}  \& {de
  Vries}}{{Hodge} et~al.}{2008}]{hodge08}
{Hodge} J.~A.,  {Becker} R.~H.,  {White} R.~L.,   {de Vries} W.~H.,  2008,
  \mn@doi [\aj] {10.1088/0004-6256/136/3/1097}, \href
  {https://ui.adsabs.harvard.edu/abs/2008AJ....136.1097H} {136, 1097}

\bibitem[\protect\citeauthoryear{{Hogg}, {Bovy}  \& {Lang}}{{Hogg}
  et~al.}{2010}]{hogg10}
{Hogg} D.~W.,  {Bovy} J.,   {Lang} D.,  2010, arXiv e-prints, \href
  {https://ui.adsabs.harvard.edu/\#abs/2010arXiv1008.4686H} {p.
  arXiv:1008.4686}

\bibitem[\protect\citeauthoryear{{Ibar} et~al.,}{{Ibar} et~al.}{2008}]{ibar08}
{Ibar} E.,  et~al., 2008, \mn@doi [\mnras] {10.1111/j.1365-2966.2008.13077.x},
  \href {http://adsabs.harvard.edu/abs/2008MNRAS.386..953I} {386, 953}

\bibitem[\protect\citeauthoryear{{Ivison} et~al.,}{{Ivison}
  et~al.}{2010a}]{ivison10a}
{Ivison} R.~J.,  et~al., 2010a, \mn@doi [\mnras]
  {10.1111/j.1365-2966.2009.15918.x}, \href
  {http://adsabs.harvard.edu/abs/2010MNRAS.402..245I} {402, 245}

\bibitem[\protect\citeauthoryear{{Ivison} et~al.,}{{Ivison}
  et~al.}{2010b}]{ivison10b}
{Ivison} R.~J.,  et~al., 2010b, \mn@doi [\aap] {10.1051/0004-6361/201014552},
  \href {http://adsabs.harvard.edu/abs/2010A%26A...518L..31I} {518, L31}

\bibitem[\protect\citeauthoryear{{Jarrett}, {Chester}, {Cutri}, {Schneider}  \&
  {Huchra}}{{Jarrett} et~al.}{2003}]{jarrett2003}
{Jarrett} T.~H.,  {Chester} T.,  {Cutri} R.,  {Schneider} S.~E.,   {Huchra}
  J.~P.,  2003, \mn@doi [\aj] {10.1086/345794}, \href
  {https://ui.adsabs.harvard.edu/abs/2003AJ....125..525J} {125, 525}

\bibitem[\protect\citeauthoryear{{Jarvis} et~al.,}{{Jarvis}
  et~al.}{2010}]{jarvis10}
{Jarvis} M.~J.,  et~al., 2010, \mn@doi [\mnras]
  {10.1111/j.1365-2966.2010.17772.x}, \href
  {http://adsabs.harvard.edu/abs/2010MNRAS.409...92J} {409, 92}

\bibitem[\protect\citeauthoryear{{Jarvis} et~al.,}{{Jarvis}
  et~al.}{2016}]{jarvis16}
{Jarvis} M.,  et~al., 2016, in MeerKAT Science: On the Pathway to the SKA. p.~6
  (\mn@eprint {arXiv} {1709.01901})

\bibitem[\protect\citeauthoryear{{Johnston} et~al.,}{{Johnston}
  et~al.}{2007}]{johnston07}
{Johnston} S.,  et~al., 2007, \mn@doi [\pasa] {10.1071/AS07033}, \href
  {https://ui.adsabs.harvard.edu/abs/2007PASA...24..174J} {24, 174}

\bibitem[\protect\citeauthoryear{{Kaplan} \& {Meier}}{{Kaplan} \&
  {Meier}}{1958}]{kaplan58}
{Kaplan} E.~L.,  {Meier} P.,  1958, Journal of the American Statistical
  Association, 53, 457

\bibitem[\protect\citeauthoryear{{Kauffmann} et~al.,}{{Kauffmann}
  et~al.}{2003}]{kauffmann03}
{Kauffmann} G.,  et~al., 2003, \mn@doi [\mnras]
  {10.1111/j.1365-2966.2003.07154.x}, \href
  {http://adsabs.harvard.edu/abs/2003MNRAS.346.1055K} {346, 1055}

\bibitem[\protect\citeauthoryear{{Kellermann}}{{Kellermann}}{1964}]{kellermann64}
{Kellermann} K.~I.,  1964, Publications of the Owens Valley Observatory, \href
  {https://ui.adsabs.harvard.edu/abs/1964POVRO...1....1K} {1, 1}

\bibitem[\protect\citeauthoryear{{Kennicutt}}{{Kennicutt}}{1998}]{kennicutt98}
{Kennicutt} Jr. R.~C.,  1998, \mn@doi [\araa] {10.1146/annurev.astro.36.1.189},
  \href {http://adsabs.harvard.edu/abs/1998ARA%26A..36..189K} {36, 189}

\bibitem[\protect\citeauthoryear{{Kewley}, {Dopita}, {Sutherland}, {Heisler}
  \& {Trevena}}{{Kewley} et~al.}{2001}]{kewley01}
{Kewley} L.~J.,  {Dopita} M.~A.,  {Sutherland} R.~S.,  {Heisler} C.~A.,
  {Trevena} J.,  2001, \mn@doi [\apj] {10.1086/321545}, \href
  {http://adsabs.harvard.edu/abs/2001ApJ...556..121K} {556, 121}

\bibitem[\protect\citeauthoryear{{Kewley}, {Groves}, {Kauffmann}  \&
  {Heckman}}{{Kewley} et~al.}{2006}]{kewley06}
{Kewley} L.~J.,  {Groves} B.,  {Kauffmann} G.,   {Heckman} T.,  2006, \mn@doi
  [\mnras] {10.1111/j.1365-2966.2006.10859.x}, \href
  {http://adsabs.harvard.edu/abs/2006MNRAS.372..961K} {372, 961}

\bibitem[\protect\citeauthoryear{{Kimball} \& {Ivezi{\'c}}}{{Kimball} \&
  {Ivezi{\'c}}}{2008}]{kimball08}
{Kimball} A.~E.,  {Ivezi{\'c}} {\v Z}.,  2008, \mn@doi [\aj]
  {10.1088/0004-6256/136/2/684}, \href
  {http://adsabs.harvard.edu/abs/2008AJ....136..684K} {136, 684}

\bibitem[\protect\citeauthoryear{{Kimball} \& {Ivezi{\'c}}}{{Kimball} \&
  {Ivezi{\'c}}}{2014}]{kimball14}
{Kimball} A.~E.,  {Ivezi{\'c}} {\v Z}.,  2014, preprint, \href
  {http://adsabs.harvard.edu/abs/2014arXiv1401.1535K} {} (\mn@eprint {arXiv}
  {1401.1535})

\bibitem[\protect\citeauthoryear{{Lacki}, {Thompson}  \& {Quataert}}{{Lacki}
  et~al.}{2010}]{lacki10a}
{Lacki} B.~C.,  {Thompson} T.~A.,   {Quataert} E.,  2010, \mn@doi [\apj]
  {10.1088/0004-637X/717/1/1}, \href
  {http://adsabs.harvard.edu/abs/2010ApJ...717....1L} {717, 1}

\bibitem[\protect\citeauthoryear{{Lang}, {Hogg}  \& {Schlegel}}{{Lang}
  et~al.}{2014}]{lang14}
{Lang} D.,  {Hogg} D.~W.,   {Schlegel} D.~J.,  2014, preprint, \href
  {http://adsabs.harvard.edu/abs/2014arXiv1410.7397L} {} (\mn@eprint {arXiv}
  {1410.7397})

\bibitem[\protect\citeauthoryear{Lang, Hogg  \& Schlegel}{Lang
  et~al.}{2016}]{lang16}
Lang D.,  Hogg D.~W.,   Schlegel D.~J.,  2016, \mn@doi [The Astronomical
  Journal] {10.3847/0004-6256/151/2/36}, 151, 36

\bibitem[\protect\citeauthoryear{{Lauer}, {Tremaine}, {Richstone}  \&
  {Faber}}{{Lauer} et~al.}{2007}]{lauer07}
{Lauer} T.~R.,  {Tremaine} S.,  {Richstone} D.,   {Faber} S.~M.,  2007, \mn@doi
  [\apj] {10.1086/522083}, \href
  {https://ui.adsabs.harvard.edu/abs/2007ApJ...670..249L} {670, 249}

\bibitem[\protect\citeauthoryear{{Leslie}, {Kewley}, {Sanders}  \&
  {Lee}}{{Leslie} et~al.}{2016}]{leslie16}
{Leslie} S.~K.,  {Kewley} L.~J.,  {Sanders} D.~B.,   {Lee} N.,  2016, \mn@doi
  [\mnras] {10.1093/mnrasl/slv135}, \href
  {https://ui.adsabs.harvard.edu/abs/2016MNRAS.455L..82L} {455, L82}

\bibitem[\protect\citeauthoryear{{Leslie} et~al.,}{{Leslie}
  et~al.}{2020}]{leslie20}
{Leslie} S.~K.,  et~al., 2020, \mn@doi [\apj] {10.3847/1538-4357/aba044}, \href
  {https://ui.adsabs.harvard.edu/abs/2020ApJ...899...58L} {899, 58}

\bibitem[\protect\citeauthoryear{{Madau} \& {Dickinson}}{{Madau} \&
  {Dickinson}}{2014}]{madau14}
{Madau} P.,  {Dickinson} M.,  2014, \mn@doi [\araa]
  {10.1146/annurev-astro-081811-125615}, \href
  {https://ui.adsabs.harvard.edu/abs/2014ARA&A..52..415M} {52, 415}

\bibitem[\protect\citeauthoryear{{Magnelli} et~al.,}{{Magnelli}
  et~al.}{2015}]{magnelli15}
{Magnelli} B.,  et~al., 2015, \mn@doi [\aap] {10.1051/0004-6361/201424937},
  \href {http://adsabs.harvard.edu/abs/2015A%26A...573A..45M} {573, A45}

\bibitem[\protect\citeauthoryear{{Mao}, {Huynh}, {Norris}, {Dickinson},
  {Frayer}, {Helou}  \& {Monkiewicz}}{{Mao} et~al.}{2011}]{mao11}
{Mao} M.~Y.,  {Huynh} M.~T.,  {Norris} R.~P.,  {Dickinson} M.,  {Frayer} D.,
  {Helou} G.,   {Monkiewicz} J.~A.,  2011, \mn@doi [\apj]
  {10.1088/0004-637X/731/2/79}, \href
  {http://adsabs.harvard.edu/abs/2011ApJ...731...79M} {731, 79}

\bibitem[\protect\citeauthoryear{{Marcillac}, {Elbaz}, {Chary}, {Dickinson},
  {Galliano}  \& {Morrison}}{{Marcillac} et~al.}{2006}]{marcillac06}
{Marcillac} D.,  {Elbaz} D.,  {Chary} R.~R.,  {Dickinson} M.,  {Galliano} F.,
  {Morrison} G.,  2006, \mn@doi [\aap] {10.1051/0004-6361:20054035}, \href
  {https://ui.adsabs.harvard.edu/abs/2006A&A...451...57M} {451, 57}

\bibitem[\protect\citeauthoryear{{Marton} et~al.,}{{Marton}
  et~al.}{2017}]{marton17}
{Marton} G.,  et~al., 2017, arXiv e-prints, \href
  {https://ui.adsabs.harvard.edu/abs/2017arXiv170505693M} {p. arXiv:1705.05693}

\bibitem[\protect\citeauthoryear{{Moln{\'a}r} et~al.,}{{Moln{\'a}r}
  et~al.}{2018}]{molnar18}
{Moln{\'a}r} D.~C.,  et~al., 2018, \mn@doi [\mnras] {10.1093/mnras/stx3234},
  \href {http://adsabs.harvard.edu/abs/2018MNRAS.475..827M} {475, 827}

\bibitem[\protect\citeauthoryear{{Mori{\'c}}, {Smol{\v c}i{\'c}}, {Kimball},
  {Riechers}, {Ivezi{\'c}}  \& {Scoville}}{{Mori{\'c}} et~al.}{2010}]{moric10}
{Mori{\'c}} I.,  {Smol{\v c}i{\'c}} V.,  {Kimball} A.,  {Riechers} D.~A.,
  {Ivezi{\'c}} {\v Z}.,   {Scoville} N.,  2010, \mn@doi [\apj]
  {10.1088/0004-637X/724/1/779}, \href
  {http://adsabs.harvard.edu/abs/2010ApJ...724..779M} {724, 779}

\bibitem[\protect\citeauthoryear{{Murphy} et~al.,}{{Murphy}
  et~al.}{2011}]{murphy11}
{Murphy} E.~J.,  et~al., 2011, \mn@doi [\apj] {10.1088/0004-637X/737/2/67},
  \href {http://adsabs.harvard.edu/abs/2011ApJ...737...67M} {737, 67}

\bibitem[\protect\citeauthoryear{{Nemmen}, {Georganopoulos}, {Guiriec},
  {Meyer}, {Gehrels}  \& {Sambruna}}{{Nemmen} et~al.}{2012}]{nemmen12}
{Nemmen} R.~S.,  {Georganopoulos} M.,  {Guiriec} S.,  {Meyer} E.~T.,  {Gehrels}
  N.,   {Sambruna} R.~M.,  2012, \mn@doi [Science] {10.1126/science.1227416},
  \href {http://adsabs.harvard.edu/abs/2012Sci...338.1445N} {338, 1445}

\bibitem[\protect\citeauthoryear{{Neugebauer} et~al.,}{{Neugebauer}
  et~al.}{1984}]{neugebauer84}
{Neugebauer} G.,  et~al., 1984, \mn@doi [\apjl] {10.1086/184209}, \href
  {http://adsabs.harvard.edu/abs/1984ApJ...278L...1N} {278, L1}

\bibitem[\protect\citeauthoryear{{Noll}, {Burgarella}, {Giovannoli}, {Buat},
  {Marcillac}  \& {Mu{\~n}oz-Mateos}}{{Noll} et~al.}{2009}]{noll09}
{Noll} S.,  {Burgarella} D.,  {Giovannoli} E.,  {Buat} V.,  {Marcillac} D.,
  {Mu{\~n}oz-Mateos} J.~C.,  2009, \mn@doi [\aap]
  {10.1051/0004-6361/200912497}, \href
  {http://adsabs.harvard.edu/abs/2009A%26A...507.1793N} {507, 1793}

\bibitem[\protect\citeauthoryear{{Norris} et~al.,}{{Norris}
  et~al.}{2011}]{norris11}
{Norris} R.~P.,  et~al., 2011, \mn@doi [\pasa] {10.1071/AS11021}, \href
  {https://ui.adsabs.harvard.edu/abs/2011PASA...28..215N} {28, 215}

\bibitem[\protect\citeauthoryear{{Novak} et~al.,}{{Novak}
  et~al.}{2017}]{novak17}
{Novak} M.,  et~al., 2017, preprint, \href
  {http://adsabs.harvard.edu/abs/2017arXiv170309724N} {} (\mn@eprint {arXiv}
  {1703.09724})

\bibitem[\protect\citeauthoryear{{Nyland} et~al.,}{{Nyland}
  et~al.}{2017}]{nyland17}
{Nyland} K.,  et~al., 2017, \mn@doi [\mnras] {10.1093/mnras/stw2385}, \href
  {http://adsabs.harvard.edu/abs/2017MNRAS.464.1029N} {464, 1029}

\bibitem[\protect\citeauthoryear{{Pannella} et~al.,}{{Pannella}
  et~al.}{2015}]{pannella15}
{Pannella} M.,  et~al., 2015, \mn@doi [\apj] {10.1088/0004-637X/807/2/141},
  \href {http://adsabs.harvard.edu/abs/2015ApJ...807..141P} {807, 141}

\bibitem[\protect\citeauthoryear{{Perley}, {Chandler}, {Butler}  \&
  {Wrobel}}{{Perley} et~al.}{2011}]{perley11}
{Perley} R.~A.,  {Chandler} C.~J.,  {Butler} B.~J.,   {Wrobel} J.~M.,  2011,
  \mn@doi [\apjl] {10.1088/2041-8205/739/1/L1}, \href
  {https://ui.adsabs.harvard.edu/abs/2011ApJ...739L...1P} {739, L1}

\bibitem[\protect\citeauthoryear{{Pilbratt} et~al.,}{{Pilbratt}
  et~al.}{2010}]{pilbratt10}
{Pilbratt} G.~L.,  et~al., 2010, \mn@doi [\aap] {10.1051/0004-6361/201014759},
  \href {http://adsabs.harvard.edu/abs/2010A%26A...518L...1P} {518, L1}

\bibitem[\protect\citeauthoryear{{Poglitsch} et~al.,}{{Poglitsch}
  et~al.}{2010}]{poglitsch10}
{Poglitsch} A.,  et~al., 2010, \mn@doi [\aap] {10.1051/0004-6361/201014535},
  \href {http://adsabs.harvard.edu/abs/2010A%26A...518L...2P} {518, L2}

\bibitem[\protect\citeauthoryear{{Read} et~al.,}{{Read} et~al.}{2018}]{read18}
{Read} S.~C.,  et~al., 2018, \mn@doi [\mnras] {10.1093/mnras/sty2198}, \href
  {https://ui.adsabs.harvard.edu/abs/2018MNRAS.480.5625R} {480, 5625}

\bibitem[\protect\citeauthoryear{{Renzini} \& {Peng}}{{Renzini} \&
  {Peng}}{2015}]{renzinipeng15}
{Renzini} A.,  {Peng} Y.-j.,  2015, \mn@doi [\apjl]
  {10.1088/2041-8205/801/2/L29}, \href
  {https://ui.adsabs.harvard.edu/abs/2015ApJ...801L..29R} {801, L29}

\bibitem[\protect\citeauthoryear{{Rigby} et~al.,}{{Rigby}
  et~al.}{2011}]{rigby11}
{Rigby} E.~E.,  et~al., 2011, \mn@doi [\mnras]
  {10.1111/j.1365-2966.2011.18864.x}, \href
  {https://ui.adsabs.harvard.edu/abs/2011MNRAS.415.2336R} {415, 2336}

\bibitem[\protect\citeauthoryear{{Roychowdhury} \& {Chengalur}}{{Roychowdhury}
  \& {Chengalur}}{2012}]{roychowdhury12}
{Roychowdhury} S.,  {Chengalur} J.~N.,  2012, \mn@doi [\mnras]
  {10.1111/j.1745-3933.2012.01273.x}, \href
  {https://ui.adsabs.harvard.edu/abs/2012MNRAS.423L.127R} {423, L127}

\bibitem[\protect\citeauthoryear{{Salim} et~al.,}{{Salim}
  et~al.}{2016}]{salim16}
{Salim} S.,  et~al., 2016, \mn@doi [\apjs] {10.3847/0067-0049/227/1/2}, \href
  {http://adsabs.harvard.edu/abs/2016ApJS..227....2S} {227, 2}

\bibitem[\protect\citeauthoryear{{Salpeter}}{{Salpeter}}{1955}]{salpeter55}
{Salpeter} E.~E.,  1955, \mn@doi [\apj] {10.1086/145971}, \href
  {https://ui.adsabs.harvard.edu/abs/1955ApJ...121..161S} {121, 161}

\bibitem[\protect\citeauthoryear{{Sargent} et~al.,}{{Sargent}
  et~al.}{2010a}]{sargent10a}
{Sargent} M.~T.,  et~al., 2010a, \mn@doi [\apjs] {10.1088/0067-0049/186/2/341},
  \href {http://adsabs.harvard.edu/abs/2010ApJS..186..341S} {186, 341}

\bibitem[\protect\citeauthoryear{{Sargent} et~al.,}{{Sargent}
  et~al.}{2010b}]{sargent10b}
{Sargent} M.~T.,  et~al., 2010b, \mn@doi [\apjl]
  {10.1088/2041-8205/714/2/L190}, \href
  {https://ui.adsabs.harvard.edu/abs/2010ApJ...714L.190S} {714, L190}

\bibitem[\protect\citeauthoryear{{Schleicher} \& {Beck}}{{Schleicher} \&
  {Beck}}{2013}]{schleicher2013}
{Schleicher} D.~R.~G.,  {Beck} R.,  2013, \mn@doi [\aap]
  {10.1051/0004-6361/201321707}, \href
  {http://adsabs.harvard.edu/abs/2013A%26A...556A.142S} {556, A142}

\bibitem[\protect\citeauthoryear{{Schreiber} et~al.,}{{Schreiber}
  et~al.}{2015}]{schreiber15}
{Schreiber} C.,  et~al., 2015, \mn@doi [\aap] {10.1051/0004-6361/201425017},
  \href {https://ui.adsabs.harvard.edu/abs/2015A&A...575A..74S} {575, A74}

\bibitem[\protect\citeauthoryear{{Smith} et~al.,}{{Smith}
  et~al.}{2014}]{smith14}
{Smith} D.~J.~B.,  et~al., 2014, \mn@doi [\mnras] {10.1093/mnras/stu1830},
  \href {http://adsabs.harvard.edu/abs/2014MNRAS.445.2232S} {445, 2232}

\bibitem[\protect\citeauthoryear{{Valiante} et~al.,}{{Valiante}
  et~al.}{2016}]{valiante16}
{Valiante} E.,  et~al., 2016, \mn@doi [\mnras] {10.1093/mnras/stw1806}, \href
  {http://adsabs.harvard.edu/abs/2016MNRAS.462.3146V} {462, 3146}

\bibitem[\protect\citeauthoryear{{Voelk}}{{Voelk}}{1989}]{voelk89}
{Voelk} H.~J.,  1989, \aap, \href
  {http://adsabs.harvard.edu/abs/1989A%26A...218...67V} {218, 67}

\bibitem[\protect\citeauthoryear{{Wang}, {Rowan-Robinson}, {Norberg}, {Heinis}
  \& {Han}}{{Wang} et~al.}{2014}]{wang14}
{Wang} L.,  {Rowan-Robinson} M.,  {Norberg} P.,  {Heinis} S.,   {Han} J.,
  2014, \mn@doi [\mnras] {10.1093/mnras/stu915}, \href
  {http://adsabs.harvard.edu/abs/2014MNRAS.442.2739W} {442, 2739}

\bibitem[\protect\citeauthoryear{{Whitaker}, {van Dokkum}, {Brammer}  \&
  {Franx}}{{Whitaker} et~al.}{2012}]{whitaker12}
{Whitaker} K.~E.,  {van Dokkum} P.~G.,  {Brammer} G.,   {Franx} M.,  2012,
  \mn@doi [\apjl] {10.1088/2041-8205/754/2/L29}, \href
  {https://ui.adsabs.harvard.edu/abs/2012ApJ...754L..29W} {754, L29}

\bibitem[\protect\citeauthoryear{{Wright} et~al.,}{{Wright}
  et~al.}{2010}]{wright10}
{Wright} E.~L.,  et~al., 2010, \mn@doi [\aj] {10.1088/0004-6256/140/6/1868},
  \href {http://adsabs.harvard.edu/abs/2010AJ....140.1868W} {140, 1868}

\bibitem[\protect\citeauthoryear{{Yun}, {Reddy}  \& {Condon}}{{Yun}
  et~al.}{2001}]{yun01}
{Yun} M.~S.,  {Reddy} N.~A.,   {Condon} J.~J.,  2001, \mn@doi [\apj]
  {10.1086/323145}, \href {http://adsabs.harvard.edu/abs/2001ApJ...554..803Y}
  {554, 803}

\bibitem[\protect\citeauthoryear{{de Jong}, {Klein}, {Wielebinski}  \&
  {Wunderlich}}{{de Jong} et~al.}{1985}]{Jong85}
{de Jong} T.,  {Klein} U.,  {Wielebinski} R.,   {Wunderlich} E.,  1985, \aap,
  \href {http://adsabs.harvard.edu/abs/1985A%26A...147L...6D} {147, L6}

\bibitem[\protect\citeauthoryear{{van Haarlem} et~al.,}{{van Haarlem}
  et~al.}{2013}]{harleem13}
{van Haarlem} M.~P.,  et~al., 2013, \mn@doi [\aap]
  {10.1051/0004-6361/201220873}, \href
  {http://adsabs.harvard.edu/abs/2013A%26A...556A...2V} {556, A2}

\bibitem[\protect\citeauthoryear{{van der Kruit}}{{van der
  Kruit}}{1971}]{kruit71}
{van der Kruit} P.~C.,  1971, \aap, \href
  {http://adsabs.harvard.edu/abs/1971A%26A....15..110V} {15, 110}

\bibitem[\protect\citeauthoryear{{van der Kruit}}{{van der
  Kruit}}{1973}]{kruit73}
{van der Kruit} P.~C.,  1973, \aap, \href
  {http://adsabs.harvard.edu/abs/1973A%26A....29..263V} {29, 263}

\makeatother
\end{thebibliography}

%%%%%%%%%%%%%%%%% APPENDICES %%%%%%%%%%%%%%%%%%%%%

\appendix

\section{Determining SPSC and PPSC matching radii and estimating contamination from spurious matches}
\label{app::contamination}

Since neither the SPSC nor the PPSC had already established optical counterparts with high accuracy positions (as opposed to, e.g., H-ATLAS), we determined band-by-band matching radii between Herschel and SDSS DR12 positions with the aim of minimising the spurious fraction while maximising the number of counterparts found.

\begin{figure*}
\centering
\includegraphics[width=0.45\textwidth]{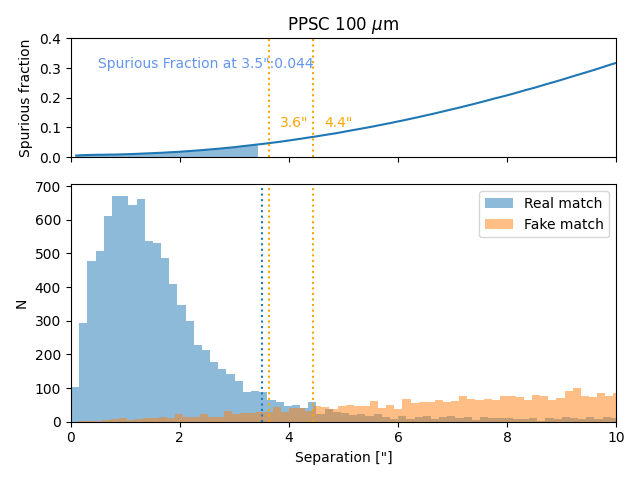}
\includegraphics[width=0.45\textwidth]{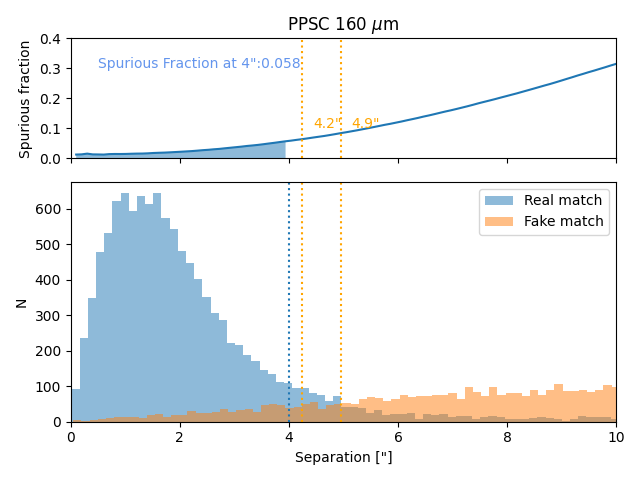}
\includegraphics[width=0.45\textwidth]{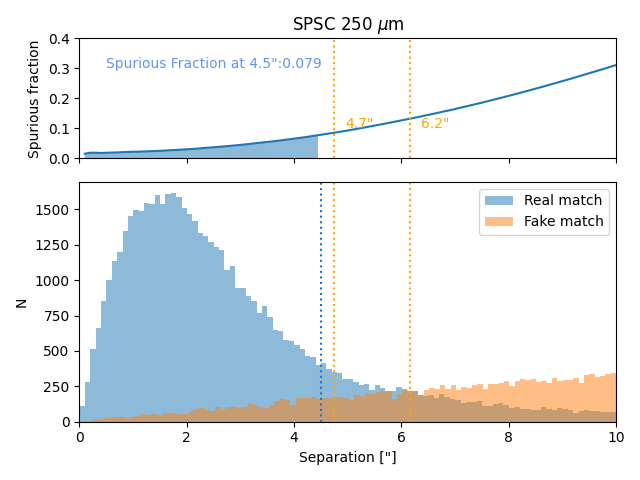}
\includegraphics[width=0.45\textwidth]{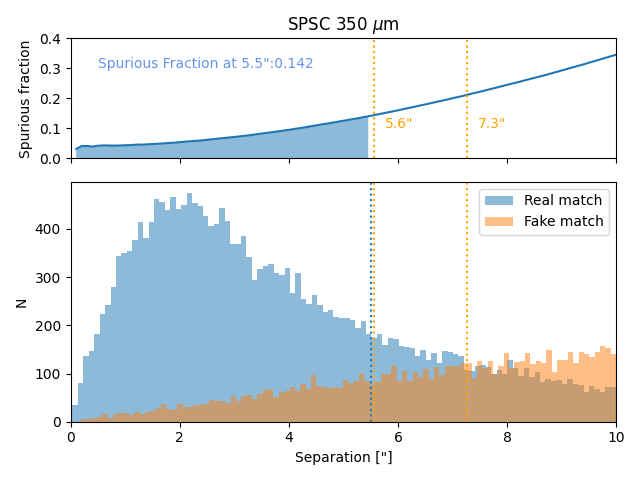}
\includegraphics[width=0.45\textwidth]{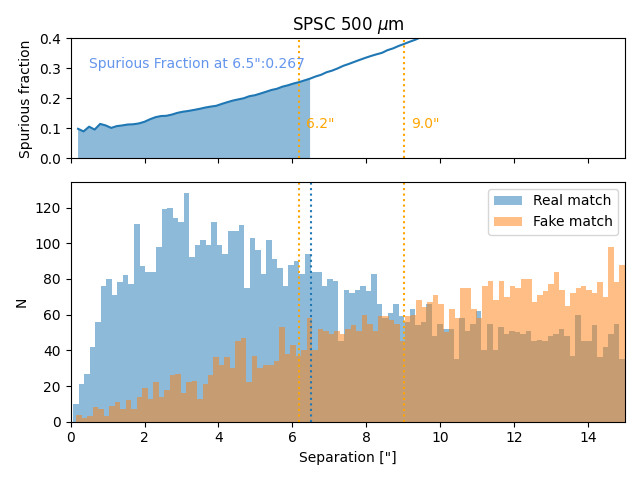}

\caption{Estimated contamination fraction from spurious matches as a function of matching radii for the PPSC and SPSC data. Blue histograms in each panel represent the distribution of optical -- IR angular separation between SDSS DR12 positions in our parent catalogue and the Herschel point source catalogue positions of a given band, while the orange ones are the number of matches as a function of separation between the optical and one of the mock IR catalogue positions. The estimated contamination fractions, calculated as the ratio of the real matches and the average separation distribution of all 15 mock matches, is shown in the smaller panels above. Vertical dashed orange lines mark the radial slices in which the real catalogue is expected to consist of 50\% and 100\% fake sources, while blue dashed lines represent the search radius we adopted for our catalogue. These radii alongside the estimated contamination fractions are listed in Table \protect\ref{tab::spur_frac}.}
\label{fig::psc_separations}
\end{figure*}

To this end, we estimated the level of contamination as a function of matching radius in each band independently. First, we generated 15 mock IR catalogues for each band. Due to the non-contiguous coverage of the SPSC and PPSC, we used the positions of the real source IR catalogues as a starting point to simply create mocks that mimic the sky coverage of the real sources. The RA and Dec of each mock IR source was calculated by adding uniformly drawn random numbers between $\pm18$\,arcsec to the real IR source positions.
We then cross-matched the mock IR catalogues with the SDSS parent sample using \textsc{TOPCAT} positional cross match, taking the closest match out to a maximum matching radius of 15\,arcsec.
For these fake matches, it is more likely to find sources with larger separations, because a larger separation, or search radius, corresponds to a larger search circumference, consequently larger search area, and thus more random associations are possible.
The distribution of optical--IR source separations for all 5 PSC bands is shown in Fig. \ref{fig::psc_separations}.
The blue histogram is the distribution of separations resulting from cross-matching our low-$z$ parent catalogue with the real IR catalogues, and the orange histogram shows the result from one of the fake IR catalogue matches.
For the spurious fraction estimations, we use the average distribution of all 15 fake catalogues. To calculate the contamination, we divide the number of fake sources (from the average of the fake catalogues) with the number of sources in the matched real catalogue that lie within our chosen search radius. The resulting curves are shown in the panels above each separation distribution.
The search radii adopted for our final catalog cross-match corresponds to the radial-slice (rounded to the nearest half arcsecond) at which 50\% of sources are spurious matches. The expected spurious fraction for each catalogue is $<$10\% except for Spire 350 and 500 micron that are 14\% and 27\%, respectively.

While the level of contamination among the 500\,$\mu$m data is formally high, we note that spurious matches likely result in unphysical SED shapes, that are flagged by our method for filtering out poor SED models (see Sect. \ref{sect::sed_fits}). Indeed, $\sim$16\% of all 500\,$\mu$m-detected sources were identified as having unreliable model fits. Furthermore, $<1$\% of the 500\,$\mu$m detections have only one other photometric point, and on average, these sources were observed in 6 bands, lowering the chances of a spurious data point significantly biasing their $L_{\rm TIR}$ estimate. Finally, due to 500\,$\mu$m observations being relatively rare ($\sim$10\% of our combined sample), even if most sources with spurious 500\,$\mu$m matches enter our depth-matched SFG sample ($\sim$100 sources), they are highly unlikely to distort our IRRC statistics in any meaningful way.

\begin{table}
\centering
\caption{Matching radii used in each band to incorporate PPSC and SPSC sources into our joint catalogue, and the resulting contamination fractions estimated using Fig. \protect\ref{fig::psc_separations}.}
\label{tab::spur_frac}
\begin{tabular}{ccc}
band & match radius & contamination \\
 & [arcsec] & [\%] \\ \hline \hline
PPSC100 & 3.5 & 4\\
PPSC160 & 4 & 6\\
SPSC250 & 4.5 & 8\\
SPSC350& 5.5 & 14\\
SPSC500 & 6.5 & 27\\
\end{tabular}
\end{table}

\section{Consistency of flux measurements from different catalogues}
\label{app::flux_consistency}

In order to improve on the sensitivity and photometric coverage of previous studies, and maximize the IRRC parameter space probed, we have drawn observations from a variety of catalogues, as described in Sect \ref{sect::dat}. The main drawback of this approach is that the resulting heterogeneous dataset involves sources detected and characterised via differing methodologies. In particular, the H-ATLAS and PPSC/SPSC catalogues were produced using different flux extraction techniques. Thus we compared these data in regions of overlapping coverage and applied corrections when necessary, as described in this Appendix. We note that these adjustments translated to only minor changes in the IRRC statistics, which remained qualitatively consistent with the results obtained using the original catalogue fluxes.

\begin{figure*}
\centering
\includegraphics[width=0.85\textwidth]{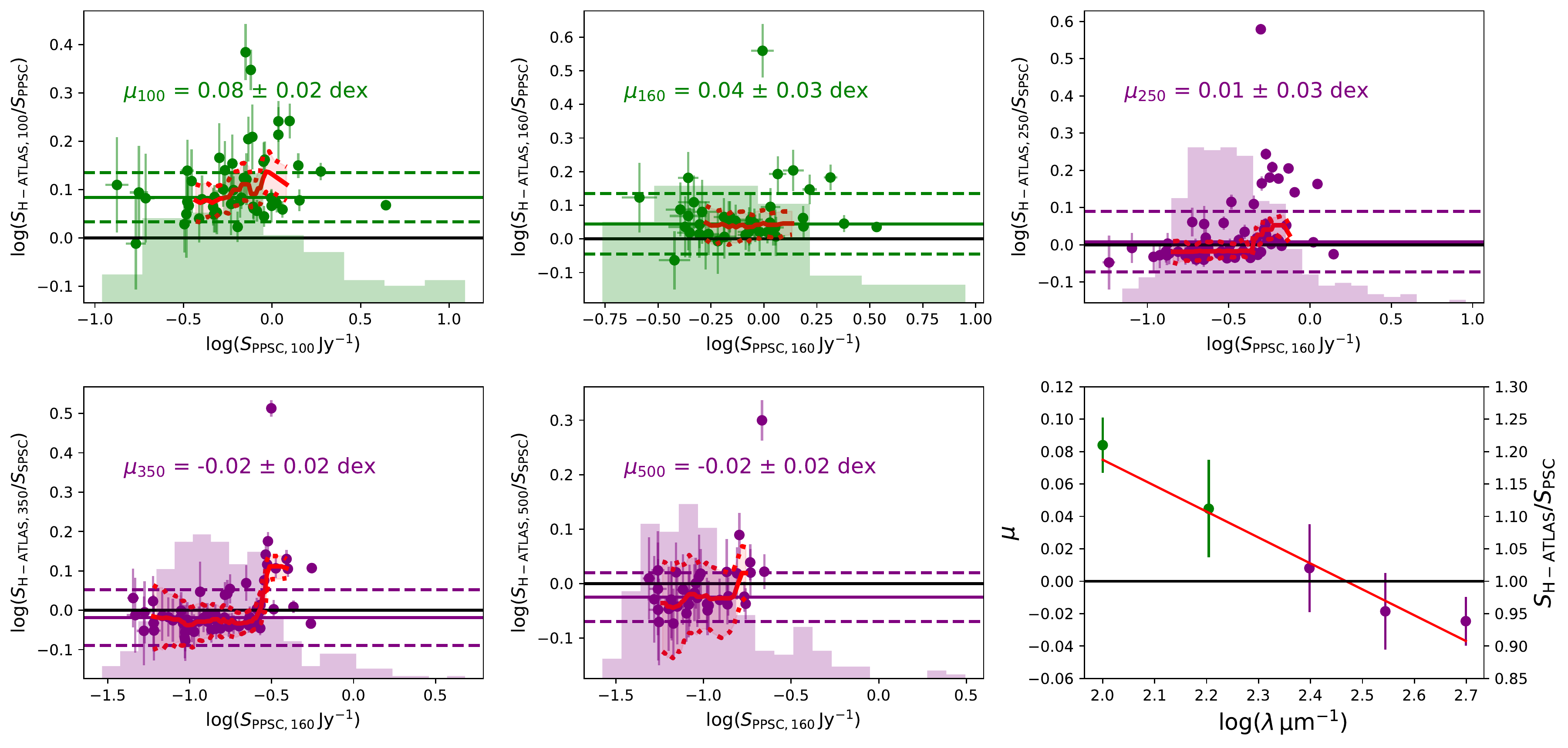}
\caption{Band-by-band flux density ratios of measurements from the PACS (green) and SPIRE (purple) Point Source Catalogs and photometry from the H-ATLAS survey as a function of H-ATLAS flux densities using star forming galaxies in our combined sample. The median logarithmic H-ATLAS to PSC flux density ratio of any given photometric band is denoted by $\mu$ in each panel, and it is represented by a green (purple) horizontal line for PACS (SPIRE) panels. Horizontal dashed lines of the same colour show the 3\,$\sigma$ uncertainties of the median value. The 1:1 relation, i.e. a logarithmic flux density ratio of 0, is shown as a black line. Red curves are the running median values. Dashed red lines indicate the 3\,$\sigma$ range around the running median. The final panel shows the median logarithmic flux ratio values $\mu$ as a function of wavelength, alongside our best-fit relation, Eq. \protect\ref{eq::off_vs_lambd}, used to derive flux corrections in the 100 and 160 $\mu$m bands.}
\label{fig::hatlas_vs_psc_fluxes}
\end{figure*}

The H-ATLAS survey identified extended sources and extracted their fluxes using apertures of appropriate sizes \citep{rigby11}. On the other hand, the PACS and SPIRE Point Source catalogues assumed, as their name indicates, appearance similar to the instrument point spread function for all sources. Indeed, a comparison of H-ATLAS and PACS Point Source fluxes\footnote{From the SPSC catalogues, we use the \textsc{timelinefitter} (TML) fluxes, which are the most accurate for point sources. PPSC fluxes were measured using the \textsc{AnnularSkyAperturePhotometry} task and apertures out to a radius of 18 and 22\,arcsec for the 100\,$\mu$m and 160\,$\mu$m bands, respectively.} using our depth-matched SFG sample, as shown in Fig \ref{fig::hatlas_vs_psc_fluxes}, reveals some 10-20 \% average excess flux in H-ATLAS measurements at PACS wavelengths. Below $z\,{<}$\,0.2, many galaxies are expected to be resolved by the $\sim$10\,arcsec Herschel PACS beam, suggesting that in our set of galaxies flux extraction with the point source assumption misses some emission at 100 and 160\,$\mu$m wavelengths\footnote{On the other hand, we have removed the likely most inaccurate SPSC measurements by omittig sources with the e.g. the blend-flag and large galaxy flag from \citet{jarrett2003}.}. On average, the lower resolution SPIRE measurements appear consistent between H-ATLAS and the SPSC within 1 and 2\,$\sigma$ below and above 400\,$\mu$m, respectively.

Due to their treatment of extended emission, we considered H-ATLAS measurements the gold standard for our Herschel data, and as the simplest approach to mitigate the differences between the two catalogues, we scaled up our PACS Point Source Catalog measurements to match the average flux levels of H-ATLAS. To calculate the scaling values, we fitted the band-by-band median log-space offsets $\mu$ as a function of central wavelength for all five Herschel bands with a linear model:

\begin{equation}
\label{eq::off_vs_lambd}
    \mu = 0.16 \cdot \log(\lambda / \mu \mathrm{m}) + 0.39.
\end{equation}

\noindent
Substituting 100 and 160 $\mu$m into this equation yields a correction factor (in linear space) of 1.19 and 1.10 for fluxes measured in the corresponding band. These were applied to all PPSC fluxes in our catalogue. Since the measured flux offsets for SPIRE data were consistent with 0 within 1-2\,$\sigma$, we did not adjust SPSC measurements. However, to retain information on the uncertainties of the PSC correction/scaling factors (both for PACS and SPIRE) in our SED modelling, we added the error on the median in each band in quadrature to the tabulated PSC flux uncertainties. These were 50, 80, 60, 50, and 30\,mJy in the 100, 160, 250, 350, 500\,$\mu$m bands, respectively. These are a factor of 2--4 larger than the typical 1\,$\sigma$ flux uncertainties of PSC measurements (see Table \ref{tab::sensitivity}), and they thus dominate the error budget of PSC data used for SED modelling.

A more detailed inspection of Fig. \ref{fig::hatlas_vs_psc_fluxes} shows that despite the generally weak dependence of flux density ratios on the intrinsic flux density (traced by H-ATLAS measurements), in the highest H-ATLAS flux regime there is an upturn of flux density ratios in almost all bands, possibly related to aperture effects. This suggests that a flux density dependent correction factor for the brightest sources could provide a more accurate correction for PPSC and HPSC flux densities than a simple scalar multiplication. To asses the impact of the inconsistency between H-ATLAS and PSC fluxes, as well as the effect of our scalar correction, we obtained $L_{\mathrm{TIR}}$ estimates via SED fitting using SPSC and PPSC fluxes with and  without our scaling applied and using only H-ATLAS photometry. We found that all three are consistent with one another within 10\%, which is well within the typical 20-30 \% uncertainties of our SED models. Ultimately, a homogeneous flux measurement approach would be preferable, but according to our assessment above the impact of the combining fluxes from different catalogues constructed with different methods in practice is not large Therefore, considering that the impact of the combining fluxes from different catalogues constructed with different methods is in practice not large, and the fact that 65 -- 80 \% of all PSC measurements lie in the constant flux density ratio regime (as evidenced by the  histograms in PPSC and HPSC sources in Fig. \ref{fig::hatlas_vs_psc_fluxes}), for the sake of simplicity we decided to apply the correction to PPSC flux densities in the form of a single value, as described above.

\section{Public data release}
\label{app::pr_descr}
In order to aid future studies on the dependence of the IRRC of low-$z$ galaxies on various galaxy properties, we make our data publicly available. Tabs. \ref{tab::PR_examp1} and \ref{tab::PR_examp2} show a small section of our data release to illustrate its content, while the online supplementary material, as well as the Zenodo data repository (at \url{https://zenodo.org/badge/latestdoi/344211798}), contain our entire combined sample of the jointly IR- and radio-detected galaxies. Its content is as follows:

\bigbreak

\noindent Column (1): SDSS ObjID -- Galaxy IDs from SDSS DR12 (corresponding to the SDSS spectroscopic ID when available, otherwise it is the SDSS photometric ID; see SpecFlag in column 5).
\bigbreak
\noindent Column (2): R.A. [J2000] -- Right ascension in degrees from SDSS DR12.
\bigbreak 
\noindent Column (3): Dec. [J2000] -- Declination in degrees from SDSS DR12.
\bigbreak
\noindent Column (4): z -- spectroscopic or photometric redshift from SDSS DR12, see column (5) detected.
\noindent Column (5): Specflag -- Set to 1 if a source was spectroscopically detected, otherwise 0.
\bigbreak
\noindent Column (6): BPTflag -- Flag values of 0, 1 and 2 denote SFG, composite and AGN sources, respectively, as described in Sect. \ref{sect::bpt}. A value of -1 indicates the lack of high quality spectral line detection, or have no $H_{\alpha}$ detections at the SNR $>$ 3 level in the MPA/JHU value-added catalogue, and consequently in SDSS DR 8.\\We note that our redshifts were drawn from SDSS DR 12, therefore a flag of 0 in column (5) in a handful of cases does not correspond to a flag of -1 in this column and vice versa.
\bigbreak
\noindent Column (7): MIRflag -- Set to 1 if a source was labelled as an AGN based on its MIR colours (for details see Sect. \ref{sect::mir_agn}), otherwise 0.\\We caution users against including sources with MIR flag $=$1 in their analyses, since our SED templates assumed pure SF across the entire IR wavelength regime. As a result, their $\log(L_{\rm TIR})$ (and by extension, $q_{\rm TIR}$) values are likely overestimated. We note that galaxies labelled as AGN based on optical emission lines but not on their MIR colours are still considered to have robust $\log(L_{\rm TIR})$ estimates.
\bigbreak
\noindent Column (8): FIRSTflag -- Set to 1 for galaxies in covered by the FIRST survey, otherwise 0. It is required to reproduce our depth-matched sample (Sect. \ref{sect::fluxmatch_samp}).
\bigbreak
\noindent Column (9): SEDflag -- Set to 1 for galaxies with robust IR SED models, otherwise 0.
\bigbreak
\noindent Column (10 - 37): $S_{\rm band}$ and $E_{\rm band}$ -- flux densities and their errors as tabulated in the various archival catalogues used to assemble our data (see Sect \ref{sect::dat}) in Jy. These bands are $22\,\mu$m WISE, $60\,\mu$m IRAS, $100\,\mu$m IRAS, $100\,\mu$m H-ATLAS, $100\,\mu$m PPSC, $160\,\mu$m H-ATLAS, $160\,\mu$m PPSC, $250\,\mu$m H-ATLAS, $250\,\mu$m SPSC, $350\,\mu$m H-ATLAS, $350\,\mu$m SPSC, $500\,\mu$m H-ATLAS, $500\,\mu$m SPSC, $1.4$GHz flux density (either NVSS or FIRST, see Sect. \ref{sect::rad_data}), respectively. We note that the IR flux values published here are \textit{not} re-scaled/corrected\footnote{Before SED fitting we carried out the following modifications: (i) we multiplied our $100$ and $160\,\mu$m PPSC data by 1.19 and 1.10, respectively; (ii) we increased the uncertainty of all PPSC and HPSC measurements by adding 0.05, 0.08, 0.06, 0.05 and 0.03 Jy in quadrature to the tabulated flux density uncertainties of 100, 160, 250, 350 and 500 $\mu$m data.}
\bigbreak
\noindent Column (38): logL1.4 -- 1.4\,GHz radio luminosity in units of $\log(\mathrm{W Hz^{-1}})$.
\bigbreak
\noindent Column (39): logdL1.4 -- 1.4\,GHz radio luminosity uncertainty.
\bigbreak
\noindent Column (40): logLTIR -- Total logarithmic IR luminosity from IR SED fitting (see Sect. \ref{sect::sed_fits}) in units of $\log(L_{\odot})$.
\bigbreak
\noindent Column (41): logLFIR -- Logarithmic far-IR luminosity from IR SED fitting (see Sect. \ref{sect::sed_fits}) in units of $\log(L_{\odot})$.
\bigbreak
\noindent Column (42): logdLIR\_upp -- Upper $\log(L_{\rm TIR})$ luminosity uncertainty corresponding to the difference of the 84th percentile and the median of the marginalized $\log(L_{\rm TIR})$ posterior distributions from our MCMC fits. We note that the errors on $\log(L_{\rm FIR})$ and $\log(L_{\rm TIR})$ are considered to be equal.
\bigbreak
\noindent Column (43): logdLIR\_low -- Lower $\log(L_{\rm TIR})$ luminosity uncertainty corresponding to the difference of the median and the 16th percentile of the marginalized $\log(L_{\rm TIR})$ posterior distributions from our MCMC fits. We note that the errors on $\log(L_{\rm FIR})$ and $\log(L_{\rm TIR})$ are considered to be equal.
\bigbreak
\noindent Column (44): qTIR -- $q_{\rm TIR}$ calculated with Columns (38) and (40).
\bigbreak
\noindent Column (45): qFIR -- $q_{\rm FIR}$ calculated with Columns (38) and (41). By definition $q_{\rm FIR}$ is lower than $q_{\rm TIR}$ by $\log(L_{\rm TIR}) - \log(L_{\rm FIR})$.
\bigbreak
\noindent Column (46): dqTIR\_upp -- Upper error on $q_{\rm TIR}$, calculated by propagating the uncertainties in Columns (39) and (42)).
\bigbreak
\noindent Column (47): dqTIR\_low -- Lower error on $q_{\rm TIR}$, calculated by propagating the uncertainties in Columns (39) and (43).

\bigbreak

To select our depth-matched SFGs with reliable SED models one has to select sources with $\mathrm{FIRSTflag} = 1$, $\mathrm{BPTflag} = 0$, $\mathrm{MIRflag} = 0$ and $\mathrm{SEDflag} = 1$ and $S22_{\rm WISE} > 0.0195$ Jy.

\begin{table*}
\scriptsize
\caption{First 13 columns of our publicly available low-z IRRC catalogue. The full catalogue, containing all 9,645 sources in our combined sample, is available as online supplementary material. For a description of all columns, see Appendix \ref{app::pr_descr}.}
\begin{tabular}{cccccccccccccc}
\hline
ObjID & RA & Dec & z & Specflag & BPTflag & MIRflag & FIRSTflag & SEDflag & $S22_{\mathrm{unWISE}}$ & $E22_{\mathrm{unWISE}}$ & $S60_{\mathrm{IRAS}}$ & $E60_{\mathrm{IRAS}}$ & ... \\
- & [deg] & [deg] & - & - & - & - & - & - & [Jy] & [Jy] & [Jy] & [Jy] & ...\\ \hline
1237653651309002970 & 25.76 & 13.65 & 0.0029 & 1 & 2 & 0 & 0 & 0 & 0.1931 & 0.0006 & 69.05 & 0.05 & ...\\
1237648720159965190 & 186.73 & -0.88 & 0.0071 & 1 & 2 & 1 & 1 & 0 & 12.744 & 0.002 & 40.68 & 0.041 & ... \\
1237657628439543935 & 135.11 & 39.06 & 0.0582 & 1 & 1 & 1 & 1 & 0 & 1.7040 & 0.0007 & 7.43 & 0.04 & ... \\
1237661815485104153 & 189.14 & 11.24 & 0.0074 & 1 & 0 & 0 & 1 & 0 & 0.2148 & 0.0008 &  &  & ... \\
1237664853107933399 & 217.02 & 32.89 & 0.1829 & 0 &  & 0 & 1 & 1 & 0.0045 & 0.0005 &  &  & ... \\
1237662225149722677 & 209.64 & 37.45 & 0.0115 & 0 & 0 & 0 & 1 & 1 & 1.0615 & 0.0009 &  &  & ... \\
1237664336632021061 & 149.17 & 30.75 & 0.0526 & 1 & 0 & 0 & 1 & 1 & 0.0276 & 0.0009 & 0.32 & 0.04 & ... \\
1237666245748064333 & 314.35 & 17.13 & 0.0883 & 0 &  & 1 & 0 & 1 & 1.178 & 0.001 &  &  & ... \\
\label{tab::PR_examp1}
\end{tabular}
\end{table*}

\begin{table*}
\scriptsize
\caption{The last 10 columns of our publicly available low-z IRRC catalogue. The full catalogue, containing all 9,645 sources in our combined sample, is available as online supplementary material. For a description of all columns, see Appendix \ref{app::pr_descr}.}
\begin{tabular}{ccccccccccc}
\hline
... & $\log(L_{\rm 1.4})$ & $\log(\Delta L_{\rm 1.4})$ & $\log(L_{\rm TIR})$ & $\log(L_{\rm FIR})$ & $\log(\Delta L_{\rm TIR,upp})$ & $\log(\Delta L_{\rm TIR,low})$ &  $\log(q_{\rm TIR})$ & $\log(q_{\rm FIR})$ & $\log(\Delta q_{\rm TIR,upp})$ & $\log(\Delta L_{\rm TIR,low})$ \\

... & [$\log(\mathrm{W Hz^{-1}})$] & [$\log(\mathrm{W Hz^{-1}})$] & [$\log(L_{\odot})$] & [$\log(L_{\odot})$] & [$\log(L_{\odot})$] & [$\log(L_{\odot})$] &- & - & - & - \\  \hline

... & 21.83 & 0.02 & 10.5106 & 10.2103 & 0.0003 & 0.0003 & 2.69 & 2.39 & 0.02 & 0.02 \\
... & 21.66 & 0.01 & 11.19857 & 10.91486 & 0.00001 & 0.00001 & 3.55 & 3.27 & 0.01 & 0.01 \\
... & 22.54 & 0.05 & 12.1011 & 11.8174 & 0.0001 & 0.0001 & 3.57 & 3.29 & 0.04 & 0.04 \\
... & 22.184 & 0.02 & 10.7 & 10.5 & 0.7 & 0.3 & 2.6 & 2.3 & 0.3 & 0.2 \\
... & 22.89 & 0.06 & 11.08 & 10.76 & 0.09 & 0.09 & 2.20 & 1.88 & 0.07 & 0.07 \\
... & 22.01 & 0.01 & 10.7724 & 10.5129 & 0.0002 & 0.0002 & 2.77 & 2.51 & 0.01 & 0.01 \\
... & 22.19 & 0.08 & 10.80 & 10.49 & 0.09 & 0.11 & 2.63 & 2.31 & 0.09 & 0.09 \\
... & 23.91 & 0.02 & 12.567 & 12.307 & 0.001 & 0.001 & 2.66 & 2.40 & 0.02 & 0.02 \\
\label{tab::PR_examp2}
\end{tabular}
\end{table*}

%%%%%%%%%%%%%%%%%%%%%%%%%%%%%%%%%%%%%%%%%%%%%%%%%%

% Don't change these lines
\bsp	% typesetting comment
\label{lastpage}
\end{document}